\documentclass[a4paper,11pt]{article}
\usepackage{heppub}
\pdfoutput=1
\usepackage{braket}
\usepackage{xspace}
\usepackage{placeins}
\usepackage{wasysym}
\usepackage{slashed}
\usepackage{empheq}
\usepackage{xcolor}
\usepackage[mathscr]{eucal}  

\newcommand{\abs}[1]{\lvert#1\rvert}

\newcommand{\ord}[1]{\mathcal{O}(#1)}
\newcommand{\Ord}[1]{\mathcal{O}\bigl(#1\bigr)}
\newcommand{\ORd}[1]{\mathcal{O}\Bigl(#1\Bigr)}
\newcommand{\ordsq}[1]{\mathcal{O}[#1]}
\newcommand{\Ordsq}[1]{\mathcal{O}\bigl[#1\bigr]}
\newcommand{\ORdsq}[1]{\mathcal{O}\Bigl[#1\Bigr]}

\newcommand{\df}{\mathrm{d}}
\newcommand{\img}{\mathrm{i}}
\newcommand{\eps}{\epsilon}

\newcommand{\nn}{\nonumber}

\newcommand{\MeV}{\,\mathrm{MeV}}
\newcommand{\GeV}{\,\mathrm{GeV}}
\newcommand{\TeV}{\,\mathrm{TeV}}

\newcommand{\bt}{{\vec b}_T}

\newcommand{\kt}{{\vec k}_T}
\newcommand{\btcut}{b_T^\mathrm{cut}}
\newcommand{\btope}{b_T^\mathrm{OPE}}
\newcommand{\ktcut}{k_T^\mathrm{cut}}

\newcommand{\C}{\Lambda}
\newcommand{\laa}{\lambda_1}
\newcommand{\lbb}{\lambda_2}

\newcommand{\cusp}{\mathrm{cusp}}
\newcommand{\cut}{\mathrm{cut}}
\newcommand{\NP}{\mathrm{NP}}

\newcommand{\as}{\alpha_s}

\newcommand{\bmax}{b_\mathrm{max}}
\newcommand{\lqcd}{\Lambda_\mathrm{QCD}}
\newcommand{\MSbar}{$\overline{\text{MS}}$\xspace}
\newcommand{\Ecm}{E_\mathrm{cm}}

\newcommand{\WidthTwoSubfigs}{0.48\textwidth}

\allowdisplaybreaks[3]

\title{\boldmath Disentangling Long and Short Distances \\in Momentum-Space TMDs}

\author[a]{Markus A.~Ebert,\hspace{-0.2ex}}
\emailAdd{ebert@mpp.mpg.de}

\author[b]{Johannes K.~L.~Michel,\hspace{-0.2ex}}
\emailAdd{jklmich@mit.edu}

\author[b]{Iain W.~Stewart,\hspace{-0.2ex}}
\emailAdd{iains@mit.edu}

\author[b]{and Zhiquan Sun}
\emailAdd{zqsun@mit.edu\!}

\affiliation[a]{Max Planck Institut f\"ur Physik, F\"ohringer Ring 6, 80805 Munich, Germany}
\affiliation[b]{Center for Theoretical Physics,\,Massachusetts Institute of Technology,\,Cambridge,\,MA\,02139,\,USA}

\abstract{%
The extraction of nonperturbative TMD physics is made challenging
by prescriptions that shield the Landau pole,
which entangle long- and short-distance contributions in momentum space.
The use of different prescriptions then makes the comparison of fit results
for underlying nonperturbative contributions not meaningful on their own.
We propose a model-independent method to restrict momentum-space observables to the perturbative domain. 
This method is based on a set of integral functionals
that act linearly on terms in the conventional position-space operator product expansion (OPE).
Artifacts from the truncation of the integral can be systematically pushed to higher powers in $\Lambda_{\rm QCD}/k_T$.
We demonstrate that this method can be used to compute the cumulative integral
of TMD PDFs over $k_T \le k_T^\mathrm{cut}$ in terms of collinear PDFs,
accounting for both radiative corrections and evolution effects.
This yields a systematic way of correcting the naive picture where the TMD PDF integrates to a collinear PDF,
and for unpolarized quark distributions we find that when renormalization scales are chosen near $k_T^\mathrm{cut}$,
such corrections are a percent-level effect.
We also show that, when supplemented with experimental data and improved perturbative inputs,
our integral functionals will enable model-independent limits to be put on the nonperturbative OPE contributions to the Collins-Soper kernel and intrinsic TMD distributions.
}

\date{January 18, 2022}

\preprint{\vbox{%
\hbox{MIT--CTP 5391}
\hbox{MPP--2021--217}
}}

\begin{document}

\maketitle

\section{Introduction}
\label{sec:intro}

Transverse momentum-dependent parton distribution functions (TMD PDFs)
that encode the longitudinal and transverse partonic structure of hadrons
enter factorization theorems for many physical processes at hadron colliders in a universal way.
Their knowledge is essential to the ongoing research program at the Large Hadron Collider (LHC)
searching for hints of new physics in precision tests of the Standard Model,
see e.g.\ \refscite{Aad:2011fp, Aad:2014xaa, Aad:2015auj, Aaboud:2017svj, Aaboud:2017ffb, Aad:2019wmn, Chatrchyan:2011wt, Khachatryan:2015oaa, Khachatryan:2016nbe, Sirunyan:2017igm, Sirunyan:2019bzr, LHCb:2016fbk, LHCb:2021bjt},
but TMD PDFs are also of great interest on their own as they provide insight into the strongly coupled dynamics of QCD that bind the proton together~\cite{Accardi:2012qut,Gao:2018grv,Liu:2019ntk,AbdulKhalek:2021gbh}.

The momentum-space TMD PDF $f_i(x, \kt)$ encodes 
the distribution of parton $i$ in an energetic hadron
at a longitudinal momentum fraction $x$ and transverse momentum $\kt$.
It is also often helpful to consider its description in the Fourier-conjugate position space,  $f_i(x,\bt)$.
A common challenge in the extraction and interpretation of TMD PDFs
is that the most intuitive way to think of them is as a distribution in $\kt$ space,
while the evolution of the TMD PDFs and their nonperturbative structure
is mathematically most transparent in $\bt$ space.
An additional complication is that physical TMD cross sections in momentum space,
where experimental data is taken,
involve a convolution of at least two TMD PDFs $f_{i}(x,\kt)$ (or TMD fragmentation or jet functions)
due to momentum conservation between the two scattered partons and the color-singlet probe, see  
\refcite{Collins:2011zzd} for a detailed review of the classic TMD processes and \refscite{Kang:2017glf,Kang:2017btw,Makris:2017arq,Gutierrez-Reyes:2018qez,Gutierrez-Reyes:2019vbx,Gutierrez-Reyes:2019msa} for examples of more recent applications with jets.
This again only turns into a transparent product in position space.
Another important momentum-space quantity
is the cumulative distribution of a single TMD PDF over transverse momenta $\abs{\kt} \leq \ktcut$.
For $\ktcut$ much larger than the QCD confinement scale $\lqcd$,
the naive expectation is that this integral should include partons of ``all'' possible
transverse momenta in the hadron,
and therefore recover the corresponding longitudinal parton distribution function.
Interestingly, the formal theoretical status of this relation is unclear to date,
and clarifying it is a major goal of this paper.
We review the current status of the relation in \sec{review_tmd_pdf_and_normalization}.

\begin{figure}
\centering
\includegraphics[width=\WidthTwoSubfigs]{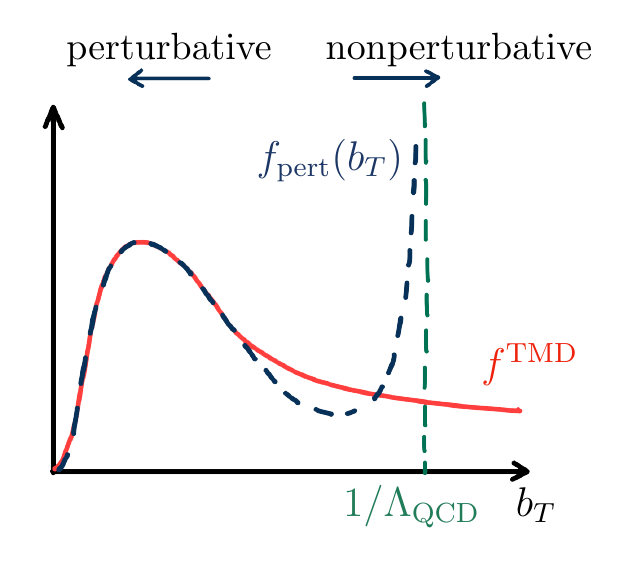}%
\hfill
\includegraphics[width=\WidthTwoSubfigs]{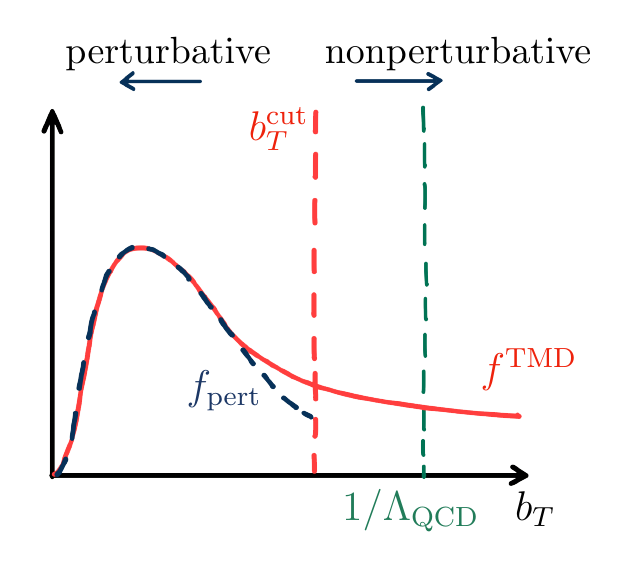}%
\caption{%
Left: A perturbative calculation of the TMD PDF (dashed blue) blows up at the Landau pole $\mu \sim 1/b_T \sim \lqcd$.
Right: In our construction in this paper, we introduce a cutoff on the Fourier integral at $\btcut$,
below which the perturbative calculation is still a good approximation of the full TMD PDF (red).
We then show that by supplementing the result with higher-order contact terms on the boundary,
a purely perturbative baseline for the Fourier transform at transverse momenta $\abs{\kt} \gg \lqcd$ can be computed
and systematically supplemented by higher-order nonperturbative corrections.
}
\label{fig:btcut_drawing}
\end{figure}

The formal advantage of a position-space description of TMD PDFs is that point by point in $b_T \equiv \abs{\bt}$,
one may formally show that it receives contributions from the running coupling $\as(\mu)$ exclusively at energy scales $\mu \gtrsim 1/b_T$.
This makes it straightforward to delineate regions in $b_T$
where the description of the transverse structure of the hadron is perturbative, $b_T \ll 1/\lqcd$,
or necessarily nonperturbative, $b_T \gtrsim 1/\lqcd$,
as illustrated in the left panel of \fig{btcut_drawing}.
The same distinction is much harder to make in momentum-space,
because at any given value of $\kt$, contributions from all scales $\mu$ are present.
A perturbative calculation of the position-space TMD PDF $f^\mathrm{pert}_i(x,\bt)$ can be described by an operator-product expansion (OPE), whose leading term involves a longitudinal momentum convolution between collinear densities $f_j(y)$ and a perturbative kernel $C_{ij}$ that is expanded in the strong coupling $\alpha_s(\mu)$, schematically $f^\mathrm{pert}_i(x,\bt) = C_{ij}\otimes f_j(x)$. (See \eq{ope_tmdpdf_leading_term} below for a complete formula.)
From \fig{btcut_drawing} it is clear that the perturbative calculation of the position-space TMD PDF $f^\mathrm{pert}_i(x,\bt)$ (dashed blue),
becomes ill-behaved in the region $1/b_T \sim \lqcd$ near the Landau pole $\mu \sim \lqcd$ of QCD.

A common way to achieve a description of the TMD PDF $f_{i/h}$ across all values of $b_T$
is to evaluate the perturbative part of the TMD PDF $f^\mathrm{pert}_{i/h}$ at some $b^*(b_T)$ instead of $b_T$~\cite{Collins:1981va},
where $b^*(b_T) \to b_\mathrm{max} < 1/\lqcd$ for $b_T \to 1/\lqcd$. 
This causes the coupling to saturate at a perturbative scale, and avoids hitting the Landau pole.  The method for implementing this cutoff is not unique, and a number of different $b^*$ prescriptions have been proposed in the literature~\cite{Collins:1981va, Bacchetta:2017gcc, Scimemi:2018xaf, Lustermans:2019plv}.
The perturbative part of the TMD PDF is then multiplied with a nonperturbative model function $F_{b^*}^\NP(b_T)$
that behaves as $1 + \ord{\lqcd^2 b_T^2}$ in the perturbative region. Thus in this setup the TMD PDF is written as
\begin{align} \label{eq:tmd_model}
 f_{i/h}(x,b_T,\mu,\zeta) &=  f_{i/h}^{\rm pert}(x,b^*(b_T),\mu,\zeta)\: F_{b^*}^\NP(b_T)
 \,, \end{align}
where $\mu$ is the renormalization scale, and $\zeta$ is the Collins-Soper scale.%
\footnote{For definiteness, in this paper we will only consider rapidity renormalization schemes
in which the hard factor $H(Q^2,\mu)$ in the factorized TMD cross section in \eq{factorized_cross_section} is renormalized in \MSbar and independent of additional parameters.
This fixes the renormalization scheme for the product of the two TMD PDFs, and fixes the scheme for individual TMDs that depend on one rapidity renormalization parameter $\zeta$, up to rescalings of $\zeta_{a,b}$ that leave $\zeta_a \zeta_b = Q^4$ invariant.  This class of equivalent TMDs includes the definition of Collins~\cite{Collins:2011zzd}, the EIS scheme~\cite{GarciaEchevarria:2011rb,Echevarria:2012js}, the CJNR scheme~\cite{Chiu:2011qc,Chiu:2012ir}, and various others, see  \refcite{Ebert:2019okf} for further details.
Our notation excludes the schemes in \refscite{Soper:1979fq, Collins:1981uk, Ji:2004wu}, which depend on an additional parameter
in the hard factor and TMD PDF. However these schemes can be perturbatively matched to the TMDs considered here, see~\cite{Ebert:quasiTMDproof} for a derivation of the  all-orders relation.
}
This result can then be Fourier transformed
to compute the desired quantity in momentum space. This procedure is used, for example, in the global fit analyses for TMDs~\cite{Scimemi:2019cmh, Bacchetta:2019sam} which fit for a number of parameters in their $F_{b^*}^\NP$ model functions.
A problem with this setup is that it does not cleanly separate the perturbative and non-perturbative contributions to the TMD. In particular the interpretation of fit results for $F_{b^*}^\NP$ intrinsically depends on the choice of the $b^*$ cutoff prescription, as we have indicated by the subscript. Thus, in  mixing up perturbative and nonperturbative effects,
$b^*$ (or similar) prescriptions make it challenging to directly compare nonperturbative model functions obtained in different fits to experimental data.

This conclusion is a bit counterintuitive. Physical intuition about the behavior of QCD and power counting
tell us that at sufficiently large transverse momenta $\abs{\kt} \gg \lqcd$
(or after integrating up to $\ktcut \gg \lqcd$),
the result for $f_i(x,\kt)$ should be dominated by the short-distance region $b_T \ll 1/\lqcd$.
In particular, deviations of $\ord{\lqcd^2 b_T^2}$ from the perturbative result should have the
scaling $\ord{\lqcd^2/\abs{\kt}^2}$ expected from dimensional analysis (or a direct momentum space OPE). However, when taking
a standard Fourier transform of $f^\mathrm{pert}_i(x,\bt)$ over all values of $\bt$,
a purely perturbative baseline is not even defined due to the presence of the Landau pole
or, equivalently, the need for a $b^*$ prescription.
A main goal of this work is to develop a method for treating momentum-space TMDs where this intuition is restored.

In this paper, we give a set of integral functionals
that only require the TMD PDF on an interval 
$0\le b_T \leq \btcut < 1/\lqcd$
as input, where the perturbative description is sufficient,
and prove that for large $\abs{\kt} \gg 1/\btcut$ these functionals asymptote to the full momentum-space spectrum.
Our construction is based on placing a hard cutoff $\btcut$ on the Fourier integral,
which makes the contributing regions in $b_T$ space completely transparent,
as illustrated in the right panel of \fig{btcut_drawing}.
We then show that the result can be systematically improved
by including higher-order contact terms on the boundary.
Importantly, by pushing the functional to higher orders in $1/(k_T \btcut)$ in this way,
the effect of nonperturbative corrections of $\ord{b_T^2}$ in $b_T$ space can be isolated,
and the correction in momentum space can be shown to be linear in the coefficient.
This provides a completely model-independent way to compute the leading effect of nonperturbative dynamics on momentum-space quantities,
while the evolution and convolution of TMD PDFs can still comfortably be performed in position space at $b_T \leq \btcut$ in our setup.
We note that a hard $b_T$-space cutoff on Fourier integrals has previously been considered in \refscite{Qiu:2000hf, Berger:2002ut}
as a check on model-specific results, and the setup there amounts to the leading term in our construction.
We also show that our construction immediately carries over to the cumulative distribution,
and allows us to produce model-independent predictions for the transverse-momentum integral of the TMD PDF.
As a byproduct of our analysis, we analyze the scaling of artifacts induced by $b^*$ prescriptions in momentum space,
including analyzing schemes whose artifacts scale with different powers or formally fall faster than any power.

Finally, we propose an application to Drell-Yan transverse-momentum ($q_T$) spectra measured at the LHC.
We are motivated here by the fact that the most recent measurements by CMS and ATLAS~\cite{CMS:2019raw, ATLAS:2019zci}
have reached sub-percent precision, making them far more precise than other data sets sensitive to TMD physics.
Model-based global TMD PDF fits typically only consider the subdominant experimental uncertainty when computing goodness-of-fit tests, which carries the risk of overconstraining certain flavor combinations and $x$ regions.
At the same time, the LHC data sets are strongly dominated
by perturbative physics $\lqcd \ll q_T \ll Q$.
For this reason it is desirable to perform a model-independent fit to LHC data
that explicitly incorporates this hierarchy by restricting to the leading nonperturbative effect,
which is made possible by our setup.
This approach is complementary to model-based fits.
As an immediate physics goal, the fit we propose would directly constrain the $\ord{b_T^2}$ coefficient
in the universal Collins-Soper kernel that governs the evolution of TMD PDFs with energy.
Existing model-dependent fits to data~\cite{Scimemi:2019cmh, Bacchetta:2019sam} and extractions on the lattice~\cite{Shanahan:2019zcq,Shanahan:2020zxr,Schlemmer:2021aij,LatticeParton:2020uhz,Li:2021wvl,Shanahan:2021tst}
both prefer a surprisingly small value for this coefficient,
and it would be valuable to confront the extremely precise LHC data with this observation in a model-independent way.

This paper is structured as follows:
In \sec{review_tmd_pdf_and_normalization},
we review the status of the relation between TMD and longitudinal PDFs,
and point out why considering the cumulative distribution for the TMD PDF is helpful.
In \sec{truncated_functionals} we introduce the truncated integral functionals
that allow us to compute Fourier transforms in terms of test function data on $0 \leq b_T \leq \btcut$,
both for transverse-momentum spectra and cumulative distributions.
We also discuss the consequences of our construction for $b^*$ prescriptions.
In \sec{single_tmdpdf}, we apply our setup to the cumulative distribution of a single (unpolarized) TMD PDF
and compare the result to the one-dimensional longitudinal PDF.
In \sec{data} we outline the proposed application of our setup to Drell-Yan transverse-momentum spectra at the LHC.
We conclude in \sec{conclusions}.

\subsection{Review of TMD PDFs and their cumulative integral}
\label{sec:review_tmd_pdf_and_normalization}

An interesting question is how a TMD PDF $f_i(x, \kt)$, 
which encodes information of both the longitudinal momentum fraction $x$ 
\emph{and} the transverse momentum $\kt$,
relates to the corresponding collinear PDF $f_i(x)$,
which only encodes the longitudinal information.
For definiteness, we will consider $f_i$ to be the distribution
of unpolarized partons within an unpolarized hadron or nucleon in this paper.
The extension of this review and of our results to polarized distributions is straightforward,
and we return to it in \sec{conclusions}.
In a naive interpretation of both distributions as probability densities, one might expect
\begin{align} \label{eq:tmd_pdf_integral_naive_expectation}
\int \! \df^2 \kt \, f_i(x, \kt) \stackrel{?}{=} f_i(x)
\,,\end{align}
i.e., the integral of the more differential TMD PDF should exactly recover the one-dimensional distribution.
Using the standard identity $\int \! \df^2 \kt \, e^{\img \bt \cdot \kt} = (2\pi)^2 \delta(\bt)$,
one can equivalently ask about the value of the Fourier transform $f_i(x, b_T)$ of the TMD PDF at $b_T = 0$,
\begin{align} \label{eq:tmd_pdf_integral_naive_expectation_fourier_space}
\int \! \df^2 \kt \, f_i(x, \kt)
= f_i(x, b_T = 0) \stackrel{?}{=} f_i(x)
\,,\end{align}
where $\bt$ is the position-space variable conjugate to $\kt$ and $b_T = \abs{\bt}$.%
\footnote{We will continue to use the same symbol $f_i$ for the unpolarized TMD PDF, its Fourier transform, and the unpolarized collinear PDF in the following as the meaning will always be clear from the arguments or the context.
}

Indeed, it is easy to work out from their operator definitions~\cite{tmd_handbook} 
that \eq{tmd_pdf_integral_naive_expectation_fourier_space} is formally satisfied at the level of the bare (unrenormalized) TMD and collinear PDF,
\begin{align} \label{eq:tmd_pdf_integral_bare_fourier_space}
f_i^\mathrm{bare}(x, b_T = 0, \eps, \zeta) = f_i^\mathrm{bare}(x, \eps)
\,,\end{align}
where dimensional regularization $d = 4 - 2\eps$ is used to treat ultra-violet (UV) divergences,
and $\zeta$ denotes the so-called Collins-Soper scale, which appears due to the need to regulate and subtract rapidity divergences in the construction of the TMD PDF.
Specifically, the bare position-space TMD PDF operator is defined as a correlator of parton fields
that are separated by some lightlike (longitudinal) distance and by $\bt$ in the transverse direction,
and that are connected to each other by a staple-shaped Wilson line extending to lightlike infinity.
For $b_T = 0$, only the lightlike separation remains
and the Wilson line collapses onto a single straight path between the two fields,
which precisely recovers the operator definition of the collinear PDF.
(Similarly, an additional soft component in the definition of the TMD PDF collapses to unity for $b_T \to 0$,
and the dependence on the rapidity regulator drops out.)
\Eq{tmd_pdf_integral_bare_fourier_space} has been explicitly verified
to hold in momentum space at one-loop level in \refcite{Echevarria:2011epo}
by performing the $\kt$ integral in $2 - 2\eps$ dimensions.

However, after renormalization, the naive result \eq{tmd_pdf_integral_naive_expectation} must be broken,
\begin{align} \label{eq:breaking_TMD_PDF_relation}
\int\df^2\kt \, f_i(x, \kt, \mu , \zeta)
\neq f_i(x, \mu)
\,,\end{align}
where $\mu$ is the \MSbar scale.
This is immediately clear from comparing the renormalization group evolution
of the TMD and collinear PDF with respect to $\mu$,
\begin{subequations} \label{eq:mu_rge_pdf_tmdpdf}
\begin{align} \label{eq:mu_rge_tmdpdf}
\mu \frac{\df}{\df\mu} f_i(x, \kt, \mu , \zeta)
&= \gamma_\mu^i(\mu, \zeta) \, f_i(x, \kt, \mu , \zeta)
\\ \label{mu_rge_pdf}
\mu \frac{\df}{\df\mu} f_i(x, \mu)
&= \sum_j \int_x^1 \frac{\df z}{z} P_{ij}(z, \mu)  f_j \Bigl(\frac{x}{z}, \mu\Bigr)
\,.\end{align}
\end{subequations}
Importantly, the $\mu$ evolution of the TMD PDF is diagonal both in flavor $i$
and momentum fraction $x$, while for the PDF it sums over all flavors $j$ and
involves a convolution in $x$. Even more strikingly, only the TMD PDF
depends on the CS scale $\zeta$ through the Collins-Soper kernel $\gamma_\zeta^i$,
also known as the rapidity anomalous dimension,
\begin{align} \label{eq:zeta_rge_tmdpdf}
\zeta \frac{\df}{\df\zeta} f_i(x, b_T, \mu , \zeta) &
= \frac{1}{2} \gamma_\zeta^i(b_T, \mu) \, f_i(x, b_T, \mu , \zeta)
\end{align}
while the PDF is independent of $\zeta$.
More generally, the two-dimensional structure of the renormalization in the TMD case,
with the two directions tied together by a closure condition involving the cusp anomalous dimension $\Gamma_\cusp^i$,
\begin{align} \label{eq:rge_rapidity_anom_dim}
\mu \frac{\df}{\df\mu} \gamma_\zeta^i(b_T, \mu)
&= 2\zeta \frac{\df}{\df \zeta} \gamma^i_\mu(\mu, \zeta) = - 2\Gamma_\cusp^i[\as(\mu)]
\,,\end{align}
has no analog in the collinear case.
Clearly, these observations forbid a simple relation between the TMD and collinear PDF at the renormalized level,
and thus break \eq{tmd_pdf_integral_naive_expectation} in general.
Only the renormalized distributions can be extracted from measurements in $d = 4$ dimensions,
so this is unsatisfactory.

At distances $b_T \ll 1/\lqcd$ that are short compared to the QCD confinement scale $\lqcd$,
another relation between the position-space TMD and collinear PDFs exists.
In this regime, TMD PDFs obey an operator product expansion in terms of local operators at $b_T = 0$, i.e., in terms of collinear PDFs~\cite{Collins:1981uw},
\begin{align} \label{eq:ope_tmdpdf_leading_term}
f_i(x, b_T, \mu , \zeta)
= \sum_j \int_x^1 \frac{\df z}{z} C_{ij}\bigl(z, b_T, \mu, \zeta\bigr) \, f_j \Bigl(\frac{x}{z}, \mu\Bigr)
+ \Ord{\lqcd^2 b_T^2}
\,.\end{align}
Here the matching coefficients $C_{ij}$ carry the dependence
on the short-distance physics at the scale $\mu \sim 1/b_T$.
They can be computed perturbatively in terms of $\as(\mu) \ll 1$
and precisely absorb the mismatch in renormalization to the working order in $\lqcd$.

Because the total integral of the TMD PDF over all $\kt$ is a UV quantity
that one expects to be dominated by large, perturbative momenta,
one may be tempted to evaluate \eq{ope_tmdpdf_leading_term} at $b_T = 0$
to relate the integral to the collinear PDF.
In this case, one option is to take $b_T \to 0$ while holding the renormalization scale $\mu$ fixed.
This expresses the TMD PDF in terms of the PDF at the scale $\mu$
and perturbative corrections starting at $\ordsq{\as(\mu)}$.
However, the perturbative corrections contain
logarithms, $\ln(b_T \mu)$,  that prohibit setting $b_T = 0$ to compute the total integral.
Alternatively, one can evaluate the TMD PDF at $\mu_0 \sim 1/b_T$
and use the evolution equations in \eqs{mu_rge_tmdpdf}{zeta_rge_tmdpdf} to evolve it back to the overall $\mu$.
In this case the perturbative corrections at the scale $\as(\mu_0)$ actually vanish for $b_T \to 0$
due to asymptotic freedom.
However, this produces a relation with the PDF at asymptotically large scales $\mu_0 \to \infty$,
which is hard to interpret.

A physically intuitive way of dealing with these issues,
which also produces a practically useful result at finite scales,
is to instead consider the cumulative TMD PDF in momentum space up to some finite cutoff,
i.e., the distribution integrated over all $\kt$ with $\abs{\kt} \leq \ktcut$,
where we can think of $\ktcut$ as an additional UV cutoff for high-energy modes.
This quantity is straightforward to compute in terms of the position-space TMD PDF,
\begin{align} \label{eq:cumulative_tmdpdf_in_terms_of_bT_space}
\int_{\abs{\kt} \leq \ktcut} \! \df^2 \kt \, f_i(x, \kt, \mu, \zeta)
&=\int_{\abs{\kt} \leq \ktcut} \! \df^2 \kt \, \int \frac{\df^2 \, \bt}{(2\pi)^2} \, e^{+i \kt \cdot \bt} f_i(x, b_T, \mu, \zeta)
\nn \\
&= \int_0^{\ktcut} \! \df k_T \, k_T \int_0^\infty \! \df b_T \, b_T J_0(b_T k_T) \, f_i(x, b_T, \mu, \zeta)
\nn \\
&=\ktcut \int_0^\infty \! \df b_T J_1(b_T \ktcut) \, f_i(x, b_T, \mu, \zeta)
\,.\end{align}
Here we inserted the Fourier transform on the first line,
performed the trivial integral over the angle of $\bt$ on the second line,
resulting in a zeroth-order Bessel function $J_0$ of the first kind.
On the third line we interchanged the order of integration
to perform the $k_T$ integral, resulting in the first-order Bessel function.

\Eq{cumulative_tmdpdf_in_terms_of_bT_space} is a typical example
of a momentum-space quantity that receives contributions from all $b_T$,
including a long-distance contribution from $b_T \gtrsim 1/\lqcd$.
Unlike the full integral over all $k_T$,
it depends on a physical scale $\ktcut$,
which can guide the choice of the overall scales $\mu, \zeta$
that the TMD PDF is evolved to,
and the behavior of the cumulative distribution for large $\ktcut$ can be studied explicitly.
In \refcite{Bacchetta:2013pqa}, cumulative distributions of the unpolarized, helicity, and transversity TMD PDFs at $\mu = \ktcut$
were studied for a specific choice of nonperturbative model at long distances (with perturbative results at next-to-leading order),
and found to be consistent with their collinear counterparts.
A primary goal of this paper, which originally motivated developing the general formalism we present in \sec{truncated_functionals},
is to compute the cumulative distribution in \eq{cumulative_tmdpdf_in_terms_of_bT_space} at perturbative $\ktcut \gg \lqcd$
in a model-independent way in terms of the leading OPE contribution in $b_T$ space and quantifiable corrections.

\section{Truncated functionals}
\label{sec:truncated_functionals}

Motivated by the physical applications discussed in \sec{intro}, in sections~\ref{sec:truncated_functionals_setup_and_assumptions}--\ref{sec:cumulative_functionals} we setup a mathematical formalism for systematically obtaining momentum space distributions from an underlying Fourier distribution that obeys an OPE. We then consider the relation to $b^*$ prescriptions in \sec{bstar_prescriptions} and the nature of an asymptotic series which arises in the construction in \sec{stability_near_landau_pole}.

\subsection{Setup and assumptions}
\label{sec:truncated_functionals_setup_and_assumptions}

We consider an integral functional
\begin{align} \label{eq:S_functional}
S[f](k_T) = k_T \int_0^\infty \! \df b_T \, b_T J_0(k_T b_T) \, f(b_T)
\,,\end{align}
for an arbitrary function $f$ that satisfies the following two assumptions:
\begin{align} \label{eq:S_assumptions}
&(1) \quad f(b_T) ~ \text{is continuously differentiable for all}~b_T > 0
\,, \nn \\
&(2)\quad \abs{f(b_T\to \infty)} \lesssim b_T^{-\rho}
~\text{and}~
\abs{f'(b_T\to \infty)} \lesssim b_T^{-\rho-1},
\quad \rho > \frac{1}{2}
\,.\end{align}
Here we take the $\lesssim$ sign to mean
that there exists a constant $K > 0$ and an exponent $\rho > 1/2$
such that the bound $\abs{f(b_T)} < K b_T^{-\rho}$ is satisfied for all sufficiently large $b_T$,
and similarly for the first derivative with $|f'(b_T)|< K b_T^{-\rho-1}$.
In our application, $S[f](k_T)$ represents the radial momentum spectrum
of any position-space function $f(b_T) = f(\abs{\vec{b}_T})$ that respects azimuthal symmetry,
see the first two lines of \eq{cumulative_tmdpdf_in_terms_of_bT_space}.
The physical interpretation of assumption~(2)
is that correlation functions of parton fields (and their spatial variations) should vanish far outside the hadron.
If $f$ is not oscillating but becomes monotonic at large $b_T$,
the second part of assumption~(2) concerning $f'$ follows from the first.%
\footnote{In addition, we assume that the original integral in \eq{S_functional} \emph{exists}
for any $k_T > 0$ in the sense of an improper Riemann integral.
Assumption~(2) can be made sufficient to guarantee its existence for any $k_T$ in the Lebesgue sense by requiring $\rho > 3/2$,
but this is not required for our manipulations below. 
We also have not made assumptions about the behavior of $f(b_T)$ as $b_T \to 0$ yet.} 

We introduce a cutoff, $\btcut$, which divides the total functional $S[f]$ into two parts:
\begin{align} 
S[f](k_T) &= S_<[f](k_T) + S_>[f](k_T) 
\,,
\end{align}
where
\begin{align} \label{eq:S_divided}
S_<[f](k_T) &= k_T \int_0^{\btcut} \! \df b_T \, ~b_T J_0(k_T b_T) f(b_T)
\,, \nn \\
S_>[f](k_T) &= k_T \int_{\btcut}^\infty \! \df b_T \, ~b_T J_0(k_T b_T) f(b_T)
\,.\end{align}
In our applications, $\btcut < 1/\lqcd$ should be chosen such that
for $S_<[f]$ perturbation theory can be used to compute $f(b_T \leq \btcut)$ at short distances,
implying that nonperturbative physics only becomes important in the long-distance contribution $S_>[f]$.

We now focus on developing a formalism to obtain $S[f](k_T)$ for \emph{perturbative} momenta $k_T \gg 1/\btcut \sim \lqcd$.
Our goal is to approximate the total functional $S[f](k_T)$ using only information from the region $b_T \leq \btcut$.
To the zeroth order, we define
\begin{align}
 S^{(0)}[f](k_T,\btcut) \equiv S_<[f](k_T) \,.
\end{align}
In our application, this ensures that all information about the perturbative domain is already included
in the zeroth-order functional, including in particular the resummed dependence on any high scales
such as the invariant mass of a color-singlet final state or the Collins-Soper scale.%
\footnote{%
An alternative would be to construct approximate Fourier functionals
in terms of the behavior of the function as $b_T \to 0$,
which has previously been considered in the applied mathematics literature~\cite{Soni:1982xxx},
see the note added at the end of this manuscript.
For our use case, this would require including also the resummed perturbative information
only as part of an asymptotic series, and would produce impractical results
in terms of the strong coupling and collinear PDFs at asymptotic scales $\mu \to \infty$ as discussed in \sec{review_tmd_pdf_and_normalization},
so we do not consider this option further.
}
In order to account for the leading contribution from $S_>[f]$, we compute the integral
\begin{align} \label{eq:int_by_part}
S_>[f](k_T) &= k_T \int_{\btcut}^\infty \! \df b_T \, b_T J_0(k_T b_T) \, f(b_T)
\nn \\
&= -b_T^{\rm cut} J_1(k_T b_T^{\rm cut}) \, f(b_T^{\rm cut})
   - \int_{b_T^{\rm cut}}^\infty \! \df b_T \, b_T J_1(k_T b_T) \, f'(b_T)
\nn \\
&= \sqrt{\frac{2b_T^{\rm cut}}{\pi k_T}} \cos\Bigl(k_T b_T^{\rm cut}+ \frac{\pi}{4}\Bigr) \, f(b_T^{\rm cut})
+ \frac{1}{k_T} \Ordsq{(k_T b_T^{\rm cut} )^{-\frac{1}{2}}}
\,.\end{align}
Starting with the first exact equality, we used integration by parts
and the fact that the surface term for $b_T \to \infty$ vanishes by assumption~(2) in \eq{S_assumptions}.
To obtain the last line we used the asymptotic form of the Bessel $J_1$ function at large $x = k_T \btcut \gg 1$
to restrict the surface term to the current working order,
\begin{align} \label{eq:J1_asymptotic}
J_1(x\to\infty) = -\sqrt{\frac{2}{\pi x}} \cos\Big(x+\frac{\pi}{4}\Bigr) + \Ord{x^{-\frac{3}{2}}}
\,.\end{align}
We further used that assumptions~(1) and (2) together
imply that $f'(b_T)$ is bounded by $K'\, b_T^{-\rho-1}$ for \emph{all} $b_T > \btcut$ for some suitable $K' > 0$,
such that we can power count away the remaining integral,%
\footnote{The first inequality is in fact incorrect in general
if $f'$ changes sign in a correlated fashion with $J_1$.
The correct proof proceeds via a Mellin transformation of $f$ and $f'$,
as well as a more explicit evaluation of Bessel integrals in the asymptotic limit,
and is given in \sec{truncated_functionals_higher_orders}.}
\begin{align} \label{eq:int_by_part_power_count_remaining_integral}
\int_{b_T^{\rm cut}}^\infty \! \df b_T \, b_T J_1(k_T b_T) \, f'(b_T)
&\lesssim \int_{b_T^{\rm cut}}^\infty \df b_T ~b_T J_1(k_T b_T) \, b_T^{-\rho-1}
= \frac{(\btcut)^{-\rho}}{k_T} \Ordsq{(k_T \btcut)^{-\frac{1}{2}}}
\,.\end{align}
This scaling again follows from the asymptotic form in \eq{J1_asymptotic},
where the oscillations of the cosine term suppress the value of the integral in \eq{int_by_part_power_count_remaining_integral} by one power of $1/(k_T \btcut)$.

With this leading correction from $S_>[f]$ at hand, we can define
\begin{align} \label{eq:S_1}
S^{(1)}[f] (k_T, \btcut) = S^{(0)}[f] (k_T, \btcut) + \sqrt{\frac{2 \btcut}{\pi k_T}}  \cos\Bigl(k_T\btcut  + \frac{\pi}{4}\Bigr) ~f(\btcut)
\,,\end{align}
which is an improved approximation of $S[f]$.
Importantly, the leading contribution from long distances
to the spectrum at perturbative $k_T$
is obtained from the value of the function at the boundary $b_T = \btcut$.

\subsection{Full proof and higher-order results for the long-distance contributions}
\label{sec:truncated_functionals_higher_orders}

We wish to define an asymptotic series of functionals $S^{(n)}[f]$ 
for $S[f]$ which have successively improving error terms for $k_T \btcut \to \infty$.
In these functionals, we aim to only use information about the test function from the region $b_T \leq \btcut$,
but systematically account for the contributions from $S_>[f]$
by including higher-order surface terms.
In order to prove the form of the surface terms up to and including order $n$,
we need the function $f(b_T)$ to satisfy stronger assumptions:
\begin{align} \label{eq:S_assumptions_higher_orders}
&(1^*) \quad f(b_T) ~ \text{is $n$ times differentiable for all}~b_T > 0
\, , \nn \\
&(2^*) \quad \frac{\df^n f(b_T)}{\df b_T^n} \lesssim b_T^{-\rho-n}, \quad \rho > 1/2
\, .\end{align}
Assumption $(2^*)$ again follows from the simpler assumption $f(b_T) \lesssim b_T^{-\rho}$
if $f(b_T)$ and its derivatives become monotonic power laws (or exponentials) for sufficiently large $b_T$.
The reverse is always true, i.e., $\df f^{k}(b_T)/\df b_T^k \lesssim b_T^{-\rho-k}$ follows from assumption~$(2^*)$
for all $0 \leq k \leq n$ by bounding the iterated integral.

\paragraph{Mellin transform.}
Our strategy is to decompose $f(b_T)$ in terms of basis functions of the form $b_T^{-N}$,
with $N > \rho$ a continuous parameter,
to make the power counting of individual terms transparent.
This amounts to taking the Mellin transform,%
\begin{align} \label{eq:Mellin_trans}
\phi(N) = \int_0^\infty \! \df b_T \, b_T^{N-1} f(b_T)
\,,\end{align}
so that the function $f(b_T)$ can be written using the inverse Mellin transform
\begin{align} \label{eq:inverse_Mellin_trans}
f(b_T) = \frac{1}{2 \pi \img} \int_{c - \img\infty}^{c + \img\infty}  \! \df N \,
b_T^{-N}  \phi(N)
\,.\end{align}
Here the real part $c$ of the chosen contour satisfies
\begin{align} \label{eq:condition_existence_Mellin_transform}
\int_0^\infty \! \df b_T \, b_T^{c-1} \abs{f(b_T)}
< \infty
\,.\end{align}
Together with bounded variations of $f$,
\eq{condition_existence_Mellin_transform} is sufficient
for the existence of the Mellin transform and its inverse, see e.g.\ theorem~28 in \refcite{Titchmarsh:1948xxx}.
Both conditions are satisified by assumption $(2^*)$ for $n = 0, 1$ and $c < \rho$.
Below we will need $c > 1/2$, and this is why we required a strict inequality $\rho > 1/2$.
Note that assumption~$(2^*)$ in the same way guarantees the existence of the (inverse) Mellin transform
for all derivatives $f^{(k)}$ with $k \leq n -1 $.
The relation is particularly compact for repeated derivatives with respect to $\ln b_T$,
\begin{align}\label{eq:Mellin_fn}
\Bigl( b_T \frac{\df}{\df b_T} \Bigr)^k f (b_T) = \frac{1}{2 \pi \img} \int_{c - \img\infty}^{c + \img\infty}
 \! \df N \, b_T^{-N} \, \phi(N) \, (-N)^k
\,.\end{align}

\paragraph{Bessel integral for a single power.}
To make contact with the Bessel integral above a cutoff $x = k_T \btcut \gg 1$,
let us consider
\begin{align} \label{eq:gbeta_S}
g(\beta, x)
&\equiv \int_x^\infty \! dy \, J_0(y) \, y^{1+\beta}
\nn  \\
&= \frac{2^{1+\beta} \, \Gamma \bigl( 1+ \frac{\beta}{2} \bigr) }
 { \Gamma\bigl( -\frac{\beta}{2} \bigr) }
-  \frac{x^{2+\beta}}{2+\beta} \,
\phantom{}_1F_{\,2}
\Bigl( 1+ \frac{\beta}{2}; 1, 2+ \frac{\beta}{2}; -\frac{x^2}{4} \Bigr) 
\nn  \\
&= \sqrt{\frac{2 x}{\pi}} x^\beta
\sum_{k=0}^{n-1} \frac{c_k(\beta)}{x^k}
\cos\Bigl(x + \frac{\pi}{4} - \frac{k\pi}{2} \Bigr) + \Ord{x^{\beta - n + \frac{1}{2}}}
\,.\end{align}
The integral exists for all complex parameters $\beta$ with $\Re(\beta) < -1/2$ and $\phantom{}_1 F_{\, 2}$ is the hypergeometric function. 
To obtain the last equality we used
the asymptotic expansion of
$\phantom{}_1 F_{\, 2} \bigl( a_1;  b_1, b_2; -x^2/4\bigr)$ for $x \gg 1$~\cite{NIST:DLMF},
which generalizes the asymptotic expansion of the Bessel function in \eq{J1_asymptotic}.
The coefficients $c_k(\beta)$ are polynomials in $\beta$ of order $k$
and are collected in \app{1F2_asymptotics_spectrum} up to $k=9$.
The first few coefficients read
\begin{align}
c_0 = 1
\,, \qquad
c_1(\beta) = -\frac{3}{8} - \beta
\,, \qquad
c_2(\beta) = -\frac{15}{128} - \frac{\beta}{8} + \beta^2
\,.\end{align}

\paragraph{Assembling the long-distance contribution.}
The above definitions allow us to compute
the long-distance contribution to $S[f]$
in terms of the Mellin transform $\phi$ of $f$,
\begin{align} \label{eq:compute_S_above}
S_>[f](k_T) 
&= k_T \int_{\btcut}^\infty \! \df b_T \, b_T J_0(k_T b_T) f(b_T)  
\nn \\
&= k_T \int_{\btcut}^\infty \! \df b_T \, J_0(k_T b_T)
\frac{1}{2\pi\img} \int_{c - \img\infty}^{c + \img\infty} \! \df N \,
\phi(N) \, b_T^{1-N}
\nn \\
&= \frac{1}{k_T} \frac{1}{2\pi\img}
\int_{c - \img\infty}^{c + \img\infty}
\! \df N \, \phi(N) \, k_T^N \, g(-N, k_T \btcut)
\nn \\
&= \sqrt{\frac{2\btcut}{\pi k_T}}
\sum_{k = 0}^{n-1} \frac{\cos \Bigl( k_T \btcut + \frac{\pi}{4} - \frac{k\pi}{2} \Bigr)}{(k_T \btcut)^k}
\frac{1}{2\pi\img} \int_{c - \img\infty}^{c + \img\infty} \! \df N \,
c_k(-N) \, \phi(N) \, (\btcut)^{-N}
\nn \\ & \qquad
+ \frac{f(\btcut)}{k_T} \Ordsq{(k_T \btcut)^{-n + \frac{1}{2}}}
\,.\end{align}
On the second line, we plugged in the inverse Mellin transform.
On the third line, we changed the order of integration
and used the definition of $g$ in \eq{gbeta_S},
which is justified because the $b_T$ integral exists for each power $b_T^{1-N}$
with $\Re(-N) = -c < 1/2$ along the contour individually.
Crucially, on the last line we performed the asymptotic expansion of $g(-N, k_T \btcut)$
and pulled the phase and the additional power suppression of each term out of the Mellin integral.
Note that naively, the scaling of the error term would be $(\btcut)^{-n}$
because the remaining $N$ integral scales as $(\btcut)^{-c} \leq (\btcut)^{-1/2}$.
However, the coefficient of this additional factor of $(\btcut)^{-1/2}$,
which is encoded in $\phi(N)$, is unrelated to $k_T$
and instead comes from the power law $K b_T^{-\rho}$ bounding the function.

We now use that the coefficients $c_k(-N)$ are polynomials in $N$,
and evaluate the remaining Mellin integral in \eq{compute_S_above} using \eq{Mellin_fn},
which turns powers of $N$ into repeated derivatives on the boundary,
\begin{align} \label{eq:compute_S_above_final_step}
S_>[f](k_T)
&=
\sqrt{\frac{2\btcut}{\pi k_T}}\:
\sum_{k = 0}^{n-1}\: \frac{\cos \Bigl( k_T \btcut + \frac{\pi}{4} - \frac{k\pi}{2} \Bigr)}{(k_T \btcut)^k} \,
c_k\Bigl( \btcut \frac{\df}{\df \btcut} \Bigr) f(\btcut)
\nn \\ & \quad
+ \frac{f(\btcut)}{k_T} \Ordsq{(k_T \btcut)^{-n + \frac{1}{2}}}
\,.\end{align}
This is our key result in this section:
At each order $n$ in the power counting, the long-distance contribution can be expressed
entirely in terms of the test function and its derivatives evaluated on the boundary.
Adding the short-distance contribution $S_<[f] = S^{(0)}[f]$, we define
\begin{align}\label{eq:S_final_series}
S^{(n)}[f](k_T, \btcut)
&\equiv k_T \int_0^{\btcut} \! \df b_T \, b_T J_0(k_T b_T) \, f(b_T)
\nn \\ & \quad
+ \sqrt{\frac{2\btcut}{\pi k_T}}\:
\sum_{k = 0}^{n-1}\: \frac{\cos \Bigl( k_T \btcut + \frac{\pi}{4} - \frac{k\pi}{2} \Bigr)}{(k_T \btcut)^k} \,
c_k\Bigl( \btcut \frac{\df}{\df \btcut} \Bigr) f(\btcut)
\,.\end{align}
We call these functionals of $f$ ``truncated functionals'',
referring both to the truncation of the $b_T$ integral and of the asymptotic series.
We reiterate that the truncated functional $S^{(n)}[f]$ approximates $S[f]$ up to
\begin{align} \label{eq:S_power_corr}
S[f](k_T) = S^{(n)}[f](k_T, \btcut) + \frac{f(\btcut)}{k_T} \Ordsq{(k_T \btcut)^{-n + \frac{1}{2}}}
\,.\end{align}
Writing out the derivatives with respect to $\ln \btcut$,
we find for the first few coefficients $0 \leq k \leq 2$, as required for $n \leq 3$:
\begin{align} \label{eq:S_trunc_functionals}
c_0\Bigl( \btcut \frac{\df}{\df \btcut} \Bigr) f(\btcut)
&= f(\btcut)
\,, \nn \\
c_1\Bigl( \btcut \frac{\df}{\df \btcut} \Bigr) f(\btcut)
&= -\frac{3}{8} f(\btcut) - \btcut f'(\btcut)
\,, \nn \\
c_2\Bigl( \btcut \frac{\df}{\df \btcut} \Bigr) f(\btcut)
&= -\frac{15}{128} f(\btcut) + \frac{7}{8} \btcut f'(\btcut) + (\btcut)^2 f''(\btcut)
\,.\end{align}
The first term reproduces the $S^{(1)}[f]$ given in \eq{S_1}.

\begin{figure}
\centering
\includegraphics[width=\WidthTwoSubfigs]{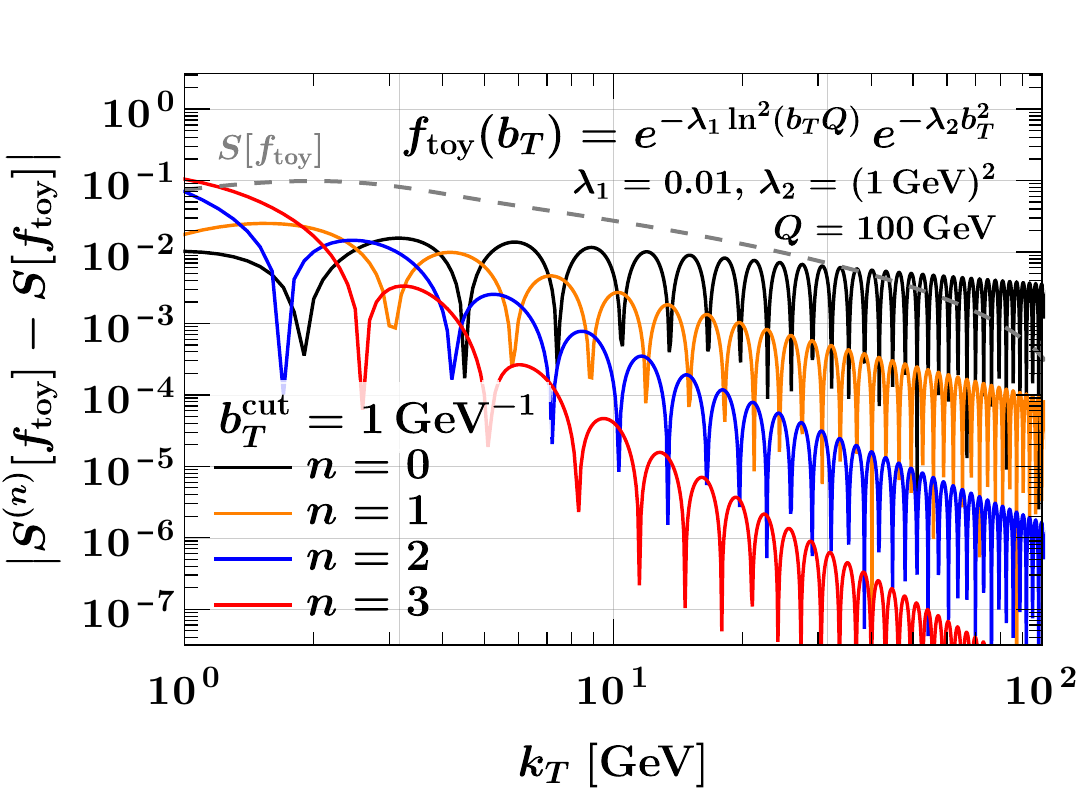}%
\hfill
\includegraphics[width=\WidthTwoSubfigs]{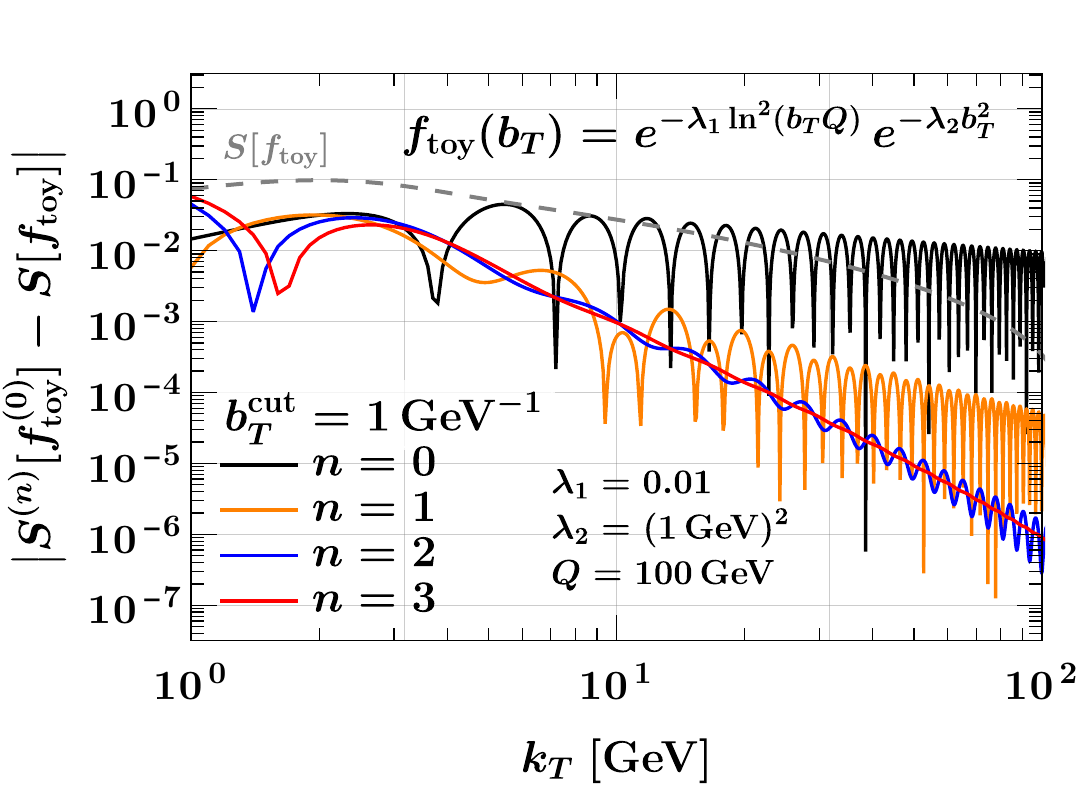}%
\caption{%
Left: Scaling of error terms of the truncated functionals
for the $k_T$ spectrum of the toy test function in \eq{toy_test_function}.
The error terms (computed as the difference to the exact result)
exactly follow the expected scaling $\sim k_T^{n-1/2}$.
Right: Same, but instead applying the truncated functionals
to the \emph{leading ``OPE'' contribution} $f^{(0)}_{\rm toy}$ for the test function.
The difference to the full result for $n \geq 2$ now clearly reveals
the presence of quadratic terms $\sim b_T^2$ in the full spectrum
that were expanded away by making use of the ``perturbative'' prediction $f^{(0)}_{\rm toy}$.
}%
\label{fig:spectrum_power_law}
\end{figure}

In the left panel of \fig{spectrum_power_law},
we test the asymptotic expansion
for a simple, but physically relevant toy test function
\begin{align} \label{eq:toy_test_function}
f_\mathrm{toy}(b_T)
&= \exp\bigl[ - \laa \ln^2(b_T Q) \big] \, \exp[- \lbb b_T^2]
\,, \nn \\[0.4em]
\laa &= 0.1
\,, \qquad
\lbb = 1 \GeV^2
\,, \qquad
Q = 100 \GeV
\,.\end{align}
Here the $\laa$ term is a proxy for the perturbative Sudakov exponential appearing in TMDs, and the $\lbb$ term is a proxy for a nonperturbative contribution.
The full spectrum $S[f_\mathrm{toy}]$ is shown by the dashed black line, while the solid lines display the absolute value of the error left over from subsequent approximations by our truncated functionals $S^{(n)}[f_\mathrm{toy}]$.
We clearly see that compared to the full spectrum $S[f_\mathrm{toy}]$,
the error term of the truncated functional exactly follows the expected scaling
given by \eq{S_power_corr} and improves as we go to higher $n$.

\subsection{Marrying the OPE to the truncated functionals}

Consider a function $f(b_T)$ which,
apart from satisfying assumptions $(1^*)$ and $(2^*)$ in \eq{S_assumptions_higher_orders},
can also be written in terms of an asymptotic series expansion,
\begin{align} \label{eq:bt_ope_assumption}
f(b_T) = \sum_{j=0}^{m -1} b_T^j f^{(j)}(b_T) + \Ordsq{(\lqcd b_T)^m}
\,,\end{align}
where $f^{(j)}(b_T) \sim \lqcd^j$ and has at most a logarithmic dependence on $b_T$.
In our physics application, the above asymptotic series arises from
an operator product expansion (OPE) that expresses the matrix element of a nonlocal operator
in terms of local operators suppressed by increasing powers of a low physical scale $\lqcd$.
We keep the Wilson coefficients of the local operators
that carry the non-analytic dependence on $b_T$ implicit in $f^{(j)}$.
The OPE may in addition have a finite radius of convergence $\btope$,
such that the limit $m \to \infty$ exists for all $b_T < \btope$,
but we do not require this in the following.
In physical terms this means we allow for the violation of quark-hadron duality~\cite{Shifman:2000jv} e.g. by terms $\sim \exp\bigl(-1/(\lqcd b_T)\bigr)$.
For simplicity we will also refer to the first OPE term $f^{(0)}(b_T)$ as the ``perturbative'' part of $f$,
since in our application it can be evaluated using pertubation theory and known,
simpler matrix elements.

Suppose we know the OPE up to order $m-1$, 
which means that the leading error term is $\sim (\lqcd b_T)^m$.
We now show that in this case,
we can approximate the total functional $S[f]$ using the truncated functionals
\emph{with only the known OPE terms} $\sum_{j=0}^{m-1} b_T^j f^{(j)}$ as input.

For simplicity let us first assume that only the leading perturbative term $f^{(0)}(b_T)$ is known,
and that the next term is suppressed by two powers of $\lqcd b_T$,
i.e., $f(b_T) = f^{(0)}(b_T) + \ordsq{(\lqcd b_T)^2}$, as is often the case.
After the Fourier transform, we have corrections
\begin{align}\label{eq:OPE_truncated_functional_corr_firstord}
S[f](k_T) 
&= S^{(n)}[f] (k_T) + \frac{f(\btcut)}{k_T} \Ordsq{(k_T \btcut)^{-n +\frac{1}{2}}} 
\\
&= S^{(n)}[f^{(0)}] + \frac{1} {k_T} \ORdsq{(k_T \btcut)^{-n +\frac{1}{2}}, \Bigl (\frac{\lqcd}{k_T} \Bigr)^2, \Bigl (\frac{\lqcd}{k_T} \Bigr)^2 (k_T \btcut)^{-n + \frac{5}{2}}}
\,, \nn \end{align}
where on the first line we account for the correction from truncating the functional,
and on the second line the three types of power corrections account for the correction from truncating the OPE.
We see that the third class of power corrections,
which is a combination of the neglected $(\btcut)^2 f^{(2)}(\btcut)$
in the surface term and the error term of the functional,
is suppressed more strongly than the intrinsic OPE correction $(\lqcd/k_T)^2$
for functionals with sufficiently high order $n > \frac{5}{2}$.
In particular, there are no corrections of $\ordsq{(\lqcd \btcut)^2}$, and we
give a short proof of this statement for the general case below.

For an illustrative example of the effect of truncating the OPE under the functional,
see the right panel of \fig{spectrum_power_law},
where we test \eq{OPE_truncated_functional_corr_firstord} with the toy function in \eq{toy_test_function}.
We put only the leading perturbative term $f_\mathrm{toy}^{(0)}$ into the truncated functionals,
where
\begin{align} \label{eq:toy_function_expansion}
f_\mathrm{toy}^{(0)} &= \exp[-\laa \ln^2 (b_T Q)] 
\, , \nn \\
f_\mathrm{toy} &= f_\mathrm{toy}^{(0)} \bigl (1 - \lbb b_T^2 + \ord{b_T^4} \bigr)
\, .\end{align}
We see that increasing the order $n$ of the truncated functionals
stops improving the result around the quadratic order,
because they become subleading to the intrinsic ``nonperturbative'' OPE correction $-\lbb b_T^2$.
This means starting at $n = 2$,
the error term of the truncated functional is ``stuck'' at a quadratic power law,
allowing us to identify the leading OPE correction directly in momentum space.
In \sec{data} we will exploit this to propose a model-independent method 
to constrain nonperturbative dynamics from momentum-space data.

Now for the more general case, 
suppose we know the OPE up to order $m-1$,
i.e. $f(b_T) = \sum_{j=0}^{m-1} b_T^j f^{(j)}(b_T) + \ordsq{(\lqcd b_T)^m}$.
We need to compute $S^{(n)}[b_T^m f^{(m)}]$
to find the leading momentum-space error term from the truncation of the OPE.
Note that for $b_T^m f^{(m)}(b_T)$ itself,
the full functional does not exist,
since $b_T^m f^{(m)}(b_T) \sim b_T^m$ with $m > 1/2$ at large $b_T$.
However, we can construct a function $f^{(m)}_*(b_T)$ 
that agrees with $b_T^m f^{(m)}(b_T)$ for all $b_T \leq \btcut$
but also satisfies assumptions $(1^*)$ and $(2^*)$ in \eq{S_assumptions_higher_orders},
e.g.\ by applying an exponential $b^*$ prescription (see \sec{bstar_prescriptions}) to $b_T^m f^{(m)}(b_T)$
and multiplying the result with a similarly constructed exponential model.
Then we have
\begin{align} \label{eq:truncation_OPE_correction}
S^{(n)}[b_T^m f^{(m)}](k_T)
&= S^{(n)}[f^{(m)}_*](k_T) 
 \\
&= \underbrace{ S[f^{(m)}_*](k_T) } +   
 \underbrace{ \frac{f^{(m)}_*(\btcut)}{k_T} \Ordsq{(k_T \btcut)^{-n + \frac{1}{2}}} }
 \,,\nn\\
&\hspace{-.4cm}
 \ORdsq{ \frac{1}{k_T}\Bigl( \frac{\lqcd}{k_T} \Bigr)^m} \qquad
  \ORdsq{\frac{1}{k_T} \Bigl(   \frac{\lqcd}{k_T} \Bigr)^m (k_T \btcut )^{ - n +m + \frac{1}{2}}} 
 \nn 
\end{align}
where in the last equality we have used that $S[f^{(m)}_*](k_T)$ is independent of $\btcut$,
and thus by dimensional analysis can only yield the intrinsic OPE correction $\ordsq{(\lqcd/k_T)^m}$.
We see that the second term, which accounts for corrections in $(\lqcd \btcut)^m$,
is suppressed compared to the first when $n > m + \frac{1}{2}$
without having to assume that $\btcut \ll 1/\lqcd$.
Combined with \eq{S_power_corr} we have
\begin{align}\label{eq:S_ope_truncated_functional_corr_higherord}
S[f](k_T)
&= \sum_{j = 0}^{m-1} S^{(n)}[b_T^j f^{(j)}] + \frac{1}{k_T}\ORdsq{ (k_T \btcut)^{-n +\frac{1}{2}},\Bigl (\frac{\lqcd}{k_T} \Bigr)^m, \Bigl( \frac{\lqcd}{k_T} \Bigr)^m (k_T \btcut )^{- n + m + \frac{1}{2}}}
\,.\end{align}
Note that each term $S^{(n)}[b_T^j f^{(j)}]$ is individually well defined
despite the growth $\sim b_T^j$ of the integrand at large $b_T$,
since it only involves a compact domain.
We could therefore exploit that the truncated functionals act linearly on the terms in the OPE, which enabled us 
to move the sum out of the functional.
Specializing to $m = 2$ and $f^{(1)} = 0$ as above, we recover \eq{OPE_truncated_functional_corr_firstord},
where power corrections of $\ordsq{(\lqcd \btcut)^m}$ are absent
and the third class of power corrections is suppressed for $n > m + \frac{1}{2} = \frac{5}{2}$.

\Eq{S_ope_truncated_functional_corr_higherord} is the main result of this section.
It allows us to compute a momentum-space quantity (a TMD spectrum) at large momenta
purely in terms of our perturbative knowledge of the position-space OPE at short distances,
without having to place any assumptions on the long-distance behavior.
The approximation is fully controlled and can be systematically improved
by either pushing the knowledge of the OPE to higher orders $m$,
or by carrying surface terms in the truncated functional to higher orders $n$.
In particular, higher-order terms in the OPE are mapped
to momentum space order by order in the power counting thanks to the linearity of the functionals.

\subsection{Extension to cumulative distributions}
\label{sec:cumulative_functionals}

In physics applications, we are often interested in a slightly different, but related integral functional,
\begin{align} \label{eq:K_functional_def}
K[f](\ktcut) 
  & \equiv \int_0^{\ktcut} \!\! \df k_T\  S[f](k_T) 
    \\
  & =  \ktcut \int_0^\infty \! \df b_T\, J_1(b_T \ktcut)\, f(b_T)
\,,\nn
\end{align}
which corresponds to the cumulative distribution over all momenta $\abs{\kt} \leq \ktcut$
in an azimuthally symmetric position-space function $f(b_T)$, see \eq{cumulative_tmdpdf_in_terms_of_bT_space}.

We may again introduce a perturbative cutoff $\btcut$ 
and divide the $K[f]$ into $K_<[f]$ and $K_>[f]$ analogous to \eq{S_divided}, 
and apply the same procedure described in \sec{truncated_functionals_higher_orders} 
to derive a series of functionals $K^{(n)}[f]$ that asymptote to $K[f]$ for large $\ktcut$,
\begin{align} \label{eq:K_final_series}
K^{(n)}[f](\ktcut, \btcut)
&\equiv \ktcut \int_0^{\btcut} \! \df b_T J_1( \ktcut b_T) f(b_T)
\\ & \quad
+ \sqrt{\frac{2}{\pi \ktcut \btcut }}
\sum_{k = 0}^{n-1} \frac{\cos \Bigl( \ktcut \btcut - \frac{\pi}{4} - \frac{k\pi}{2} \Bigr)}{(\ktcut \btcut)^k} \,
\tilde{c}_k\Bigl( \btcut \frac{\df}{\df \btcut} \Bigr) f(\btcut)
\nn \,.\end{align}
These truncated cumulative functionals $K^{(n)}[f]$ approximate $K[f]$ 
up to an error term given by
\begin{align} \label{eq:K_power_corr}
K[f](\ktcut) = K^{(n)}[f](\ktcut, \btcut) +\Ordsq{(\ktcut \btcut)^{-n - \frac{1}{2}}}
\,.\end{align}
The expressions above are analogous to \eqs{S_final_series}{S_power_corr} for the spectrum.
We give the first few coefficients needed to construct $K^{(n)}[f]$ with \eq{K_final_series} for $n \leq 3$:
\begin{align}\label{eq:K_trunc_functionals}
\tilde{c}_0\Bigl( \btcut \frac{\df}{\df \btcut} \Bigr) f(\btcut)
&= f(\btcut)
\,, \nn \\
\tilde{c}_1\Bigl( \btcut \frac{\df}{\df \btcut} \Bigr) f(\btcut)
&= -\frac{1}{8} f(\btcut) + \btcut f'(\btcut)
\,, \nn \\
\tilde{c}_2\Bigl( \btcut \frac{\df}{\df \btcut} \Bigr) f(\btcut)
&= -\frac{9}{128} f(\btcut) + \frac{5}{8} \btcut f'(\btcut) - (\btcut)^2 f''(\btcut)
\, . \end{align}
Higher-order functionals can be computed using the coefficients given in \app{1F2_asymptotics_cumulant}.
Restricting the input to the functionals to the first $m - 1$ orders in the OPE of the test function at small $b_T$,
we have
\begin{align}\label{eq:K_ope_truncated_functional_corr_higherord}
K[f](\ktcut)
&= \sum_{j = 0}^{m-1} K^{(n)}[b_T^j f^{(j)}] + \ORdsq{
   (\ktcut \btcut)^{-n - \frac{1}{2}},
   \Bigl (\frac{\lqcd}{\ktcut} \Bigr)^m,
   \Bigl( \frac{\lqcd}{\ktcut} \Bigr)^m (\ktcut \btcut )^{- n + m - \frac{1}{2}}
}
\,,\end{align}
which should be compared with \eq{S_ope_truncated_functional_corr_higherord}.
Once again the linearity of the functional $K[f]$ was used to pull the sum outside the functional.

\begin{figure}
\centering
\includegraphics[width=\WidthTwoSubfigs]{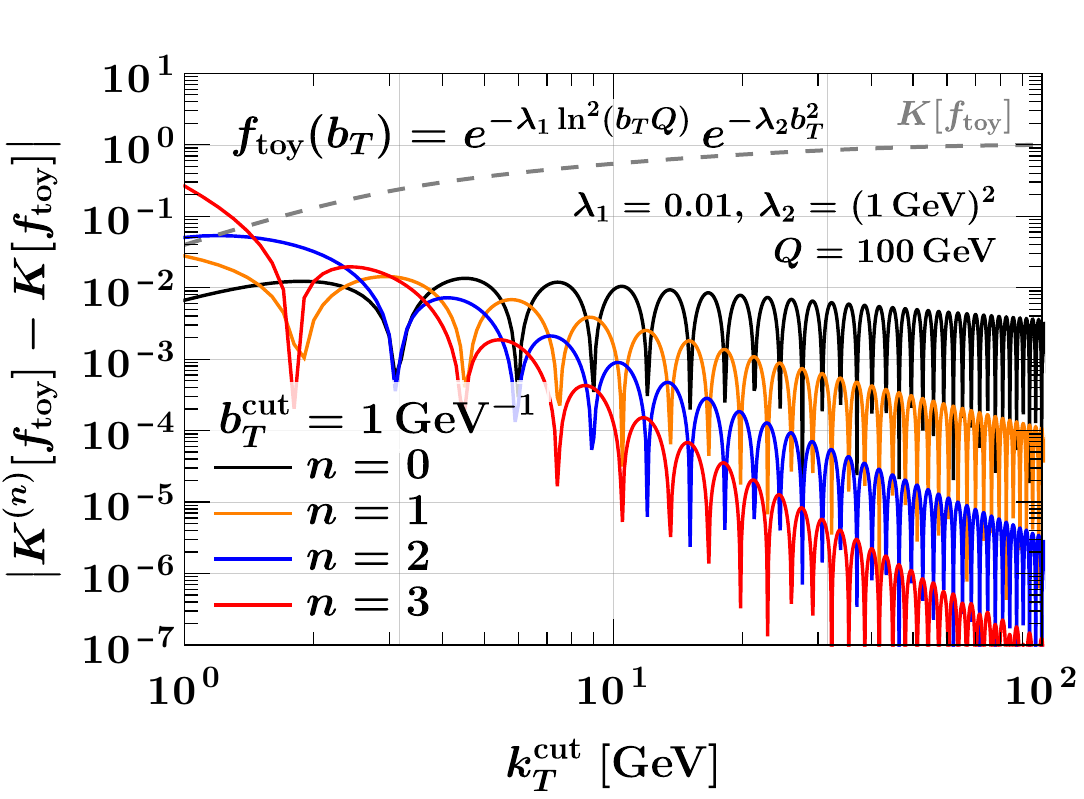}%
\hfill
\includegraphics[width=\WidthTwoSubfigs]{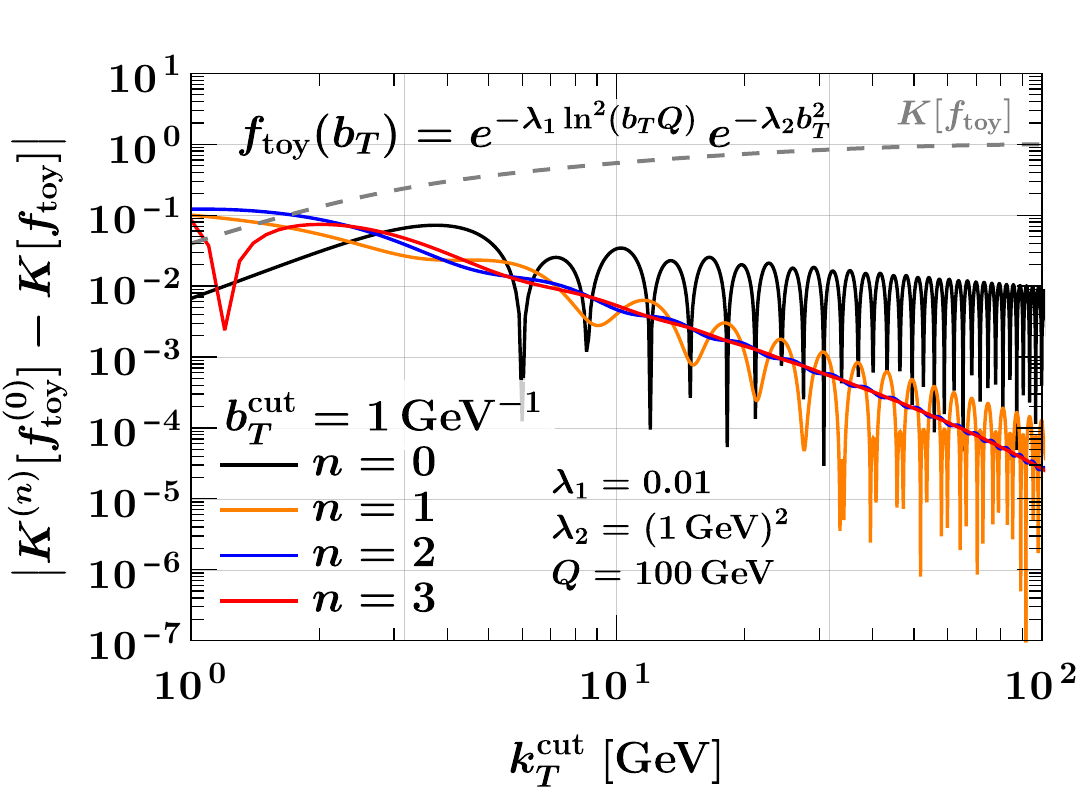}%
\caption{%
Same as \fig{spectrum_power_law}, but for the cumulative functionals $K^{(n)}[f](\ktcut)$.
}%
\label{fig:cumulant_power_law}
\end{figure}

The left panel of \fig{cumulant_power_law} shows that the cumulative functionals
exhibit the expected power-law scaling of error terms for the toy function in \eq{toy_test_function}.
In the right panel of \fig{cumulant_power_law} we again restrict the input of the functionals to the leading term at small $b_T$,
$K^{(n)}[f_\mathrm{toy}^{(0)}]$, and compute the difference to the full result $K[f_\mathrm{toy}]$.
Starting from $n = 2$, we find that the intrinsic $\ord{b_T^2}$ correction
results in a power-law behavior $\sim \ktcut$ of this difference, as expected.
Note that for the cumulative distributions, $n > m - \frac{1}{2}$ rather than $m + \frac{1}{2}$
is already sufficient to expose the quadratic behavior of the intrinsic OPE correction.

\subsection{Consequences for \texorpdfstring{$b^*$}{b*} prescriptions}
\label{sec:bstar_prescriptions}

In the $b_T$-space formalism of TMD PDF modeling, 
one often introduces a function $b^*(b_T)$ to shield the Landau pole in the perturbative computation, 
as presented in \eq{tmd_model} in the introduction.
Two common $b^*$ prescriptions in the literature are the Collins-Soper prescription~\cite{Collins:1981va}
and the Pavia prescription~\cite{Bacchetta:2017gcc}:%
\footnote{We have taken $b_\mathrm{min} \to 0$
in the original $b^*_\mathrm{Pavia}$ prescription, which as argued in \refcite{Bacchetta:2017gcc}
only amounts to a modification in the UV (beyond the validity of TMD factorization). 
This allows for a more direct comparison to other models which have $b^*(b_T \to 0) \to 0$.}
\begin{align}\label{eq:bstar_CS_Pavia}
b_\mathrm{CS}^*(b_T) &
= \frac{b_T}{\sqrt{1+(b_T/\bmax^2})} 
= b_T \Bigl( 1 - \frac{b_T^2}{2\,\bmax^2} \Bigr) + \Ord{b_T^5} 
\, ,\nn \\
b_\mathrm{Pavia}^*(b_T) &
= \bmax \biggl[ 1- \exp \Bigl( - \frac{b_T^4}{\bmax^4} \Bigr)\biggr]^\frac{1}{4}
= b_T \Bigl( 1- \frac{b_T^4}{8\,\bmax^4} \Bigr) + \Ord{b_T^9}
\, .\end{align}
When evaluating $f$ at $b^*(b_T)$ and expanding for $b_T\ll \bmax$, we have
\begin{align}\label{eq:f_bstar}
f\bigl(b_\mathrm{CS}^*(b_T)\bigr) = f(b_T) - \frac{b_T^3}{2\, \bmax^2} f'(b_T) + \Ord{b_T^5}
\,, \nn \\
f\bigl(b_\mathrm{Pavia}^*(b_T)\bigr) = f(b_T) - \frac{b_T^5}{8\,\bmax^4} f'(b_T) + \Ord{b_T^9}
\,,
\end{align}
where we note that the first subleading terms in the series differ, with  $b_T^3$ and $b_T^5$ scaling respectively.

We can formally assess the \emph{momentum-space} effect of the residual $\bmax$ dependence
by applying our truncated functionals at sufficiently high orders to $f(b^*(b_T))$.
For example, using $S^{(n)}$ with $n \geq 3$ for CS and $n \geq 5$ for Pavia,
we can show using \eq{S_ope_truncated_functional_corr_higherord}
that the leading correction terms in \eq{f_bstar}
carries over to momentum space with the scaling $\sim 1/(b_\mathrm{max} q_T)^n$ expected from naive power counting, 
where $n = 2$ for the Collins-Soper prescription and $n = 4$ for Pavia.
As we push to higher orders of $S^{(n)}$, 
we can identify the scaling of the leading corrections in momentum space using a plot similar to \fig{spectrum_power_law}.

\begin{figure}[t!]
\centering
\includegraphics[width=0.55\textwidth]{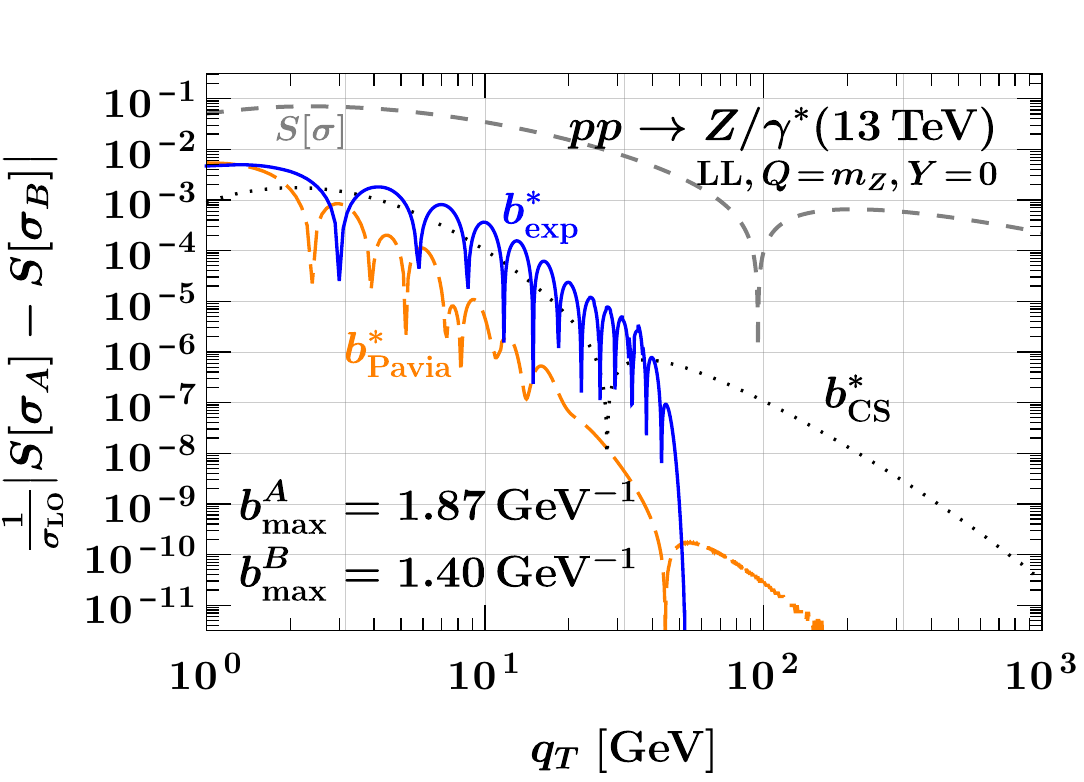}%
\caption{%
Scaling of the ambiguity from the choice of $b_\mathrm{max}$ for different $b^*$ prescriptions.
}%
\label{fig:bstar_prescriptions}
\end{figure}

To test the impact of different $b^*$ choices we consider $S[\sigma]$ where $\sigma$ is the leading (double) logarithmic (LL) inclusive $q_T$ differential cross section for $pp \to Z/\gamma^*$ at the LHC, normalized to the tree-level cross section.
We use the LL solution of the running coupling starting from $\as(m_Z) = 0.118$,
which exhibits a Landau pole at $\lqcd^\mathrm{LL} \approx 100 \MeV$.
For numerical stability, we keep the scale of the PDFs in the resummed prediction fixed at the high scale $\mu_f = Q = m_Z$, which is an NLL effect.
In \fig{bstar_prescriptions}, we show the scaling of the difference between using 
$\bmax^A = 1.87\, \mathrm{GeV}^{-1}$ and $\bmax^B = 1.40 \, \mathrm{GeV}^{-1}$
in momentum space  for different $b^*$ models. 
We also include the value of an individual $S[\sigma]$ (dashed gray curve) using $b^*_\mathrm{Pavia}$ with $\bmax=1.87 \, \mathrm{GeV}^{-1}$ as reference. 
All curves are normalized to the Born cross section $\sigma_\mathrm{LO}$.
We consider the behavior of $b^*_\mathrm{CS}$ (dotted black) and $b^*_\mathrm{Pavia}$ (dashed orange), 
as well as an exponential $b^*$ prescription (solid blue) defined as
\begin{equation}\label{eq:bstar_exponential}
b^*_\mathrm{exp}(b_T) = 
  \begin{cases}
        b_T \Bigl[1 - \exp\Bigl(-\frac{1}{b_T/\bmax - a}\Bigr) \Bigr] \,, \quad   & b_T \geq  a\, \bmax,
        \\
        b_T \,,  \quad &b_T < a \,\bmax
  \,.\end{cases}
\end{equation}
The benefit of the $b^*_\mathrm{exp}$ prescription is that it does not introduce
additional powers beyond those predicted by the OPE. 
In addition, for $b_T \leq a \bmax$ any function including the prescription is exactly equal to the input function, $f(b^*(b_T))=f(b_T)$.
In our numerics we have chosen $a = 1/2$, which is the largest value allowed such that $b^*_\mathrm{exp}$ is still monotonic.
We observe that the scaling of the $\bmax$ ambiguity of the CS and Pavia prescriptions both stabilize to their corresponding power law ($q_T^{-2}$ and $q_T^{-4}$, respectively),
as expected from our analytic arguments below \eq{f_bstar}.
For our newly introduced $b^*_\mathrm{exp}$ prescription we see that the scaling indeed falls off faster than any power for large enough $q_T$, as expected.
Unfortunately, we find that the coefficient of the exponentially suppressed ambiguity is rather large,
so that the practical usefulness of the prescription, beyond the formal improvement of preserving the OPE coefficients, may be limited.

We note that these conclusions also apply 
when the $b^*$ prescription is only implemented 
through the choice of boundary scales in resummed perturbation theory
(a so-called local $b^*$ prescription),
rather than also substituting the (global) kinematic dependence 
of the factorized cross section by $b_T\to b^*(b_T)$,
which is needed e.g.\ in the presence of additional measurements
on soft radiation to preserve refactorization relations
for the relevant matrix elements~\cite{Lustermans:2019plv}.
It also applies to the choice of TMD PDF boundary scale $\mu_\mathrm{OPE} = b_0/b_T + \Delta$ with $\Delta = 2 \GeV$ 
at which the OPE of the TMD PDF is performed in \refcite{Scimemi:2019cmh},
and which we expect to translate to linear power corrections $\ord{\Delta/q_T}$ in momentum space.
In either of these cases, the total $b_T$ derivative in \eq{f_bstar}
is replaced by partial derivatives $\partial/\partial \mu_X$ with respect to the relevant scale(s) $\mu_X$,
but the power-law scaling of the artifacts induced by the prescription
still follows from expanding the relevant profile scale function $\mu_X(b_T)$ for small $b_T$,
and will carry over to momentum space in the way described above.
We also note that in both the setup of \refcite{Lustermans:2019plv} and \refcite{Scimemi:2019cmh}, the modification of the function occurs in a region
where left-over fixed-order logarithms of $b_T \mu_X$ may be large,
so the $\mu_X$ dependence is not guaranteed to decrease as the perturbative order increases.

\subsection{Series stability near the Landau pole}
\label{sec:stability_near_landau_pole}

\begin{figure}
\centering
\includegraphics[width=\WidthTwoSubfigs]{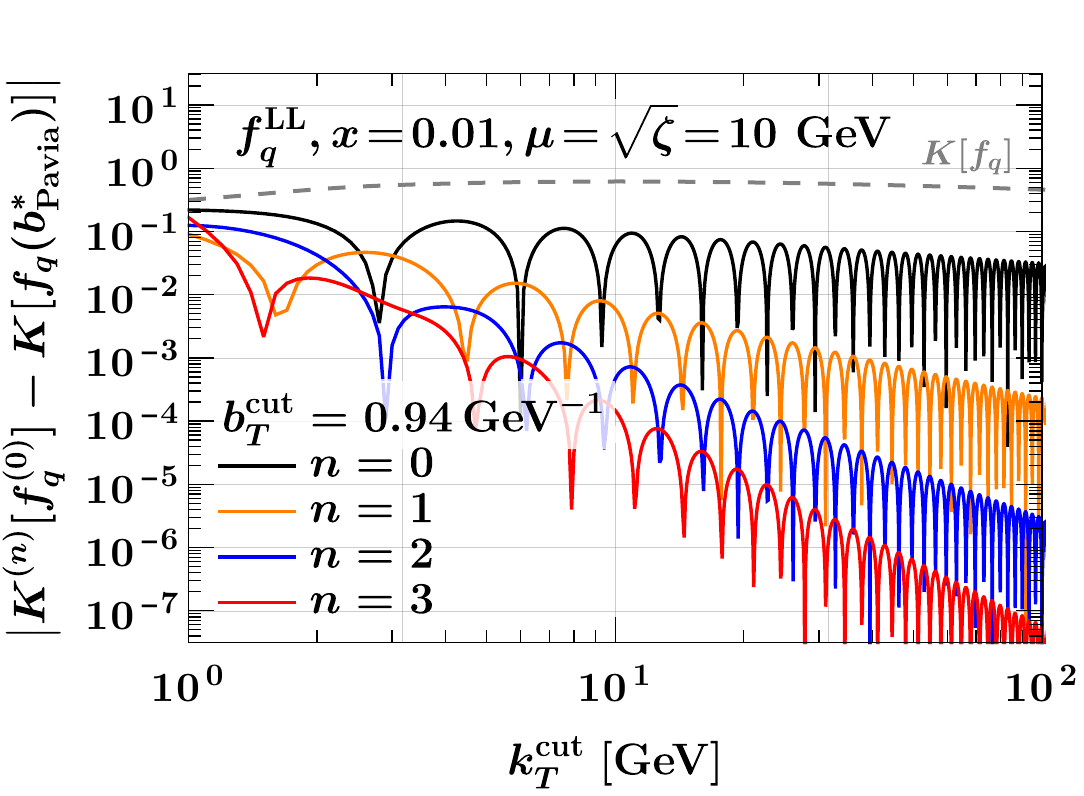}%
\hfill
\includegraphics[width=\WidthTwoSubfigs]{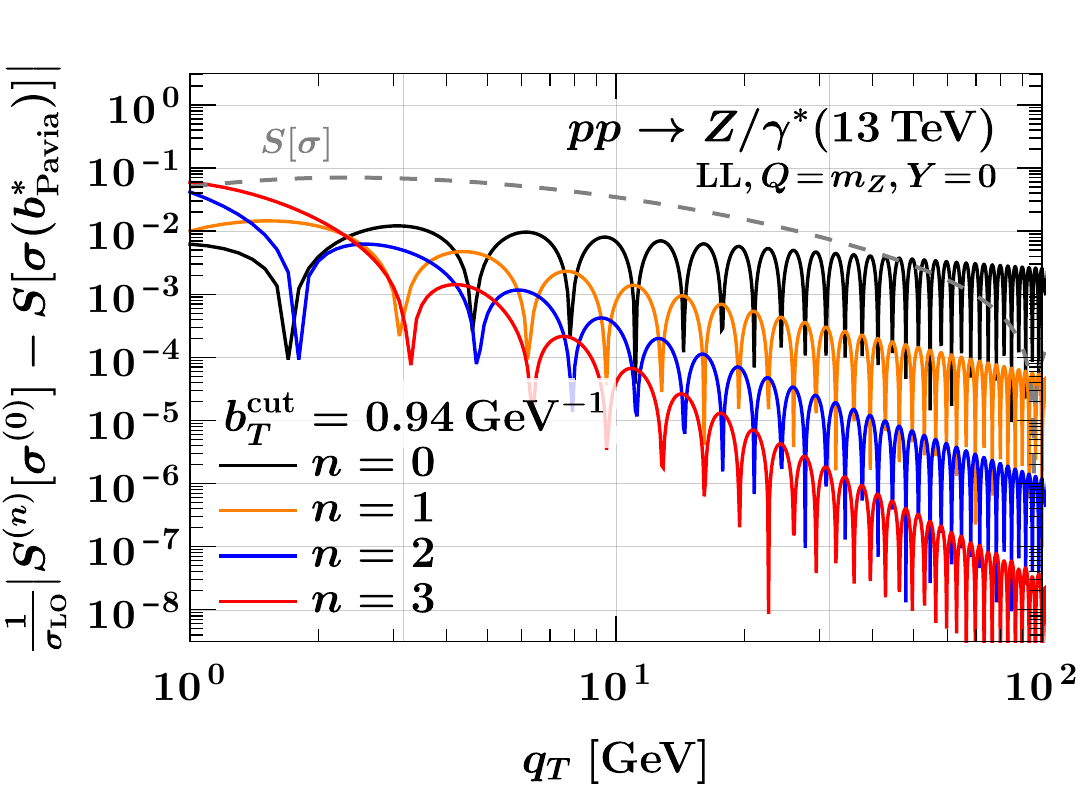}%
\caption{%
Absolute error of the truncated functionals in the presence of a Landau pole
for the cumulative distribution of a single TMD PDF (left)
and the $Z/\gamma^\ast$ transverse-momentum spectrum at the LHC (right).
}%
\label{fig:stability_near_landau_pole}
\end{figure}

As a final check on our setup, 
we consider the stability of the truncated functional series
when applied to a test function that exhibits an actual Landau pole
beyond the support of the integral functionals.
This is relevant because in the presence of a pole at $b_T = b_T^\mathrm{pole}$, 
we generically expect the derivatives of the test function to grow factorially, e.g.,
\begin{align}
f(b_T) = \frac{1}{b_T - b_T^\mathrm{pole}}
\,, \qquad 
f^{(n)}(b_T) = \frac{(-1)^n n!}{\bigl(b_T - b_T^\mathrm{pole}\bigr)^{n+1}}
\,.\end{align}
As the highest derivative $f^{(n)}$
enters the truncated functional at order $n$ with coefficient $(-1)^n$,
see \app{1F2_asymptotics},
this factorial growth feeds through into the result for the functional.
We use the same numerical setup as in the previous setup,
evaluating a single TMD PDF (or the  Drell-Yan cross section) at leading-logarithmic (LL) order with a Landau pole at $\lqcd \approx 100 \MeV$,
but without applying any $b^*$ prescription in this case.
We continue to fix the PDF scales at the high scale $\mu_f = \sqrt{\zeta}$ 
for the TMD PDF and at $\mu_f = Q$ for the cross section, for numerical stability.

In \fig{stability_near_landau_pole} LL results for the cumulant of a single quark TMD PDF (left) and the Drell-Yan transverse momentum spectrum at the LHC (right) are shown.
In both cases, while the overall coefficient of the error term is substantially larger than for the test function in \figs{spectrum_power_law}{cumulant_power_law},
we still observe a stable improvement in the error term
when going to higher orders in the functional.
This indicates that at least up to the order $n = 3$ we consider here,
the Landau pole does not deteriorate the quality of the asymptotic approximation.

\section{Calculating the TMD PDFs and their cumulative integral}
\label{sec:single_tmdpdf}

\subsection{Evolution and operator product expansion of the TMD PDF}
\label{sec:single_tmdpdf_evolution_ope}

The dependence of the TMD PDF on the renormalization scale $\mu$
and the CS scale $\zeta$ is governed by the evolution equations in \eqs{mu_rge_tmdpdf}{zeta_rge_tmdpdf}.
They can be solved in closed form to evolve the TMD PDF
from some initial values $(\mu_0, \zeta_0)$ to $(\mu, \zeta)$,
\begin{align} \label{eq:tmdpdf_evolved}
f_i(x, b_T, \mu, \zeta)
&= f_i(x, b_T, \mu_0, \zeta_0) \,
  \exp \biggl[ \int_{\mu_0}^\mu \! \frac{\df \mu'}{\mu'} \, \gamma^i_\mu(\mu', \zeta_0) \biggr] \,
  \exp \biggl[ \frac{1}{2} \gamma^i_\zeta(b_T, \mu) \ln \frac{\zeta}{\zeta_0} \biggr]
\nn \\
&= f_i(x, b_T, \mu, \zeta_0) \,
  \exp \biggl[ \frac{1}{2} \gamma^i_\zeta(b_T, \mu) \ln \frac{\zeta}{\zeta_0} \biggr]
\,.\end{align}
Here we have chosen the evolution path as $(\mu_0, \zeta_0) \to (\mu, \zeta_0) \to (\mu, \zeta)$.
We can in particular use \eq{tmdpdf_evolved} to evolve the TMD PDF away from the canonical boundary condition
$\mu_0 \,, \sqrt{\zeta_0} \sim 1/b_T$
at which the TMD PDF is a single-scale quantity and thus free of large logarithms
to all orders in perturbation theory and in $\lqcd b_T$.
In explicit numerics we choose the canonical boundary scales as
\begin{align}
\mu_0 = \sqrt{\zeta_0} = \frac{b_0}{b_T}
\,,\end{align}
where we include a conventional $\ord{1}$ numerical coefficient $b_0 = 2 \exp^{-\gamma_E} \approx 1.12292$
with $\gamma_E$ the Euler-Mascheroni constant.
We will also refer to this boundary condition
simply as the ``intrinsic'' TMD PDF.%
\footnote{Other choices of the boundary scale and evolution path are also possible;
for examples relevant to recent TMD PDF extractions see e.g.\ \refcite{Bacchetta:2017gcc},
where $\sqrt{\zeta_0}$ is held fixed at a low scale $Q_0 = 1 \GeV$,
or \refscite{Scimemi:2018xaf, Vladimirov:2019bfa},
where the boundary scales are set by a recursive procedure
taking the (perturbative and nonperturbative) rapidity anomalous dimension as input.
We stress that while the result for the rapidity anomalous dimension is unambiguous,
different choices of $\zeta_0$ amount to moving nonperturbative contributions
proportional to the rapidity anomalous dimension in and out of the intrinsic TMD PDF.}

We are interested in values of $1/b_T, \mu_0, \mu \gg \lqcd$
where the TMD PDF and the rapidity anomalous dimension $\gamma^i_\zeta$
obey an operator product expansion~\cite{Collins:1981uw, Collins:2011zzd, Collins:2014jpa, Collins:2016hqq, Vladimirov:2020umg},
\begin{align} \label{eq:tmd_rapidity_anom_dim_ope}
f_i(x, b_T, \mu, \zeta)
&= f_i^{(0)}(x, b_T, \mu, \zeta) 
  + \sum_{j = 1}^{m-1} b_T^j f_i^{(j)}(x, b_T, \mu, \zeta)
  + \Ordsq{(\lqcd b_T)^m}
 \,, \nn \\
\gamma^i_\zeta(b_T, \mu)
&= \gamma^{(0)}_{\zeta,i}(b_T, \mu)
  + \sum_{j = 1}^{m-1} b_T^j \gamma^{(j)}_{\zeta,i}(b_T) 
  + \Ordsq{(\lqcd b_T)^m}
\,,\end{align}
with $f_i^{(j)}, \gamma^{(j)}_{\zeta,i} \sim \lqcd^j$. 
Most often the terms with odd $j$ vanish. The $\mu$ anomalous dimension in \eq{mu_rge_tmdpdf} is a UV quantity and does not receive nonperturbative corrections,
so we simply wrote down the OPE at the overall scale $\mu$,
which does not need to be equal to $\mu_0$.
On the other hand, the $\zeta$ evolution mixes powers in the $b_T$ expansion.
Specifically, for the first nonvanishing orders in a TMD PDF with $f_i^{(0)} \neq 0$, $f_i^{(1)} = 0$,
\begin{align} \label{eq:rge_tmdpdf_expanded}
\zeta\frac{\df}{\df \zeta} f^{(0)}_i(x, b_T, \mu, \zeta)
&= \frac{1}{2} \gamma^{(0)}_{\zeta,i}(b_T, \mu) \, f^{(0)}_i(x, b_T, \mu, \zeta)
\,, \\
\zeta\frac{\df}{\df \zeta} f^{(2)}_i(x, b_T, \mu, \zeta)
&= \frac{1}{2} \Bigl[
     \gamma^{(0)}_{\zeta,i}(b_T, \mu) \, f^{(2)}_i(x, b_T, \mu, \zeta)
   + \gamma^{(2)}_{\zeta,i}(b_T) \, f^{(0)}_i(x, b_T, \mu, \zeta)
\Bigr]
\,. \nn \end{align}

The structure of $\gamma_\zeta^i$ at higher orders is relatively well understood.
The leading perturbative term $\gamma_{\zeta,i}^{(0)}$ is known to three loops~\cite{Luebbert:2016itl, Li:2016ctv, Vladimirov:2016dll},
and also contributes the overall additive $\mu$ renormalization of $\gamma_\zeta^i$ in \eq{rge_rapidity_anom_dim}.
Recently, \refcite{Vladimirov:2020umg} demonstrated that the $\ord{b_T^2}$ contribution can be related to a gluon vacuum condensate.
Assuming a tree-level Wilson coefficient and ignoring the running of the condensate as in \refcite{Vladimirov:2020umg},
$\gamma_{\zeta,i}^{(2)}(b_T) = \gamma_{\zeta,i}^{(2)} \sim \lqcd^2$ is a single fixed number with units of $\GeV^2$.

The entries $f^{(j)}_i$ of the OPE for a TMD PDF in general
are given by multiple convolutions of perturbative Wilson coefficients,
which encode the logarithmic $b_T$ dependence as $b_T \to 0$,
and higher-power matrix elements that provide the power suppression by $\lqcd^j$.
The leading term for the unpolarized quark or gluon TMD PDF takes the form
\begin{align} \label{eq:tmd_ope_leading_term}
f^{(0)}_i(x, b_T, \mu, \zeta)
& = \sum_j \int_x^1 \! \frac{\df z}{z} \,
C_{ij}(z, \, b_T, \, \mu, \, \zeta) \, f_j\Bigl(\frac{x}{z}, \mu \Bigr)
\nn \\
& = f_i (x, \mu) + \ordsq{\alpha_s(\mu)}
\, ,\end{align}
where $f_j(x/z, \mu)$ is the collinear PDF,
$C_{ij}(z, b_T, \mu, \zeta) = \delta_{ij} \delta(1-z) + \ordsq{\as(\mu)}$ are the matching coefficients,
and the sum runs over all partons $j$.
The linear correction $f^{(1)}_i = 0$ to the unpolarized TMD PDF vanishes by azimuthal symmetry~\cite{Collins:2014jpa, Collins:2016hqq}.

Not much is known about the OPE of the unpolarized TMD PDF beyond the linear order.
A subset of terms that multiply the leading-twist collinear PDF $f_i(x, \mu)$
was determined to all powers in $b_T$ and at tree level in $\as$ in \refcite{Moos:2020wvd}.
These correspond to target mass corrections to the TMD PDF
because only the hadron (or nucleon) mass $M^j$ can provide the power suppression at $\ord{b_T^j}$ in this case;
nevertheless, they only constitute a subset of the full OPE at any given order in $b_T$.%
\footnote{Interestingly, we find that 
the Fourier transform of the result in \refcite{Moos:2020wvd}
for the unpolarized TMD PDF in their eq.~(4.13)
actually exists (in the distributional sense), with the cumulative distribution given by
\begin{align}
K \bigl[ f_{i\,\text{tree-level}}^\text{\refcite{Moos:2020wvd}}(x, b_T) \bigr](\ktcut)
= f_i(x, \mu) - \frac{1}{[1+(\ktcut/xM)^2]} f_i\Bigl( x + \frac{(\ktcut)^2}{xM^2}, \mu \Bigr)
\,.\end{align}
In particular, the $k_T$ dependence of the TMD PDF in this approximation is cut off at $k_T^2 \leq x(1-x) M^2$,
which has a simple interpretation in terms of the remnant of the struck hadron consisting of on-shell states,
i.e., $(P - k)^2 \geq 0$ with $P^\mu$ the hadron and $k^\mu$ the parton momentum.
It would be interesting to investigate whether the results of \refcite{Moos:2020wvd}
for spin-dependent TMD PDFs can be understood in a similar physical way.}
For the purposes of this paper, we will consider the most general form that the quadratic correction
can have based on the RG properties,
\begin{align} \label{eq:tmd_ope_quadratic_term}
b_T^2 f^{(2)}_i(x, b_T, \mu, \zeta)
= b_T^2 \, \C_i^{(2)}(x, b_T, \zeta) \, f^{(0)}_i(x, b_T, \mu, \zeta) \Bigl[ 1 + \Ordsq{\as(\mu)} \Bigr]
\,,\end{align}
where $\C_i^{(2)} \sim \lqcd^2$ has units of $\GeV^2$.
To see this, recall that the $\mu$ dependence of the quadratic term has to be that of the leading-power one,
which motivates writing it as a product involving $f^{(0)}$.
By \eq{rge_tmdpdf_expanded}, the remaining coefficient $\C_i^{(2)}$ is renormalized as 
\begin{align}
\zeta\frac{\df}{\df \zeta} \C_i^{(2)}(x, b_T, \zeta)
= \frac{1}{2} \gamma_{\zeta,i}^{(2)}(b_T)
\,.\end{align}
Evaluating $\C_i^{(2)}$ at our reference boundary scale $\sqrt{\zeta_0} = b_0/b_T$,
we are left with a function of the flavor $i$, $x$, and $b_T$.
As for the rapidity anomalous dimension, we assume leading-order Wilson coefficients
and ignore the running of the higher-twist matrix elements and soft condensates from $\lqcd$ to $1/b_T$.
(The running in general would dress the leading scaling $\sim b_T^2$ with an anomalous exponent,
and similarly for Wilson coefficients that start at $\ordsq{\as(1/b_T)}$.)
This leaves us with a function $\C_i^{(2)}(x, b_T, \zeta_0) = \C_i^{(2)}(x)$
of a single variable for each flavor.

Combining both sources of $\ord{b_T^2}$ nonperturbative corrections and fixing $\zeta_0$, we have
\begin{align} \label{eq:tmdpdf_evolved_expanded}
f_i(x, b_T, \mu, \zeta)
= f_i^{(0)}(x, b_T, \mu, \zeta) \Bigl\{
   1
   +  b_T^2 \Bigl( \C_i^{(2)}(x) + \frac{1}{2} \gamma^{(2)}_{\zeta,i} L_\zeta \Bigr)
\Bigr\}
+ \Ordsq{(\lqcd b_T)^4}
\,,\end{align}
where we defined $L_\zeta \equiv \ln (b_T^2 \zeta/b_0^2)$ as a shorthand.
We also absorbed the leading-power rapidity evolution factor
$\exp \bigl[ \frac{1}{2} \gamma^{(0)}_{\zeta,i} L_\zeta \bigr]$
into $f_i^{(0)}$, which evolves it back to the overall $\zeta$.
Importantly, the impact of $\gamma^{(2)}_{\zeta,i}$ on the evolved TMD PDF
is \emph{linear} at this order thanks to the power expansion of the evolution kernel.
This is crucial, because our truncated functionals are precisely
set up to act linearly on terms in the $b_T$-space OPE, so we will be able to move
the coefficient $\gamma_{\zeta,i}^{(2)}$ out of the Fourier transform.
This setup will allow us to study the linear impact of the OPE coefficients $\C_i^{(2)}$ and $\gamma^{(2)}_{\zeta,i}$ on momentum-space observables,
either as an uncertainty on the transverse-momentum integral of a single TMDPDF in \sec{tmd_cumulant_np}
or as a signal template for our proposed fit to data in \sec{data}.

\subsection{Approximating the cumulative TMD PDF with truncated functionals}
\label{sec:tmd_cumulant_np}

The cumulative integral of an azimuthally symmetric TMD PDF over $\abs{\kt} \leq \ktcut$,
\begin{align} \label{eq:tmd_cumulant}
\int_{\abs{\kt} \leq \ktcut} \! \df^2 \kt \, f_i(x, \kt, \mu, \zeta)
&= K[f_i(x, b_T, \mu, \zeta)] (\ktcut)
\,,\end{align}
can be computed in terms of the $b_T$-space TMD PDF $f_i$
with exactly the $K$ functional we have defined in \eq{K_functional_def}.
At large $\ktcut \gg \lqcd$ we can therefore use the formalism described in \sec{truncated_functionals}
to approximate the cumulative distribution of a single TMD PDF
in order to eventually study its normalization.

Thanks to the OPE \eq{tmd_ope_leading_term},
we can calculate the resummed leading OPE term $f_i^{(0)}$ in $b_T$ space in terms of collinear PDFs.
We use the numerical implementation of the matching coefficients
and anomalous dimensions in \texttt{SCETlib}~\cite{scetlib} as also used in \refcite{Ebert:2020dfc}.
More details on our evolution and resummation setup for $f_i^{(0)}$ are given in \sec{delta_res}.
We evaluate $f_i^{(0)}$ at N$^3$LL order,
which requires the two-loop matching coefficients for the TMD PDF~\cite{Gehrmann:2012ze, Gehrmann:2014yya, Luebbert:2016itl},
the three-loop collinear quark anomalous dimension~\cite{Moch:2005id},
the perturbative Collins-Soper kernel to three loops~\cite{Li:2016ctv, Vladimirov:2016dll},
and the four-loop cusp anomalous dimension~\cite{Henn:2019swt, vonManteuffel:2020vjv},
We assume $n_f = 5$ active flavors in the matching coefficients and throughout the TMD evolution.
We use the \texttt{CT18NNLO} PDF set~\cite{Hou:2019efy}
as provided by LHAPDF~\cite{Buckley:2014ana}
together with four-loop running~\cite{vanRitbergen:1997va, Czakon:2004bu} of the coupling starting from $\as(m_Z) = 0.118$.

Applying our truncated functionals to the leading OPE term, $K^{(n)}[f_i^{(0)}]$,
we obtain a purely perturbative baseline for the $k_T$-cumulative TMD PDF.
In the following we use $n = 3$, which we find to be numerically stable
and sufficient for our purposes in this section.
We set $\btcut = b_0/(1.2 \GeV)$ such that all scales in the resummed TMD PDF
stay above $1.2 \GeV$, i.e., well away from the Landau pole.
We use standard adaptive integration tools to evaluate $K^{(0)}[f]$,
and evaluate the required derivatives for the surface terms in a numerically stable fashion
using Lanczos' method of differentiation-by-integration~\cite{Groetsch:1998xxx, Rangarajan:2005xxx}.
We find it convenient to change variables to $\ln b_T$ in the test function
before differentiating at $\ln \btcut$, which both improves the numerical stability
and makes it straightforward to push the implementation to high orders
using the coefficients in \app{1F2_asymptotics}.
When needed, e.g.\ when comparing to explicit models,
we perform Bessel integrals over the full range $0 < b_T < \infty$
using a double-exponential method for oscillatory integrals~\cite{Takahasi1974, Ooura2001, Ooura1997}
as implemented in \texttt{SCETlib}~\cite{scetlib}.

\begin{figure}
\centering
 \includegraphics[width=\WidthTwoSubfigs]{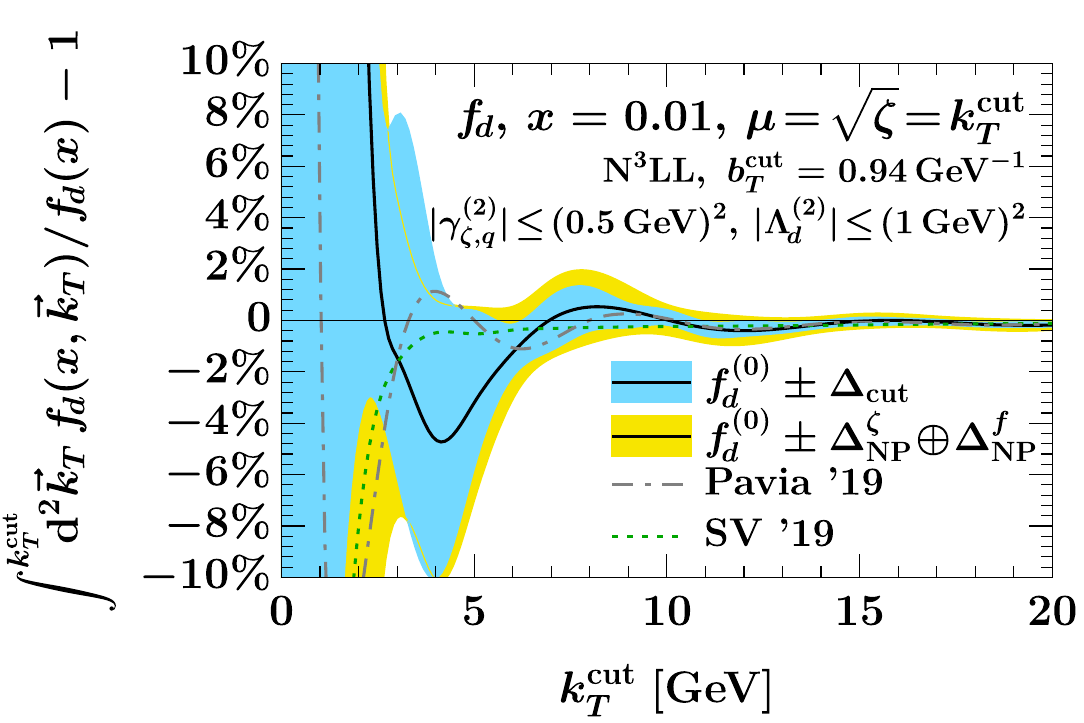}
 \hfill
 \includegraphics[width=\WidthTwoSubfigs]{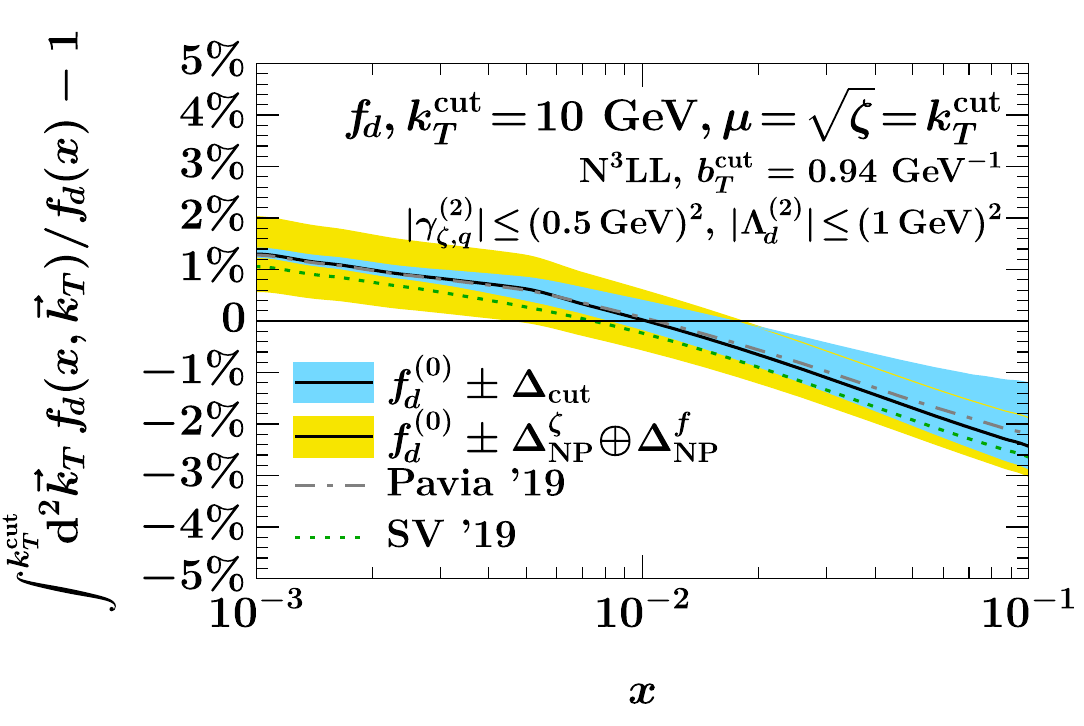}
\caption{%
The normalized deviation of the transverse-momentum integral (with an upper limit)
of the TMD PDF from the collinear PDF
as a function of the transverse-momentum cutoff $\ktcut$ (left) and 
of the momentum fraction $x$ (right).
The blue band shows the uncertainty from varying the position-space integral cutoff $\btcut$,
and the yellow band shows the uncertainty from varying
the quadratic OPE coefficients $\gamma_{\zeta,q}^{(2)}$ and $\C_q^{(2)}$.
We also include the cumulative TMD PDF as predicted by SV and Pavia global fits, 
both of which show small deviation from the collinear PDF.
We approximate the transverse-momentum integral of the TMD PDF with $K^{(3)}[f_i^{(0)}]$,
which is purely perturbative.
In both figures, the evolution scales are set to be $\mu = \sqrt\zeta = \ktcut$,
and the TMD PDF is for a down quark. 
See \fig{more_np_uncertainty} in \app{more_results_tmdpdf} for the results for other light quark flavors
including $u, \bar{d}, \bar{u}$, and $s$.
}
\label{fig:tmd_np_uncertainty_ktcut_x}
\end{figure}

We can now compare our approximated cumulant to the collinear PDF
in order to quantify the deviation from our naive expectation \eq{tmd_pdf_integral_naive_expectation}.
In the left panel of \fig{tmd_np_uncertainty_ktcut_x},
we show the normalized deviation of $K^{(3)} [f_i^{(0)}]$ from the collinear PDF
as a function of the transverse-momentum cutoff $\ktcut$
while keeping $x=0.01$ fixed.
In the right panel of \fig{tmd_np_uncertainty_ktcut_x},
we show the same deviation from the collinear PDF as a function of the momentum fraction $x$,
while keeping $\ktcut = 10 \GeV$ fixed.
In both cases we set $\mu = \sqrt\zeta = \ktcut$.
We find that the central value of our prediction (solid black)
only deviates from the collinear PDF at most at the percent level for $\ktcut \geq 5 \GeV$,
and for $x = 0.01$ is essentially equal to the PDF for $\ktcut \geq 10 \GeV$.
For $x = 10^{-3}$ ($x = 0.1$) we find a small positive (negative) deviation.

We consider two sources of nonperturbative uncertainty on our result in this section,
deferring an estimate of perturbative uncertainties to \sec{delta_res}.
The first uncertainty component $\Delta_\cut$, shown in blue in \fig{tmd_np_uncertainty_ktcut_x},
is estimated by varying $\btcut$
from $b_0/(1.5 \GeV)$ to $b_0/(1 \GeV)$.
Since the residual $\btcut$ dependence is intrinsically beyond our working order,
it parametrizes our ignorance of the long-distance behavior of the TMD PDF.

The second source of nonperturbative uncertainty is the behavior
of the TMD PDF at short to intermediate distances $\lesssim \btcut$,
where the TMD PDF is governed by the OPE.
We can assess the uncertainty related to this region
in a fully model-independent way by treating
the quadratic OPE coefficients $\gamma_{\zeta,i}^{(2)}$ and $\C_i^{(2)}$ in \eq{tmdpdf_evolved_expanded} as unknowns
and propagating them into the cumulative distribution in momentum-space
by injecting the corresponding OPE term into our truncated $K^{(3)}$ functional.
Importantly, the action of the functional on each term in brackets in \eq{tmdpdf_evolved_expanded}
is linear, i.e., we have, suppressing the arguments of the TMD PDF,
\begin{align} \label{eq:tmdpdf_evolved_expanded_transformed}
K^{(3)}[f_i](\ktcut)
= K^{(3)}[f_i^{(0)}](\ktcut)
+ \C_i^{(2)} K^{(3)}[b_T^2 f_i^{(0)}]
+ \frac{1}{2} \gamma^{(2)}_{\zeta,i} K^{(3)}[b_T^2 L_\zeta f_i^{(0)}]
+ \ORdsq{\Bigl(\frac{\lqcd}{\ktcut}\Bigr)^4}
.
\end{align}
The impact of $\gamma_{\zeta,i}^{(2)}$ and $\C_i^{(2)}$
on the predicted transverse-momentum integral of the TMD PDF is therefore also linear,
including their a priori unknown sign.
While we consider simple estimates of the unknown coefficients here
to demonstrate the method,
their linear impact makes it completely straightforward
to update the uncertainty estimate (or the central value)
to any determination of the coefficients
from model-based global fits to experimental data or a lattice calculation.
It also makes the uncertainty very cheap to evaluate
since the three functionals in \eq{tmdpdf_evolved_expanded_transformed}
only need to be evaluated once, rather than having to perform a Fourier transform
for each choice of model parameters.

For definiteness, we consider a variation of $\C_q^{(2)}(x) \in [ -1 \GeV^2, 1 \GeV^2 ]$
independent of $x$, which yields an uncertainty $\Delta_\mathrm{NP}^f$.
We stress again that this assumption is straightforward to relax
point by point in the right panel of \fig{tmd_np_uncertainty_ktcut_x} by a trivial rescaling.
To estimate the uncertainty due to higher OPE corrections to the CS kernel,
we adopt a very conservative range
of $\gamma_{\zeta,q}^{(2)} \in [ - (0.5 \GeV)^2, ( 0.5 \GeV)^2]$,
yielding the second OPE-related uncertainty component $\Delta_\mathrm{NP}^\zeta$;
recent model-based global fits~\cite{Scimemi:2019cmh, Bacchetta:2019sam} and the condensate-based estimate from \refcite{Vladimirov:2020umg}
all point to substantially lower values
of $\abs{\gamma_{\zeta,q}^{(2)}} \lesssim 0.05 \GeV^2 \approx (0.22 \GeV)^2$.
For readability, both nonperturbative uncertainty components are added in quadrature in \fig{tmd_np_uncertainty_ktcut_x},
$\Delta_\NP^\zeta \oplus \Delta_\NP^f \equiv \bigl[(\Delta_\NP^\zeta)^2 + (\Delta_\NP^f)^2\bigr]^{1/2}$,
and are shown in yellow.

We find that for $x = 0.01$ the integral of the TMD PDF is fully compatible with the collinear PDF
for all values of $\ktcut$ already within the nonperturbative uncertainties,
while for general $x$ the nonperturbative uncertainty alone only covers part of the difference.
Both uncertainty components quickly decrease as $\ktcut$ increases,
as expected from their respective power counting.

Finally we also include in \fig{tmd_np_uncertainty_ktcut_x}
the cumulative TMD PDF as predicted by the SV~\cite{Scimemi:2019cmh}
and Pavia~\cite{Bacchetta:2019sam} model-based global TMD fits.
We use \texttt{arTeMiDe}~\cite{Scimemi:2017etj} to evaluate the SV fit result in $b_T$ space,
assuming the best-fit N$^3$LL model parameters from \refcite{Scimemi:2019cmh},
and use our in-house setup to perform the Bessel transform.
In the case of the Pavia global fit,
we encounter the difficulty
that our treatment of quark mass thresholds
is different from the one in the \texttt{NangaParbat}~\cite{NangaParbatLandingPage} native fit code,
where the number $n_f$ of active massless flavors is changed discontinuously as a function of $b_T$.
This has a very large effect at large $b_T$
(about $4\%$ at $b_0/b_T \approx 2 \GeV$, and up to $10 \%$ at $b_0/b_T \approx 1 \GeV$),
which distorts the comparison of the genuine nonperturbative effect.
For this reason we have set up an independent implementation of the Pavia fit model,
using the central replica of the N$^3$LL Pavia fit as provided in \refcite{NangaParbatPV19xN3LLReport},
and combined it with our implementation of $f_i^{(0)}$ in \texttt{SCETlib}
evaluated with the $b^*_\mathrm{Pavia}$ prescription of \refcite{Bacchetta:2017gcc},
with $b_\mathrm{min} \to 0$ as in \eq{bstar_CS_Pavia}.
We have verified that above the bottom-quark threshold,
our result for $f^{(0)}$ is within a permil of the result returned by \texttt{NangaParbat} for $b_\mathrm{min} = 0$.%
\footnote{The discrepancy to \texttt{arTeMiDe} in the perturbative region,
or when turning off the \texttt{arTeMiDe} nonperturbative model,
is somewhat larger, at roughly half a percent.
We attribute this to the different choice of boundary scale within the so-called $\zeta$ prescription
used in \texttt{arTeMiDe}.}
Both global fits support our conclusion
that the deviation from the naive expectation is within percent level.
In particular, they are compatible with our perturbative baseline
well within our estimate of the nonperturbative uncertainty,
indicating that our model-independent approach is sound.

In \fig{more_np_uncertainty} in \app{more_results_tmdpdf}, 
we show the same plots as in \fig{tmd_np_uncertainty_ktcut_x} for other light quark flavors ($u, \bar{d}, \bar{u}$, and $s$).
We find that the results are very similar to the discussion for the down quark in this section.

\subsection{Perturbative uncertainty}
\label{sec:delta_res}

Since we rely on the accuracy of the perturbative input $f_i^{(0)}$,
it is also important to consider the theoretical uncertainty from the truncation of perturbation theory.
We estimate the perturbative uncertainty at different resummation orders
by varying the boundary scales entering the evolved TMD PDF.
As described below, we adapt the variation procedures developed for the resummed $q_T$ spectrum in \refcite{Ebert:2020dfc},
noting that for our case there are some simplifications compared to the original setup.

The TMD PDF at generic $\zeta$, which may be parametrically separate from $\zeta_0 \sim 1/b_T^2$, is a multi-scale problem.
One way of treating it in resummed perturbation theory
is to evaluate it at $\mu_0 \sim \sqrt{\zeta_0} \sim 1/b_T$, where it is free of large logarithms,
and use \eq{tmdpdf_evolved} to evolve it back to the overall $(\mu, \zeta)$.
The uncertainty estimate in that case could be estimated by varying the boundary scales $(\mu_0, \zeta_0)$.

Alternatively, one can consider a factorization of the TMD PDF at generic $\zeta$
into purely collinear and purely soft matrix elements,
\begin{equation} \label{eq:tmd_beam_soft}
f_i(x, b_T, \mu, \zeta) = B_i(x, b_T, \mu, \nu/\sqrt{\zeta}) \sqrt{S^i(b_T, \mu, \nu)}
\, , \end{equation}
which individually are single-scale problems and feature an additional rapidity renormalization scale $\nu$.
In the SCET literature, these matrix elements are known as beam and soft functions.%
\footnote{We use the exponential regulator of \refcite{Li:2016axz}
that renders the overlap subtraction between the beam and soft function scaleless,
see the discussion in Section~2 of \refcite{Ebert:2019okf}.
More details on rapidity regulators can be found in Appendix~B of \refcite{Ebert:2019okf}.}
In this setup, the running in $\nu$ from the canonical beam scale $\nu_B \sim \sqrt{\zeta}$
to the soft scale $\nu_S \sim 1/b_T$ resums the large logarithms of $\zeta b_T^2$.
For the central prediction, where all scales are set exactly to their canonical values,
this is exactly equivalent to running the TMD PDF as a whole as described below \eq{tmdpdf_evolved}.
The two approaches differ, however, once the possible variations of the boundary scales are considered.
In particular, the factorized form in \eq{tmd_beam_soft} allows
for varying all of $\mu_{B,S}$ and $\nu_{B,S}$ individually,
providing a more robust estimate of the perturbative uncertainty
that makes use of the full available information about the factorization structure.
(In practice, to estimate the perturbative uncertainty on the leading OPE term $f_i^{(0)}$
we of course only need to evaluate \eq{tmd_beam_soft} at leading power in $\lqcd b_T$.)

The procedure of \refcite{Ebert:2020dfc} is to consider the variations%
\footnote{Note that in \refcite{Ebert:2020dfc}, hybrid profile scales were used to turn off the resummation as $q_T \to Q$
to implement the matching to power corrections of $\ord{q_T^2/Q^2}$ (also known as the $Y$ term).
For a single TMD PDF, which is already one of the objects appearing in the leading-power factorization,
these corrections are absent, so we can keep the resummation on everywhere.
This amounts to taking $x_{1,2,3} \to \infty$ in the profile scales in \refcite{Ebert:2020dfc}.}
\begin{align}
\mu_B = 2^{v_{\nu_S}}\, \frac{b_0}{b_T}
\, , \qquad
\nu_B = 2^{v_{\nu_S}}\, \sqrt{\zeta}
\, , \qquad
\mu_S = 2^{v_{\mu_S}} \, \frac{b_0}{b_T}
\, , \qquad
\nu_S = 2^{v_{\nu_S}}\, \frac{b_0}{b_T}
\, ,\end{align}
where the four variation exponents can be $v_i = \{-1, 0, 1\}$.
The central scale choices are given by $(v_{\mu_S}, v_{\nu_S}, v_{\mu_B}, v_{\nu_B}) = (0,0,0,0)$.
A priori there are 80 possible different combined variations of the $v_i$.
Since the arguments of the resummed logarithms are ratios of scales, e.g.\ $\nu_B/\nu_S$,
some combinations of the $v_i$'s lead to variations of these arguments
by more than the conventional factor $2$ and are therefore excluded,
leaving us with 36 different combined scale variations.
We calculate $K^{(3)}[f_i^{(0)}]|_{v_i}$ for each of them
and then take the maximum envelope as our estimate of the perturbative resummation uncertainty $\Delta_\mathrm{res}$.

\begin{figure}
\centering
\includegraphics[width=\WidthTwoSubfigs]{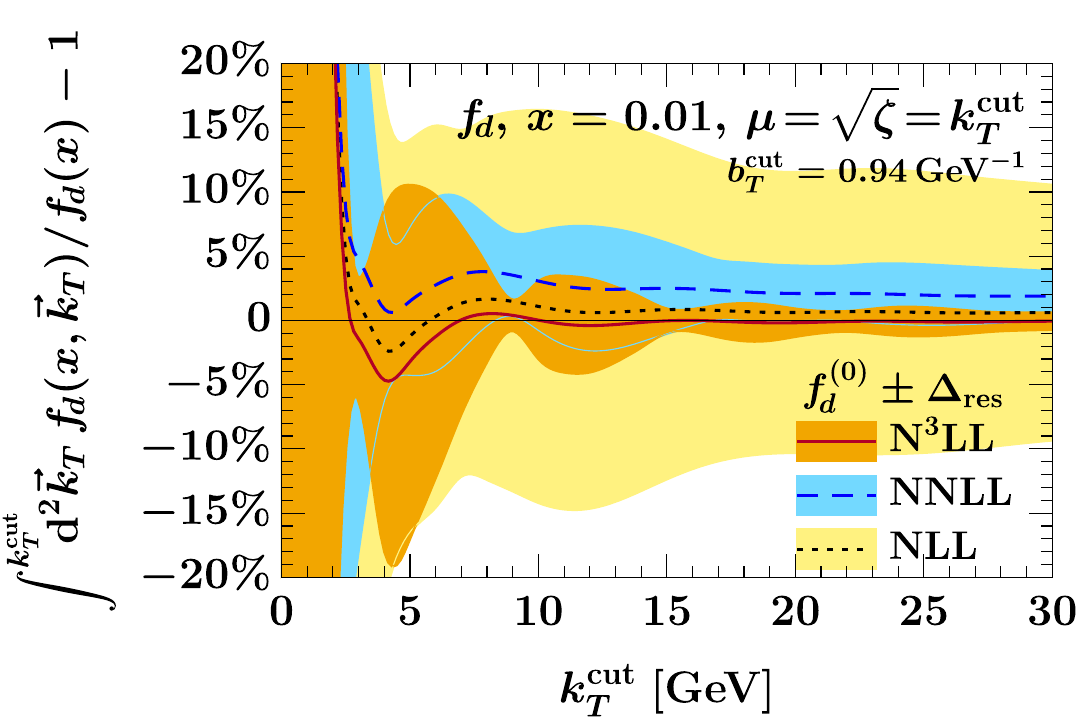}
\hfill
\includegraphics[width=\WidthTwoSubfigs]{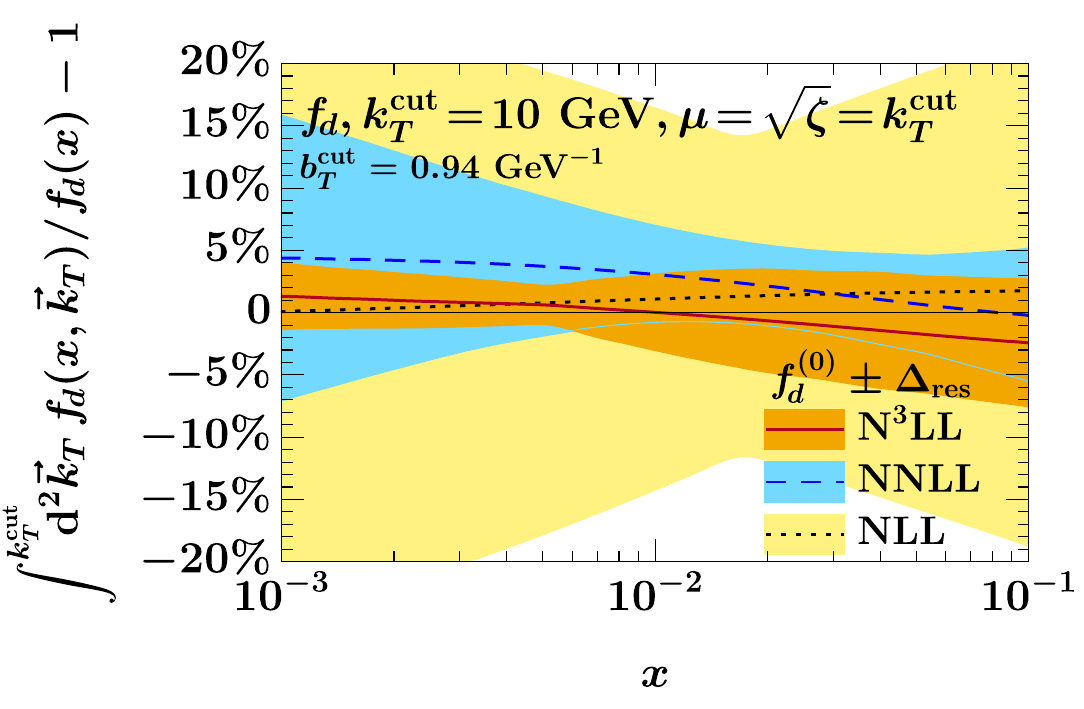}
\caption{%
Similar to \fig{tmd_np_uncertainty_ktcut_x},
but for the perturbative uncertainties from resummation orders,
which are estimated from varying the scales in the boundary condition of the TMD PDF evolution.
When increasing the resummation order, the uncertainty bands converge.
The central values at all orders are within a few percent from the collinear PDF.
See \fig{more_scale_variation} in \app{more_results_tmdpdf} for the results for other light quark flavors
including $u, \bar{d}, \bar{u}$, and $s$.
}
\label{fig:tmd_resum_uncertainty_ktcut_x}
\end{figure}

In \fig{tmd_resum_uncertainty_ktcut_x} we show $\Delta_\mathrm{res}$
for the cumulative TMD PDF calculated at NLL, NNLL, and N$^3$LL
as a function of $\ktcut$ and of $x$.
The parameter choices are the same as in \sec{tmd_cumulant_np}.
When we calculate the perturbative $f^{(0)}_i$ to higher resummation orders,
we see a convergence of the $\Delta_\mathrm{res}$ bands,
giving us confidence in the robustness of the uncertainty estimate.
The central values at all resummation orders are within a few percent from the collinear PDF for $\ktcut \gtrsim 10\GeV$,
and the cumulative TMD PDF is compatible with the collinear PDF well within the perturbative uncertainty for all $\ktcut$ and $x$.
Notice that the perturbative uncertainty in \fig{tmd_resum_uncertainty_ktcut_x}
is much larger than the nonperturbative uncertainties shown in \fig{tmd_np_uncertainty_ktcut_x}.

\Fig{more_scale_variation} in \app{more_results_tmdpdf} shows the same plots as in \fig{tmd_resum_uncertainty_ktcut_x} for other light quark flavors. 
The results are once again very similar to the discussion for the down quark in this section.

\subsection{Impact of evolution effects}
\label{sec:evolution_effects}

In the discussions above, we have chosen $\mu = \sqrt{\zeta} = \ktcut$,
which is the natural scale choice
such that the left-hand side of \eq{tmd_cumulant} only depends on one scale $\ktcut$.
In this section, we would like to consider how evolution away from these natural scales
impacts our conclusion about the transverse-momentum integral of TMD PDFs.

\begin{figure}
\centering
\includegraphics[width=\WidthTwoSubfigs]{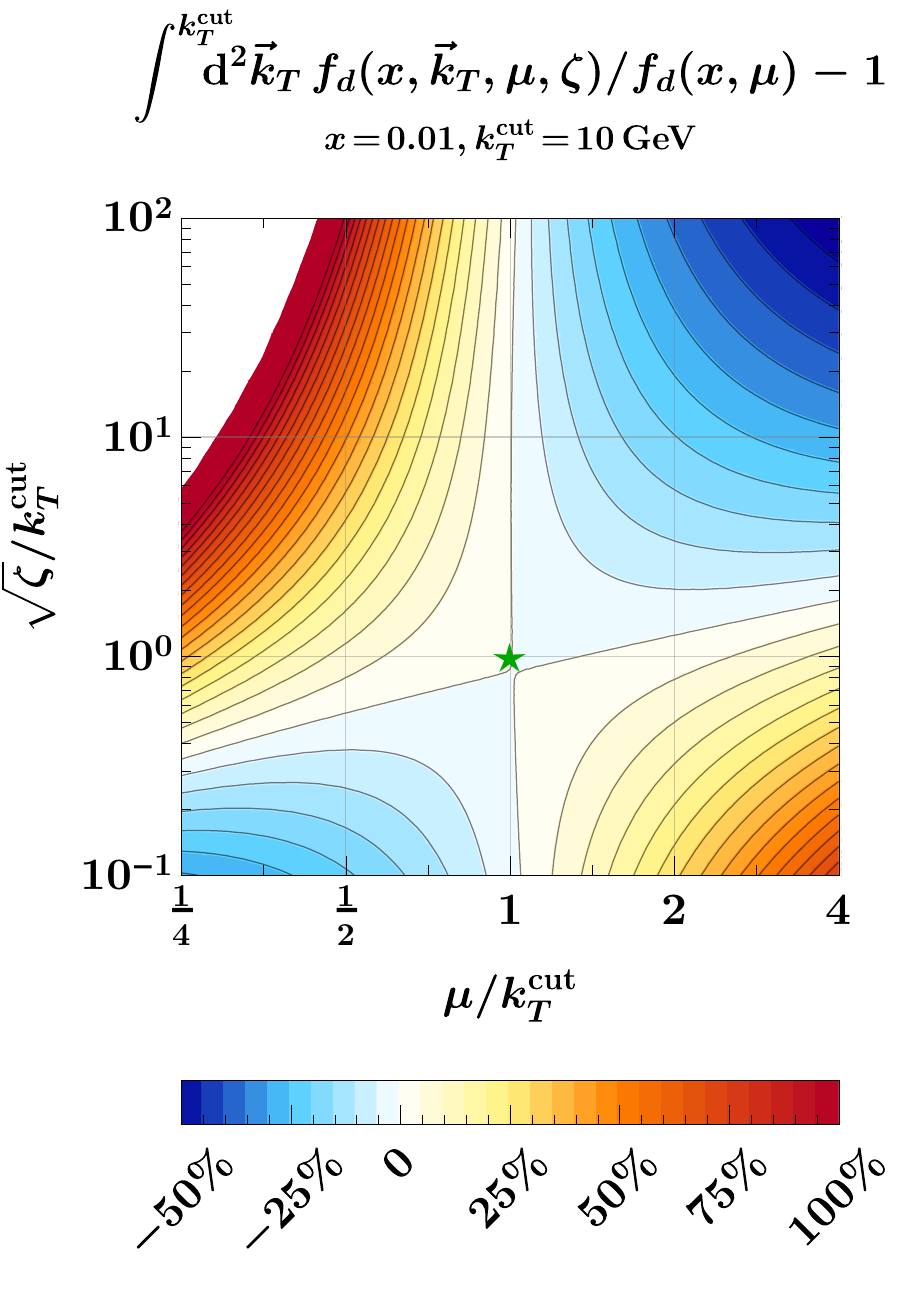}
\caption{%
The normalized deviation of the transverse-momentum integral of the TMD PDF from the collinear PDF
as a function of the evolution scales $\mu$ and $\zeta$.
The green star marks $\mu = \sqrt\zeta = \ktcut$.
The countour lines indicate increments of five percent points.
There are only percent-level deviations in the vicinity of $\mu = \sqrt\zeta = \ktcut$. 
For $\mu = \ktcut$, the evolution of $\zeta$ is negligible and only a shape effect.
There are sizable corrections from evolution away from the central regions.
We approximate the transverse-momentum integral of the TMD PDF with $K^{(3)}[f_i^{(0)}]$,
which is purely perturbative. 
In this figure, we use $\ktcut = 10 \, \mathrm{GeV}$ and $x=0.01$,
and the TMD PDF is for a down quark. 
See \fig{more_heatmap} in \app{more_results_tmdpdf} for the results for other light quark flavors
including $u, \bar{d}, \bar{u}$, and $s$.
}
\label{fig:tmd_mu_zeta_heatmap}
\end{figure}

In \fig{tmd_mu_zeta_heatmap}, we show the normalized deviation of
the cumulative TMD PDF at $\ktcut = 10 \GeV$ from the collinear PDF
as a function of both the renormalization scale $\mu$ and the CS scale $\zeta$. 
We find that, in the vicinity of the natural scale choice $\mu = \sqrt\zeta = \ktcut$,
our conclusion is robust against evolution effects
and the deviation remains within percent level. 
In particular, along the line of fixed $\mu = \ktcut$,
we find that the $\zeta$ evolution is negligible.
This has the intriguing physical interpretation
that the $\zeta$ evolution of the momentum-space TMD PDF is only a shape effect,
i.e., while it changes the shape of the TMD PDF at low $k_T$,
the transverse-momentum integral stays the same and is determined by collinear physics.

When we move away from $\mu = \ktcut$,
sizable deviations from the collinear PDF arise from the large logarithm induced by the cusp anomalous dimension.
To further understand this, consider the evolution path
$(\mu_a,\zeta_a) = (\ktcut, \ktcut) \to (\mu_a,\zeta_b) \to (\mu_b,\zeta_b)$.
Since the $\zeta$ evolution at fixed $\mu_a = \ktcut$ is only a shape effect,
the first evolution step only changes the transverse-momentum integral of the TMD PDF within a few percent.
In the second step, however, the $\mu$ evolution factor
resums large double logarithms of the form
\begin{align}\label{eq:gamma_mu_log}
\exp \biggl[ \int_{\mu_a}^{\mu_b} \! \frac{\df \mu'}{\mu'} \, \gamma^i_\mu(\mu', \zeta_b) \biggr]
= 1 + \frac{\as(\mu_a)}{2\pi} \Bigl[
   \frac{\Gamma_0^i}{4} \ln \frac{\mu_b^2}{\mu_a^2} \ln \frac{\mu_a^2}{\zeta_b}
   + \cdots
\Bigr] + \ord{\as^2}
\,,\end{align}
where the ellipses indicate single logarithms and $\Gamma_\cusp[\as] = \frac{\as}{4\pi} \Gamma_0^i + \ord{\as^2}$
is the one-loop coefficient of the cusp anomalous dimension.
Thus the effect of the $\mu$ evolution is stronger
for a larger ratio of $\sqrt{\zeta_b}/\mu_a$ at the starting point,
which is responsible for the saddle point structure observed in \fig{tmd_mu_zeta_heatmap}.
This confirms the initial observation in \sec{intro}
that an equality between cumulative TMD PDF and collinear PDF
cannot hold at all scales due to the different renormalization,
which resums double logarithms in the case of the TMD PDF.
Nevertheless, a relation in the vicinity of $\mu, \sqrt{\zeta} \sim \ktcut$
is not in contradiction with the renormalization, and is in fact
observed in \fig{tmd_mu_zeta_heatmap} to also extend
to parametrically separate values of $\zeta$ at fixed $\mu = \ktcut$.

We conclude that, even when we take the evolution effects
in the vicinity of $\mu = \ktcut$ into account,
our intuitive result that the transverse-momentum integral of TMD PDF 
is within few percent of the corresponding collinear PDF
remains robust.

\begin{figure}
\includegraphics[width=\WidthTwoSubfigs]{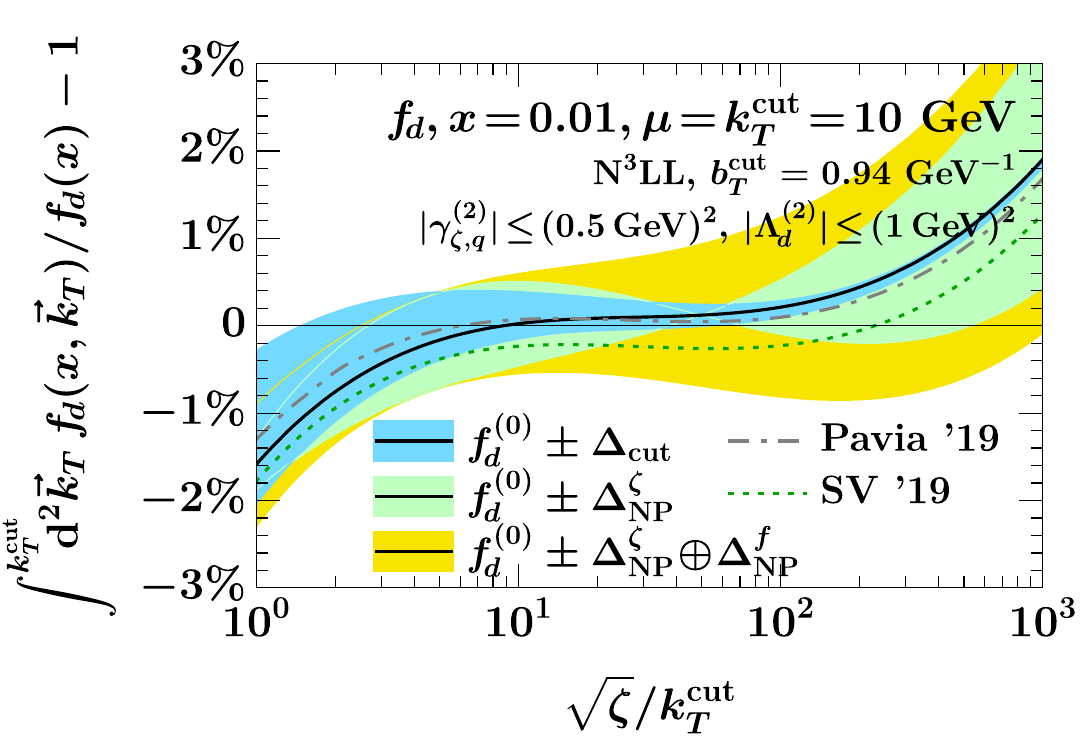}
\hfill
\includegraphics[width=\WidthTwoSubfigs]{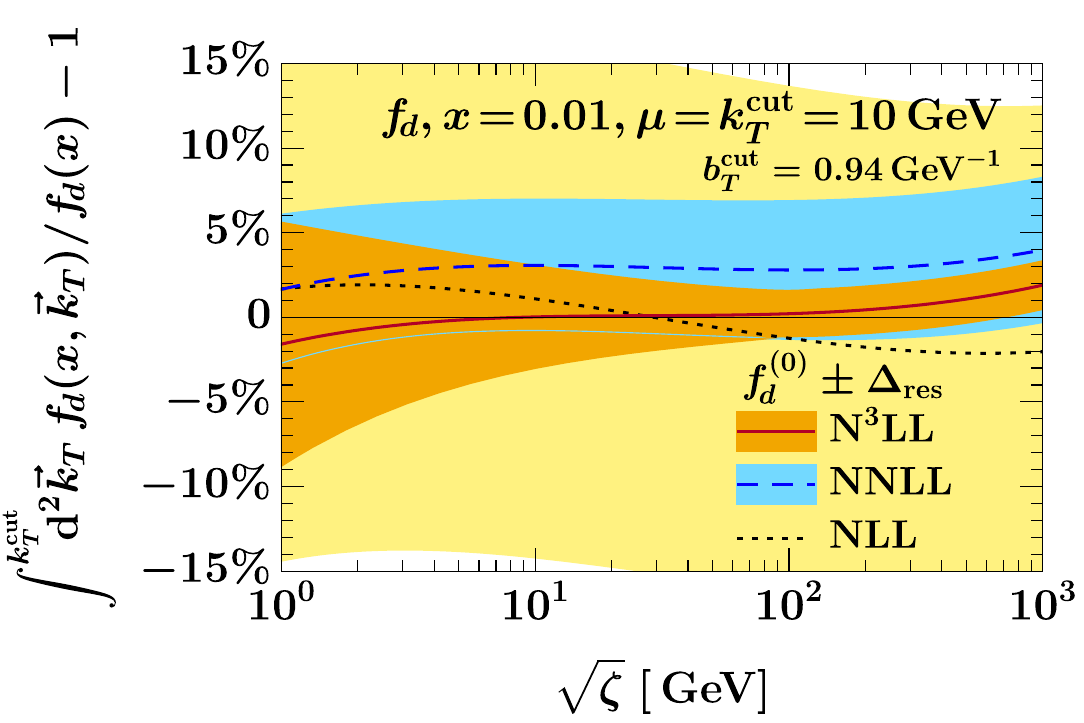}
\caption{%
The normalized deviation of the transverse-momentum integral of the TMD PDF from the collinear PDF
as a function of the evolution scale $\sqrt\zeta$.
We also include the SV and Pavia global fits in the left panel.
We approximate the transverse-momentum integral of the TMD PDF with $K^{(3)}[f_i^{(0)}]$,
which is purely perturbative.
In both figures, we use fixed $\mu = \ktcut = 10\, \mathrm{GeV}$ and $x=0.01$, 
and the TMD PDF is for a down quark.
See \fig{more_zeta_evolution} in \app{more_results_tmdpdf} for the results for other light quark flavors
including $u, \bar{d}, \bar{u}$, and $s$.
}
\label{fig:zeta_evolution_scale_np}
\end{figure}

It is also interesting to look at the $\zeta$ evolution alone.
In \fig{zeta_evolution_scale_np}, 
we show the deviation of the transverse-momentum integral of a TMD PDF from the collinear PDF
as a function of $\sqrt\zeta$,
while keeping $x=0.01$ and $\mu = \ktcut = 10\, \mathrm{GeV}$ fixed.
In the left (right) panel of \fig{zeta_evolution_scale_np},
we consider nonperturbative (perturbative) uncertainties
$\Delta_\mathrm{cut}$, $\Delta_\mathrm{NP}^\zeta$, and $\Delta_\mathrm{NP}^\zeta \oplus \Delta_\mathrm{NP}^f$ ($\Delta_\mathrm{res}$) 
as described in \sec{tmd_cumulant_np} (\sec{delta_res}). 
We see that the perturbative uncertainties from resummation still dominate over
the nonperturbative uncertainties described in \sec{tmd_cumulant_np},
but reduce as the TMD PDF is evolved to very high energies $\sqrt{\zeta} \sim 1 \TeV$.
The $\Delta_\cut$ uncertainty (light blue in the left panel) likewise reduces at large $\sqrt{\zeta}$,
which is expected as the Sudakov suppression of the long-distance region increases,
while the $\Delta_\mathrm{NP}^\zeta$ (light green) component comes to dominate the total OPE uncertainty (yellow)
both at very small, $\zeta \to 1 \GeV$, and very large $\zeta \to 1 \TeV$,
i.e., far away from the natural choice $\zeta \sim \ktcut$.

\Fig{more_heatmap} and \fig{more_zeta_evolution} in \app{more_results_tmdpdf} 
show the same plots as in \fig{tmd_mu_zeta_heatmap} and \fig{zeta_evolution_scale_np} 
for other light quark flavors. 
The results are very similar to the discussion for the down quark here.
Once again, remarkably good agreement is found for the integral of the TMD PDF and the collinear PDF with the most natural choice for fixing the evolution scales.

\FloatBarrier
\section{Model-independent constraints on nonperturbative physics from data}
\label{sec:data}

\subsection{Outline of the method}
\label{sec:data_method_outline}

\begin{figure}
\centering
\includegraphics[width=\WidthTwoSubfigs]{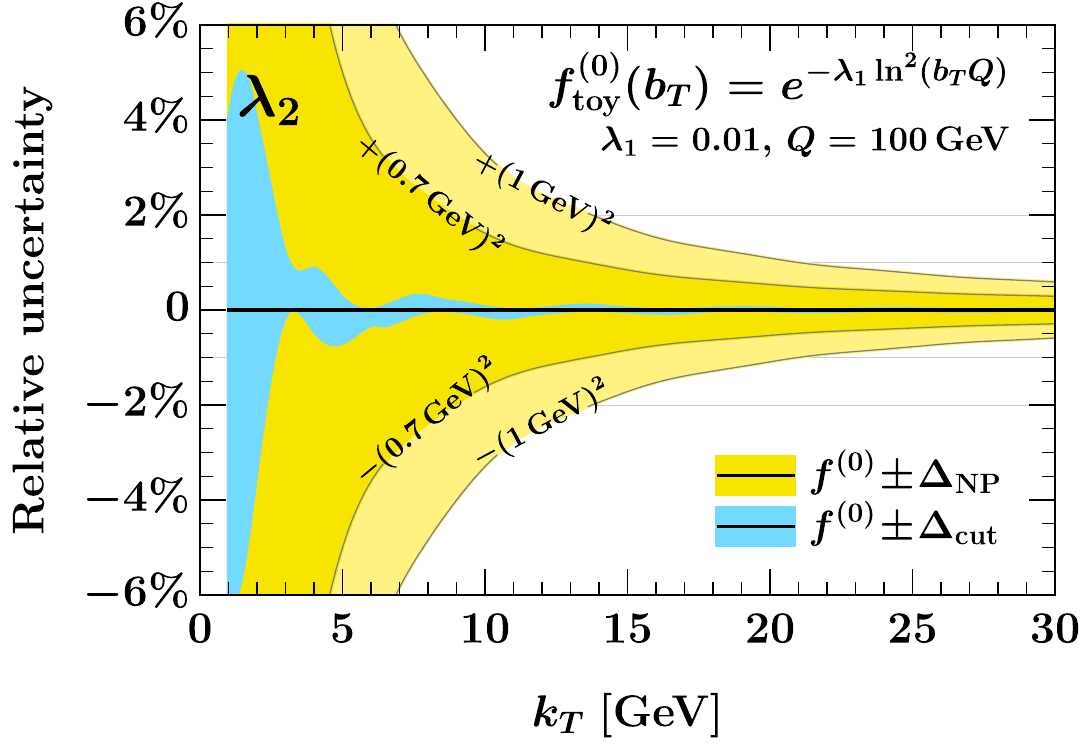}%
\hfill
\includegraphics[width=\WidthTwoSubfigs]{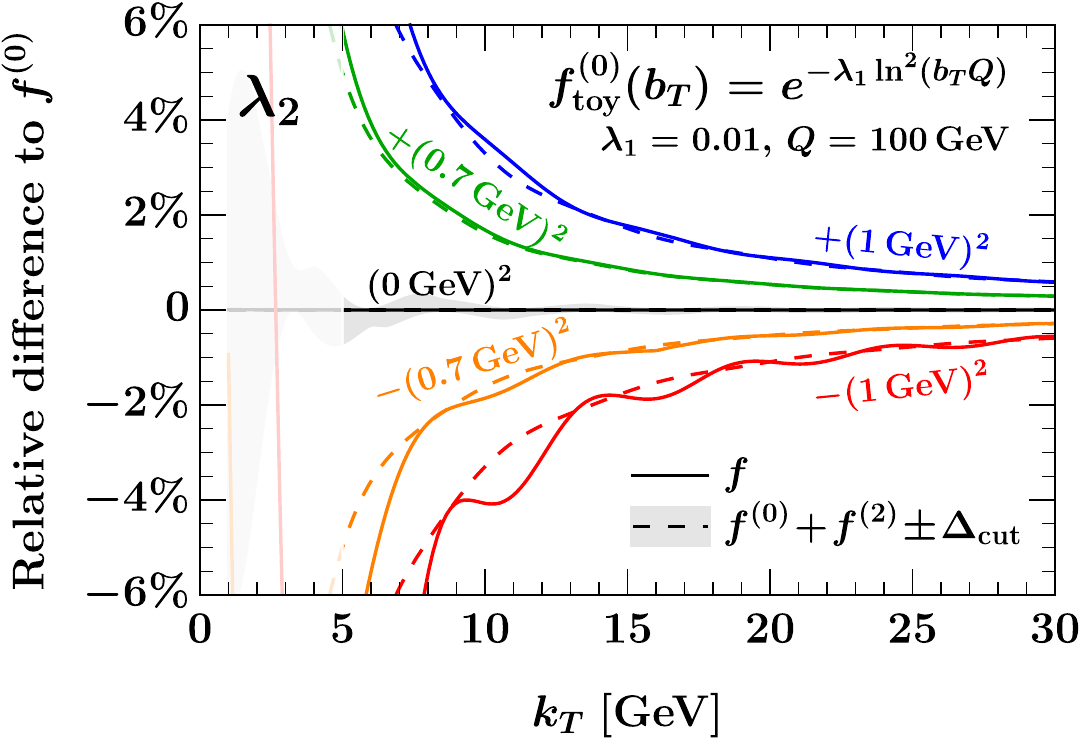}%
\caption{%
Impact plots of quadratic ``OPE'' parameters in the toy function in \eq{toy_function_expansion} interpreted
as nonperturbative uncertainties (left)
or fits to data (right).
We consider the impact of the ``nonperturbative'' parameter $\lbb$ on the toy model in \eq{toy_test_function} with the expansion in \eq{toy_function_expansion}.
To produce ``data'' (solid lines on the right) for coefficients $\lbb > 0$,
where the original definition has a divergent Fourier transform,
we have modified the toy function as described in the text, see \eq{toy_function_modified}.
We use a third-order functional $S^{(3)}$ throughout.
The innermost uncertainty bands in light blue or gray, respectively,
indicate $\Delta_\mathrm{cut}$,
which for clarity is only shown for $\lbb = 0$ in the right panel.
}%
\label{fig:toy_signal}
\end{figure}

In the previous section we used the action of our truncated functionals
on the quadratic terms in the OPE to provide a model-independent estimate of nonperturbative uncertainties
in cumulative TMD PDFs, treating the coefficients as unknowns.
This approach of course remains valid for transverse-momentum cross sections.
In this section we explain how information on nonperturbative contributions to cross sections can be isolated,
illustrating the steps with the toy model $\sigma = f_\mathrm{toy}$ in \eq{toy_function_expansion}.
In the left panel of \fig{toy_signal}
we show the transverse-momentum spectrum $S[f_\mathrm{toy}]$,
where we vary the quadratic coefficient $\lbb$ within two different ranges of values, as indicated by the yellow bands.
We normalize the impact of $\lbb$ to a baseline given by the third-order truncated functional
applied to the purely ``perturbative'' component of the function, $S^{(3)}[f^{(0)}]$.
We see that the relative impact of $\lbb$ is slightly larger than expected from power counting,
e.g., it amounts to $\sim 2 \%$ at $k_T = 10 \GeV$ for $\lbb = 1 \GeV$,
which we attribute to the falloff of the spectrum beyond the Sudakov peak
that is present also for this toy function.

There is, however, a second way to make use of the linear impact of the quadratic OPE coefficients in our setup when the transverse-momentum spectrum is known from \emph{data}.
In this case, a fit to data may be performed treating the OPE coefficients as a ``signal''
that can be discriminated from the purely perturbative ``background'' in a model-independent way.
Importantly, since the action of the functional is linear,
the fit template for the signal only needs to be computed once
and can then be trivially rescaled in the fit, e.g. for our toy function in \eq{toy_function_expansion},
\begin{align}
S^{(3)}[f_\mathrm{toy}^{(0)} + f_\mathrm{toy}^{(2)}]
= S^{(3)}[f_\mathrm{toy}^{(0)}] - \lbb S^{(3)}[b_T^2 f_\mathrm{toy}^{(0)}]
\,,\end{align}
where $\lbb$ (the combined ``OPE'' coefficient at the quadratic order)
would be the parameter of interest in the fit.
We illustrate this approach in the right panel of \fig{toy_signal},
where the dashed lines are the combined fit template
for signal and background for different values of $\lbb$.
To generate illustrative ``data'' (solid lines) for $\lbb < 0$,
where the original toy function in \eq{toy_test_function} has a divergent Fourier transform,
we modify it as
\begin{align} \label{eq:toy_function_modified}
f_\mathrm{toy}(b_T)
= f^{(0)}_\mathrm{toy}(b_T) \, \begin{cases}
   \exp(-\lbb b_T^2)
   \,, \quad &
   \lbb > 0
   \,, \\
   (1 + 2 \abs{\lbb} b_T^2) \exp(-\abs{\lbb}b_T^2)
   \,, \quad &
   \lbb < 0
   \,, \end{cases}
\end{align}
where $f_\mathrm{toy}^{(0)} = \exp[-\laa \ln^2 (b_T Q)]$ is unchanged.
This ensures that the full integral $S[f_\mathrm{toy}]$ exists for all $\lbb$
while maintaining $f^{(2)}(b_T) = - \lbb b_T^2$ for either sign.
We see that the fit template predicted by the truncated functionals
closely follows the mock data in \fig{toy_signal} for all values of $k_T$ and $\lbb$,
with the exception of $\lbb = -(1 \GeV)^2$ where our modification in \eq{toy_function_modified}
results in a strong modification of $f_\mathrm{toy}$ at large distances,
leading to a larger oscillatory error term.

\subsection{Proposed application to LHC Drell-Yan data}
\label{sec:data_proposal_drell_yan_lhc}

We next outline an application of our method aimed at constraining nonperturbative corrections using LHC data for the Drell-Yan process,
$pp \to Z/\gamma^* \to \ell^+ \ell^-$,
which are some of the most precise data taken at the LHC to date~\cite{CMS:2019raw, ATLAS:2019zci}.
At leading power in $q_T \ll Q$, but to all orders in $\lqcd / q_T$,
the fiducial cross section differential in the dilepton transverse momentum reads
\begin{align} \label{eq:fiducial_spectrum}
\frac{\df \sigma}{\df q_T}
= \int \! \df \Phi_B \, A_B(\Phi_B) \, S\bigl[\sigma(Q, Y, b_T)\bigr](q_T) \, \Bigl[ 1 + \ORd{\frac{q_T}{Q}} \Bigr]
\,,\end{align}
where $\Phi_B = \{Q, Y, \cos \theta\}$ is the Born phase-space for a dilepton pair
with total invariant mass $Q$ and total rapidity $Y$,
where the lepton is scattered at a polar angle $\theta$
measured in the Collins-Soper frame~\cite{Collins:1977iv}.
(For the Born kinematics considered here, the Collins-Soper frame coincides
with the frame boosted by $Y$ along the beam axis from the lab frame.)
In \eq{fiducial_spectrum}, we have made power corrections in $q_T/Q$ explicit,
but will suppress these in the following.
The acceptance of the fiducial volume evaluated on Born-level kinematics with $q_T = 0$
is denoted by $A_B(\Phi_B)$.
For a typical fiducial volume at the LHC,
\begin{align}
\df \Phi_B
&= \df Q \, \df Y \, \df \cos \theta \, \frac{3}{8} \bigl(1 + \cos^2 \theta\bigr)
\,, \nn \\
A_B(\Phi_B)
&= \Theta\bigl(Q_\mathrm{min} \leq Q \leq Q_\mathrm{max}\bigr) \, \Theta\bigl(p_T^{1,2} \geq p_T^\mathrm{cut}\bigr) \, \Theta\bigl(\abs{\eta_{1,2}} \leq \eta_\mathrm{cut}\bigr)
\,,\end{align}
where $Q_{\mathrm{min},\mathrm{max}}$ define an invariant-mass bin (e.g.\ around the $Z$ resonance)
and the lepton transverse momenta $p_T^1 = p_T^2 = \sin \theta \, Q/2 $ and pseudorapidities $\eta_{1,2} = Y \mp \ln \bigl( \tan \theta/2 \bigr)$
are required to be within the fiducial detector volume.
(The extension to bins of rapidity as e.g.\ in \refcite{CMS:2019raw} is trivial.)
In \eq{fiducial_spectrum} we have used our shorthand $S[\dots](q_T)$ for the integral functional in \eq{S_functional}
that performs the Fourier transform from position space.
It acts on the factorized unpolarized cross section in position space~\cite{Collins:1984kg},
\begin{align} \label{eq:factorized_cross_section}
\sigma(Q, Y, b_T)
= H(Q^2, \mu) \sum_{i,j} \sigma^B_{ij}(Q) \, f_i(x_a, b_T, \mu, \zeta_a) \, f_j(x_b, b_T, \mu, \zeta_b)
\end{align}
where $\sigma^B_{ij}(Q)$ is the Born cross section for the hard production process $ij \to Z/\gamma^*$
that includes the $Q$-dependent $Z$ lineshape,
$H(Q^2, \mu) = 1 + \ord{\as}$ is the so-called hard function that encodes virtual corrections
to the hard process,%
\footnote{Here we have neglected singlet contributions to the quark axial and vector form factors
starting at $\ord{\as^2}$ and $\ord{\as^3}$, respectively,
where the initial-state flavors $i,j$ do not couple directly to the $Z$ boson.
These contributions are straightforward to restore for the high-precision background prediction,
but can safely be neglected in the recasting step
for a measured $\overline{\C}^{(2)}$ coefficient that we outline below.}
and the sum runs over all contributing quark flavors.
The two TMD PDFs on the right-hand side of \eq{factorized_cross_section} are evaluated at
\begin{align}
x_{a,b} = \frac{Q}{\Ecm} \, e^{\pm Y}
\,, \qquad
\zeta_{a,b} = Q^2
\end{align}
with $\Ecm$ the total center-of-mass energy of the proton-proton collision.
Inserting \eq{tmdpdf_evolved_expanded},
it is straightforward to give the factorized unpolarized cross section to $\ord{\lqcd^2 b_T^2}$,
\begin{align} \label{eq:factorized_cross_section_expanded_Ci_Cj}
\sigma(Q, Y, b_T)
&= H(Q^2, \mu) \sum_{i,j} \sigma^B_{ij}(Q) \, f_i^{(0)}(x_a, b_T, \mu, \zeta_a) \, f_j^{(0)}(x_b, b_T, \mu, \zeta_b)
\nn \\ &\quad \times
\Bigl\{
   1
   +  b_T^2 \Bigl( \C_i^{(2)}(x_a) + \C_j^{(2)}(x_b) + \gamma^{(2)}_{\zeta,q} L_{Q^2} \Bigr)
\Bigr\}
+ \Ordsq{(\lqcd b_T)^4}
\,,\end{align}
where $L_{Q^2} = \ln (b_T^2 Q^2/b_0^2)$.

If we try to use \eq{factorized_cross_section_expanded_Ci_Cj} in \eq{fiducial_spectrum} directly,
we face a problem commonly encountered in extractions of (TMD)PDFs,
namely that the flavor and $x_{a,b}$ dependence of the model,
in this case the $\C_{i,j}^{(2)}(x_{a,b})$,
only enters under the sum over flavors
and the integral over hard phase space,
and enters in different ways for different center-of-mass energies.
Resolving this usually requires writing down a model-specific ansatz
and performing a global fit to a large data set at once to constrain it.
We now demonstrate that thanks to the linear impact of the OPE contribution to the TMD PDF at quadratic order,
its $x$ and flavor dependence can be condensed into a single aggregate coefficient $\overline{\C}^{(2)}$ per fiducial volume
that can be determined by a two-parameter fit with $\gamma_{\zeta,i}^{(2)}$ to the data in that single fiducial volume.
We also demonstrate that the measured central value, or limits, for $\overline{\C}^{(2)}$
can be straightforwardly recast and checked against specific models
by performing a simple weighted tree-level integral over the model function.

To set up some more notation, let us define the total fiducial cross section in $b_T$ space,
\begin{align}
\sigma(b_T) \equiv \int \! \df \Phi_B \, A(\Phi_B) \, \sigma(Q, Y, b_T)
\,,\end{align}
which is possible because we consider the $q_T$-independent Born acceptance $A(\Phi_B)$.
We then simply have $\df \sigma/\df q_T = S[\sigma(b_T)](q_T)$.
Our goal in the following is to break down the quadratic contribution $\sigma^{(2)}(b_T)$ to $\sigma(b_T)$ into two terms $\sigma^{(2,\gamma_\zeta)} + \sigma^{(2,\C)}$ with a single, overall coefficient each encoding the nonperturbative physics.
This is straightforward for the contribution from the rapidity anomalous dimension,
\begin{align}
\sigma^{(2, \gamma_\zeta)}(b_T)
= \gamma_{\zeta,q}^{(2)} b_T^2 \int \! \df \Phi_B \, L_{Q^2} \, A(\Phi_B) \, \sigma^{(0)}(Q, Y, b_T)
\,,\end{align}
because it is flavor independent and thus only involves a simple
logarithm of $Q$ as an additional weight under the hard phase-space integral
over the leading-power cross section.
To cast the contribution from the intrinsic TMD PDFs into the same form, we define
\begin{align}
\sigma^{(2, \C)}(b_T) = 2\overline{\C}^{(2)} b_T^2 \, \sigma^{(0)}(b_T)
\,,\end{align}
where to match \eq{factorized_cross_section_expanded_Ci_Cj} the average coefficient must be defined by
\begin{align} \label{eq:def_Cbar}
\overline{\C}^{(2)} \!=\! \frac{
   \int \!\! \df \Phi_{\!B}  A(\Phi_{\!B}) H(Q^2, \mu)
    \sum_{i,j}\limits \sigma^B_{ij}(Q)
    f_i^{(0)}\!(x_a, b_T, \mu, \zeta_a)
    f_j^{(0)}\!(x_b, b_T, \mu, \zeta_b)
    \bigl[ \C_i^{(2)}(x_a) \!+\! \C_j^{(2)}(x_b) \bigr]
}{
   \int \!\! \df \Phi_{\!B}  A(\Phi_{\!B}) H(Q^2, \mu)
    \sum_{i,j}\limits \sigma^B_{ij}(Q)  
    f_i^{(0)}\!(x_a, b_T, \mu, \zeta_a) 
    f_j^{(0)}\!(x_b, b_T, \mu, \zeta_b) 
}
.\end{align}
The fact that it is simply a ratio of (weighted) leading-power integrals
is due to the linear impact of $\C_{i,j}(x_{a,b})$ on the cross section point by point in phase space.
It is always possible to evaluate the right-hand side of \eq{def_Cbar}
for a given model to compare it to a measured value of $\overline{\C}^{(2)}$.

However, given the expected sensitivity to $\overline{\C}^{(2)}$ at the LHC (discussed below),
we anticipate that the right-hand side of \eq{def_Cbar} will only be required
at relatively low accuracy to compare (limits on) $\overline{\C}^{(2)}$ to models.
In this case, we can neglect logarithmic $Q$ dependence across the $Q$ bin
in several places to a good approximation,
namely in the hard function boundary condition,
in its evolution down to the TMD PDF scale $\mu_0 \sim b_0/b_T$,
and in the leading-power rapidity evolution factor.
As a consequence, all of these terms cancel in the ratio
since we can effectively evaluate them at some reference value of $Q$ within the bin.
We can also evaluate the intrinsic leading-power TMD PDF at NLL,
leaving only
\begin{align} \label{eq:recasting_Cbar_nll}
\overline{\C}^{(2)}
\stackrel{\text{NLL}}{=} \frac{
   \int \! \df \Phi_B \, A(\Phi_B) \, \sum_{i,j} \sigma^B_{ij}(Q) \, f_i^{(0)}(x_a, \mu_0) \, f_j^{(0)}(x_b, \mu_0) \bigl[ \C_i^{(2)}(x_a) + \C_j^{(2)}(x_b) \bigr]
}{
   2 \int \! \df \Phi_B \, A(\Phi_B) \, \sum_{i,j} \sigma^B_{ij}(Q) \, f_i^{(0)}(x_a, \mu_0) \, f_j^{(0)}(x_b, \mu_0)
}
\,.\end{align}
Both the numerator and the denominator are a straightforward
tree-level integral of collinear PDFs against the $Z$ line shape and the acceptance,
with an additional weight of $\C_{i,j}^{(2)}(x_{a,b})$ in the numerator.
The residual dependence on $\mu_0$ can be used as a handle on the theory uncertainty
during this recasting step, where $\mu_0 \sim b_0/b_T$ should be picked
near the $b_T$-space Sudakov peak, $b_T \sim 0.2 \GeV^{-1}$.

\begin{figure}
\centering
\includegraphics[width=\WidthTwoSubfigs]{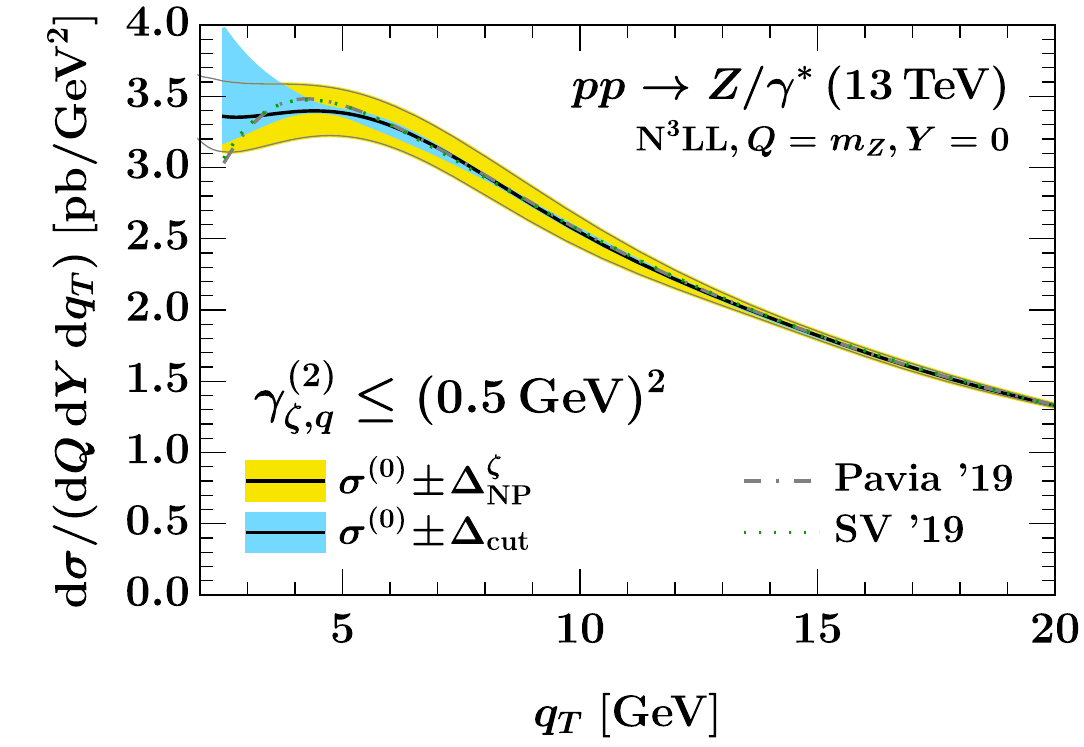}%
\hfill
\includegraphics[width=\WidthTwoSubfigs]{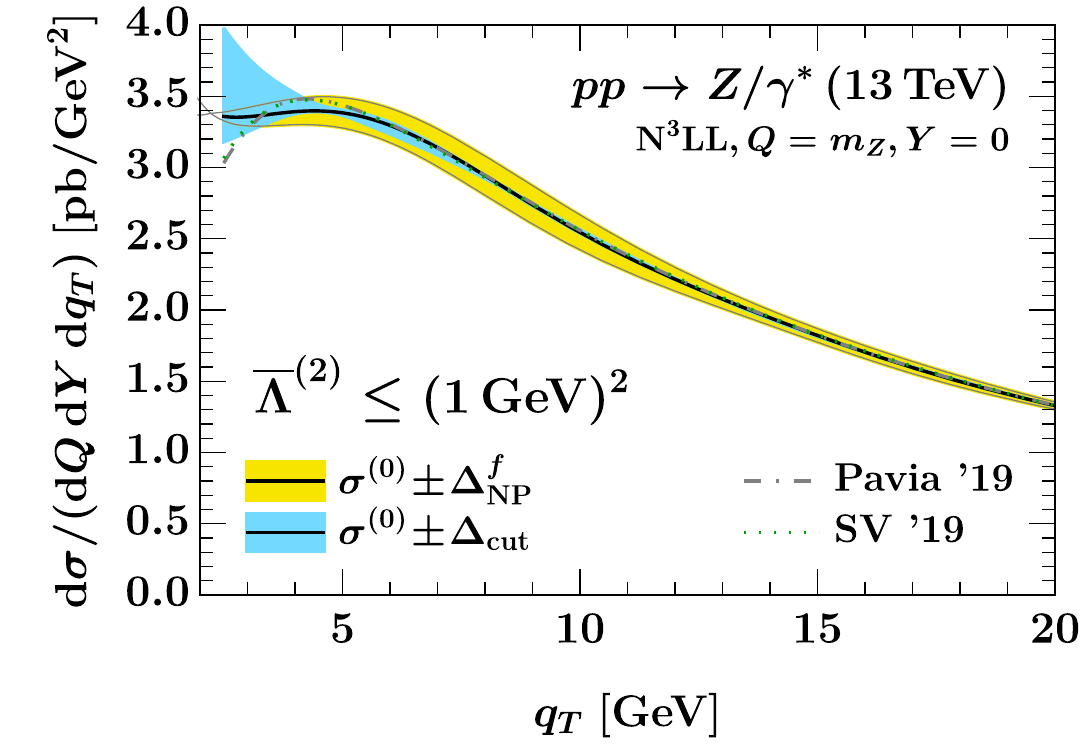}%
\\
\includegraphics[width=\WidthTwoSubfigs]{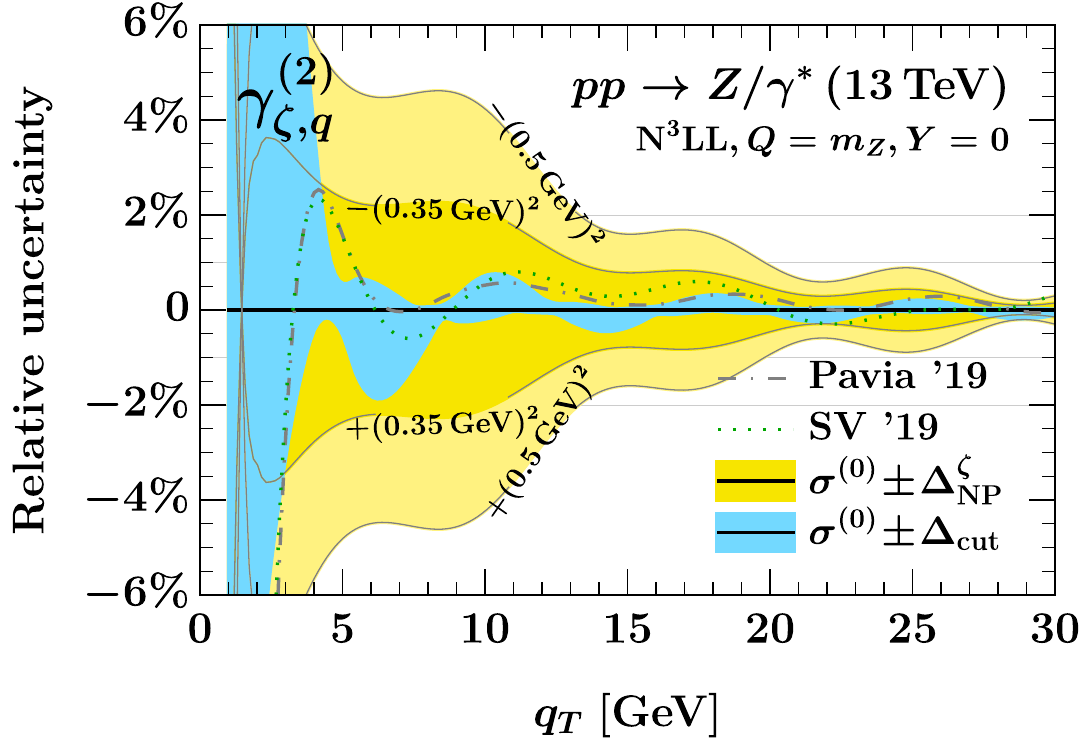}%
\hfill
\includegraphics[width=\WidthTwoSubfigs]{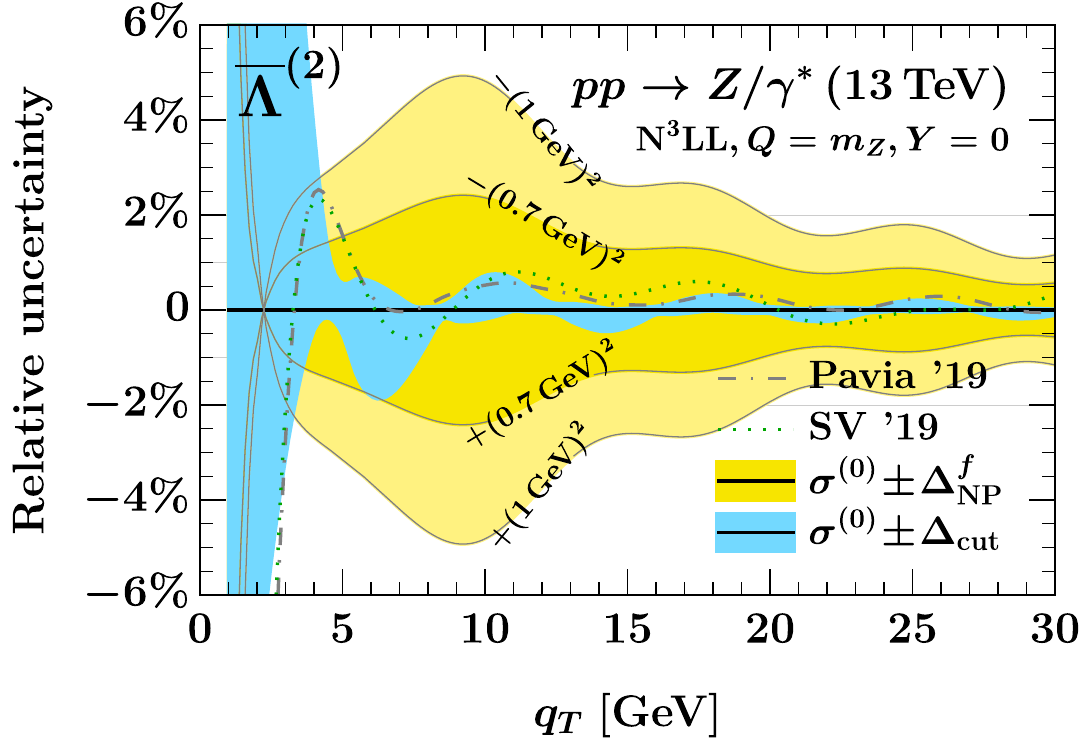}%
\\
\includegraphics[width=\WidthTwoSubfigs]{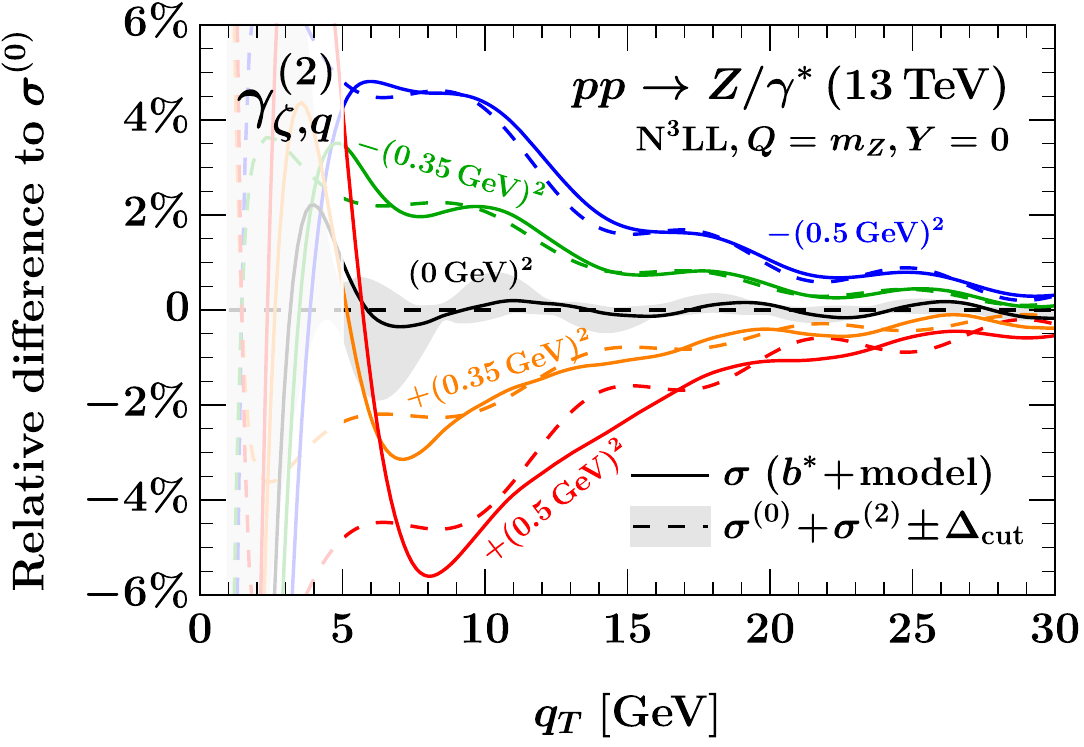}%
\hfill
\includegraphics[width=\WidthTwoSubfigs]{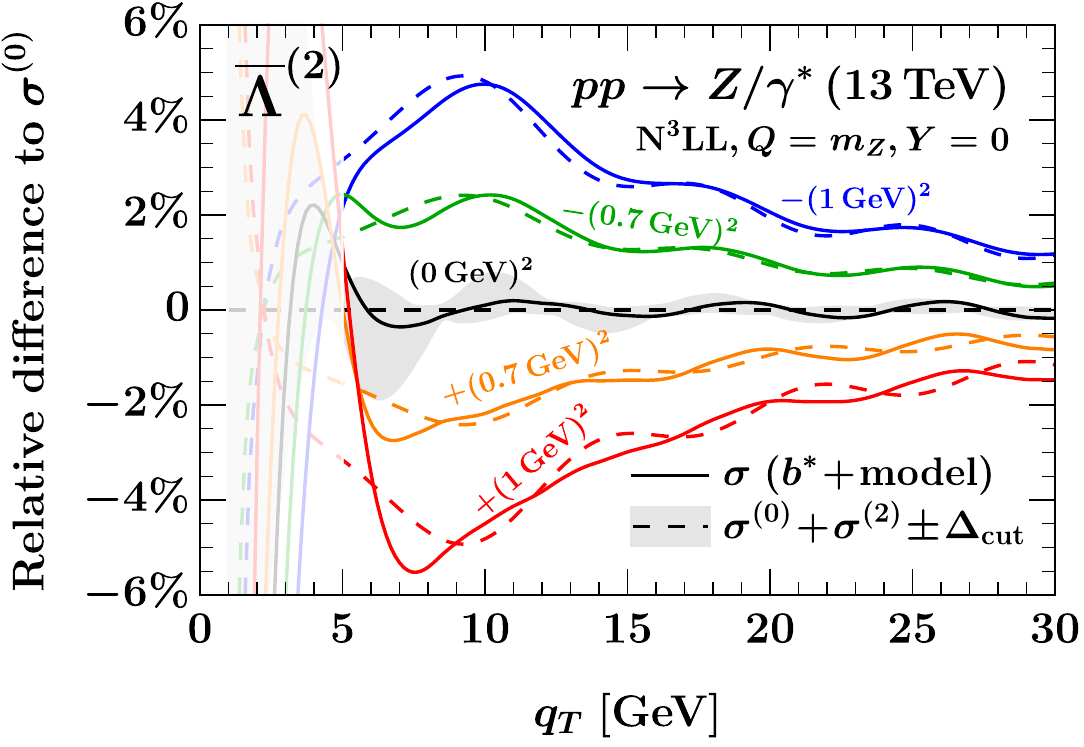}%
\caption{%
Impact plots of quadratic OPE parameters in the LHC Drell-Yan cross section interpreted
as nonperturbative uncertainties (top and middle row)
or fits to data (bottom).
We consider the impact of the quadratic coefficient in the rapidity anomalous dimension (left column)
and the intrinsic TMD PDF averaged over hard phase space (right column) on the physical N$^3$LL cross section.
Our averaging procedure for the intrinsic TMD PDF coefficient $\overline{\C}^{(2)}$
is defined in \eq{def_Cbar} and discussed there.
We generate mock data (solid lines in bottom row) as described around \eq{np_gaussian_model_functions_modified}.
For clarity the uncertainty band for $\Delta_\mathrm{cut}$
(light blue band in bottom and middle row, gray in bottom row) is only shown
for the central fit template in the bottom row.
For nonzero values of the nonperturbative parameters of interest, the band looks essentially identical.
All OPE results use a third-order functional $S^{(3)}$.
}%
\label{fig:real_signal}
\end{figure}

In total, our $b_T$-space input for a given fiducial volume integrated over hard phase space reads
\begin{align} \label{eq:factorized_cross_section_expanded_Cbar}
\sigma(b_T) = \sigma^{(0)}(b_T)
\Bigl\{
   1
   + b_T^2 \Bigl( 2 \overline{\C}^{(2)} + \gamma^{(2)}_{\zeta,q} L_{Q^2} \Bigr)
\Bigr\}
+ \Ordsq{(\lqcd b_T)^4}
\,.\end{align}
Applying our truncated functionals at order $n = 3$ and exploiting their linearity, we have,
\begin{align} \label{eq:fiducial_spectrum_final_result}
\frac{\df \sigma}{\df q_T}
&= S^{(3)}\big[\sigma^{(0)}(b_T)\big](q_T)
+ 2\overline{\C}^{(2)} \, 
  S^{(3)}\big[b_T^2 \sigma^{(0)}(b_T)\big](q_T)
+ \gamma_{\zeta,q}^{(2)} \: 
  S^{(3)}\big[b_T^2 L_{Q^2} \sigma^{(0)}(b_T)\big](q_T)
\nn \\ & \quad
+ \frac{1} {q_T} \ORdsq{(q_T \btcut)^{-\frac{5}{2}}, \Bigl (\frac{\lqcd}{q_T} \Bigr)^4\,}
\,.\end{align}
We now illustrate the numerical impact of the two remaining nonperturbative parameters,
working at fixed $Q = m_Z$ and $Y = 0$ and applying no fiducial cuts for simplicity,
i.e., we consider
\begin{align} \label{eq:np_gaussian_model_functions_modified}
\df \Phi_B
= \df Q \, \df Y \df \cos \theta \, \frac{3}{8} \bigl( 1 + \cos^2 \theta \bigr) \,
  \delta(Q - m_Z) \, \delta(Y)
\,, \qquad
A_B = 1
\,.\end{align}
The recasting relation in \eq{recasting_Cbar_nll}
in this scenario reduces to a weighted flavor average at $x_{a,b} = m_Z/\Ecm$.
We plot the impact of $\gamma_\zeta^{(2)}$ ($\overline{\C}^{(2)}$)
in the left (right) column of \fig{real_signal},
where in the first two rows we again interpret the impact
of the parameters as a model-independent nonperturbative uncertainty estimate
and compare it to two recent model-based global fits~\cite{Scimemi:2019cmh, Bacchetta:2019sam}
as described in \sec{tmd_cumulant_np}.
The top row shows our uncertainty estimates as bands around our central result
for the Drell-Yan spectrum.
The middle row shows the corresponding relative uncertainty.
We find that both global fits are roughly compatible with the perturbative baseline
already within the truncation uncertainty $\Delta_\mathrm{cut}$,
i.e., setting both nonperturbative parameters to zero.
This result implies that the fits report
a vanishing total effect of (quadratic) nonperturbative parameters for central LHC kinematics.
The impact of these terms is thus smaller than one would estimate purely from dimensional analysis~\cite{Vladimirov:2020umg}.

It would be valuable to check this conclusion with a complementary model-independent method.
This is illustrated in the plots in the bottom row of \fig{real_signal}
to which experimental data would need to be added to determine the favored values.
We plot the signal template (dashed lines)
for different values of the nonperturbative parameters of interest with $\gamma_{\zeta,q}^{(2)}$ on the bottom-left and $\overline \C^{(2)}$ on the bottom-right,
taking the respective other one to be zero.
In lieu of using experimental data, we produce corresponding mock data for the left-hand side of \eq{fiducial_spectrum_final_result},
shown by solid lines in the bottom row of \fig{real_signal}.
To do so we use a $b^*_\mathrm{Pavia}$ prescription on $\sigma^{(0)}$,
which as demonstrated in \sec{bstar_prescriptions} modifies the cross section
beyond our working order in \eq{fiducial_spectrum_final_result}.
We then combine this with a nonperturbative model function as in \eq{toy_function_modified}
that flips sign as needed and recovers the respective quadratic coefficient,
\begin{align}
f_\mathrm{NP}(Q, b_T)
= \begin{cases}
   \exp\big(-\lambda L_{Q^2}^k b_T^2\big)
   \,, \quad &
   \lambda > 0
   \,,  \\
   \big(1 + 2 \abs{\lambda} L_{Q^2}^k b_T^2\big) 
    \exp\big(-\abs{\lambda} L_{Q^2}^k b_T^2\big)
   \,, \quad &
   \lambda < 0
   \,,\end{cases}
\end{align}
where $\lambda = -\overline{\C}^{(2)}$ and $k = 0$
for the intrinsic TMD PDF, and $\lambda = -\gamma^{(2)}_{\zeta,q}$ and $k = 1$ for the rapidity anomalous dimension.

Comparing the dashed and solid lines in \fig{real_signal},
we find that the proposed fit template approximates the model results well down to $q_T \sim 8 \GeV$ before nonlinearities and oscillatory error terms kick in,
which suggests $q_T \geq 8 \GeV$ as a possible lower bound for the fit window.
This is similar in spirit to searches for nonzero $d \geq 6$ SMEFT coefficients at the LHC,
which also often involve restricting the data set to data below (here: above) a certain cut
to ensure that the fit stays in the linear region
and that limits remain model independent.
We stress again that the signal is linear in the nonperturbative rapidity anomalous dimension
after the power expansion, even though it appears in the exponent
when working to all orders in $b_T$ space.
We anticipate that this will simplify breaking the (likely) degeneracy
between the rapidity anomalous dimension and the intrinsic TMD PDF
by performing the above fit to at least two fiducial volumes
since the universal parameter $\gamma_{\zeta,q}^{(2)}$ enters linearly in both.

\subsection{Overview of theory systematics}
\label{sec:data_theory_systematics}

From \fig{real_signal},
we can read off that to be sensitive to $\abs{\gamma_{\zeta,q}^{(2)}} 
\lesssim (0.35 \GeV)^2$
and $\abs{\overline{\C}^{(2)}} \lesssim (0.7 \GeV)^2$,
we must maintain control over the perturbative baseline
at a precision level of well below $2 \%$.
For definiteness we will consider a precision goal of $1 \%$
on the perturbative baseline in the following
and review the effects relevant at this level of precision.
The experimental systematics and statistical uncertainties
on the normalized Drell-Yan spectrum are negligible at this scale~\cite{CMS:2019raw, ATLAS:2019zci},
so we do not discuss them further.
Many items on this list require a dedicated implementation effort,
or even are not yet understood theoretically.
For this reason, a complete fit to actual LHC data is beyond the scope of this paper.

\paragraph{QCD corrections to $\sigma^{(0)}(Q, Y, b_T)$.}
Based on the perturbative uncertainty estimates in \refcite{Ebert:2020dfc},
the N$^3$LL perturbative order we consider here for illustration 
are at the $3$--$4\%$ level, and hence not sufficient to reach the $1 \%$ goal.
The complete three-loop QCD boundary conditions
for the leading-power factorized cross section in \eq{factorized_cross_section_expanded_Cbar}
have been computed in the meantime~\cite{Luo:2019szz, Ebert:2020qef, Luo:2020epw}
and enable the resummation at N$^3$LL$'$ accuracy,
which we expect to lower the perturbative QCD uncertainty by a factor of $\approx 2$~\cite{Camarda:2021ict, Re:2021con, Ju:2021lah, Billis:XXX}.

\paragraph{Power corrections to TMD factorization.}
The factorization in \eq{fiducial_spectrum} receives power corrections from several sources.
The first are power corrections of $\ord{q_T/Q}$ from the full $q_T$ dependence of the acceptance,
dubbed ``fiducial'' power corrections in \refcite{Ebert:2020dfc}.
They are logarithmically divergent at small $q_T$
and thus have to be evaluated in resummed perturbation theory,
but this is straightforward since the linear power corrections
can be shown~\cite{Ebert:2020dfc} to multiply the factorized cross section in \eq{factorized_cross_section_expanded_Ci_Cj},
where the effect of $\gamma_{\zeta, q}^{(2)}$ and $\C_{i,j}(x_a, x_b)$
can be dropped as higher order cross terms.
In addition, there are genuine QCD corrections of $\ord{q_T^2/Q^2}$, often referred to as non-singular or $Y$-term corrections.
These can be incorporated by a standard fixed-order matching,
implying that the fit window would not be restricted from above
by a validity range for TMD factorization like $q_T/Q \lesssim 0.25$.

\paragraph{QED effects.}
Incorporating the effect of QED initial-state radiation
into the factorization in \eq{factorized_cross_section} is well understood~\cite{Cieri:2018sfk, Bacchetta:2018dcq, Billis:2019evv}.
It was also shown in \refcite{Ebert:2020dfc} that QED final-state radiation
preserves the form of \eq{fiducial_spectrum}
and only leads to $\ord{\alpha_\mathrm{em}}$ corrections to the Born acceptance,
as long as the experimental lepton definition is sufficiently inclusive.
The key effect of QED is thus to break the factorization
of the cross section into a leptonic and hadronic part.
Very little is known about these initial-final interference (IFI) contributions
in the regime of small transverse momentum.
In the vicinity of the $Z$ resonance they can be expected to scale as $\sim \alpha_{em} \Gamma_Z/q_T$
in order to recover the narrow-width approximation at large $q_T$.
(Since the leptons distinguish a transverse direction in this case,
one can no longer appeal to azimuthal symmetry to push the correction to $\ord{b_T^2}$.)
Understanding the interplay of IFI effects with the all-order resummation
would thus be an important improvement for the fit we are proposing.

\paragraph{Quark mass effects.}
In our analysis we used $n_f = 5$ massless active flavors throughout.
This assumption will break down at small $q_T$, close to the fit window we examined here.
Corrections to this limit from massive quarks $Q$ have been calculated~\cite{Pietrulewicz:2017gxc}
that affect both the rapidity anomalous dimension
and the TMD PDF boundary condition at $\ord{\as^2 m_Q b_T^2}$,
where $m_Q = m_c, m_b$ is the respective quark mass.
(We refer the reader to \refcite{Pietrulewicz:2017gxc} for a review of earlier work on quark mass effects in the TMD PDF boundary condition.)
The importance of treating the quark mass dependence of the CS kernel correctly
was also stressed in \refcite{Collins:2014jpa}.
As these effects have the same scaling $b_T^2$ as our signal,
they must be accounted for in the resummed background prediction
in order to not pollute our fit.
We remind the reader that in \sec{tmd_cumulant_np}, we found large numerical differences
to the \texttt{NangaParbat} code used in \refcite{Bacchetta:2019sam},
where the number of massless active flavors was changed dynamically as a function of $b_T$.
Like our approach of using a fixed $n_f = 5$, this cannot be a complete description
of the region between $m_c \lesssim 1/b_T \lesssim m_b$:
Both approaches treat all quarks as either completely massless or fully decoupled,
and thus they are both missing the exact (nonlogarithmic) dependence on $b_T m_{c,b}$,
i.e., an $\ord{1}$ effect in this region.
Nevertheless, taking the two approaches to be the extreme choices,
the spread between them can serve as an indicator of the size of these mass effects,
and we conclude that it will be critical to correctly account for them in a resummed prediction in the future.

\FloatBarrier
\section{Conclusions}
\label{sec:conclusions}

In this work, we developed a formalism to disentangle the long- and short-distance
physics in TMD objects in momentum space.
This task is challenging because taking the Fourier transform
of position-space TMD PDFs mixes up contributions from perturbative short-distance physics
with the models used to treat the Landau pole and to describe the nonperturbative long-distance behavior.

We introduced a position-space cutoff $\btcut < 1/\lqcd$ in the Fourier transform,
such that the momentum-space TMDs at perturbative momenta $k_T \gg \lqcd$
manifestly depend only on the short-distance physics.
We then defined two series of truncated integral functionals, 
$S^{(n)}[f]$ for the momentum-space spectrum and $K^{(n)}[f]$ for the cumulative momentum distribution,
which systematically account for the long-distance contributions from the region beyond $\btcut$.
The power corrections can be calculated as contact terms on the boundary of integration,
and only depend on the information of the integrand at $\btcut$.
In the short-distance region, where we calculate our integral functionals,
TMD PDFs can be perturbatively calculated using an operator product expansion (OPE).
Pairing the truncated functionals with the perturbative TMDs computed from the OPE,
we constructed a model-independent, systematically improvable perturbative baseline for momentum-space TMD quantities
and presented explicit formulas for all sources of uncertainty in the approximation.

As an important application, we showed how to determine the transverse-momentum integral of TMD PDFs by
combining the OPE with the $K^{(n)}$ functionals
to compute their cumulative distribution
up to a transverse momentum cutoff $\ktcut$.
This allows us to quantitatively assess the naive expectation that the integral over transverse momentum of a TMD PDF
equals the corresponding collinear PDF.
We first considered the natural choices for the renormalization and Collins-Soper scales 
$\mu = \sqrt\zeta = \ktcut$,
performing the renormalization group evolution of the TMD PDF to these scales in $b_T$ space at N$^3$LL order,
and found that the deviation 
of the integral of the TMD PDF from the collinear PDF is at the
percent level for perturbative momenta $\ktcut \gtrsim 10\GeV$.
In particular, the deviation as a function of the momentum fraction $x$ is within $\sim 2\%$ for $10^{-3} \leq x \leq 10^{-2}$.
These N$^3$LL results in particular include the two-loop radiative corrections to the TMD PDF
as encoded in the leading OPE,
and we find that the cumulative distribution is always compatible with the collinear PDF within our estimate of the perturbative uncertainty.
Using our truncated functional construction, we can also assess the nonperturbative uncertainty
on our results in a model-independent way, which in particular includes
a conservative estimate of the leading nonperturbative contribution from the Collins-Soper kernel.
We find that the agreement of the integral of the TMD PDF with the collinear PDF deteriorates for general $x$
if only the nonperturbative uncertainty is considered,
motivating both an extension to higher perturbative orders in the future to push down the dominant perturbative uncertainty
and dedicated model-independent studies of the leading (quadratic) contributions to the Collins-Soper kernel and the intrinsic TMD PDF.

Evolving the TMD PDF to different overall scales $\mu, \zeta$,
we find that the deviation remains within a few percent in the vicinity of $\mu = \sqrt\zeta = \ktcut$.
Furthermore, the evolution of $\zeta$ at fixed $\mu = \ktcut$ is found to only be a shape effect,
meaning that while it changes the shape of the TMD PDF at low $k_T$,
the cumulative integral stays the same and is determined by the collinear PDF.
We tested our conclusions for the unpolarized TMD PDFs for all light quark flavors.

We also proposed to apply our $S^{(n)}$ functionals to momentum-space spectra for the Drell-Yan process,
which have been measured to sub-percent precision at the LHC, in order 
to develop a method of putting model-independent constraints on the leading nonperturbative effects in TMD PDFs.
In our approach, the $\ord{b_T^2}$ nonperturbative contributions to the Collins-Soper kernel, $\gamma_{\zeta,q}^{(2)}$,
and the intrinsic TMD PDF, ${\overline\C_i}^{(2)}$,
are treated as a ``signal'',
which can be discriminated from the perturbative ``background''
by comparing to experimental data.
Since the functionals are linear in the OPE coefficients (the ``signal strengths''),
we are optimistic that the proposed fit to data will be greatly simplified.
The method we propose is model-independent and complementary to model-based global fits.
We briefly discussed the major theory systematics in the proposed fit,
such as QCD corrections to the perturbative cross section,
power corrections to the TMD factorization,
QED effects, and quark mass effects,
all of which influence the perturbative precision.
These systematics are important to consider because our estimates suggest
that the background prediction must be pushed to roughly a percent accuracy
in order to be able to see the nonperturbative signal for a generic $\ordsq{(1 \GeV)^2}$ size of the coefficients.

Our results in this paper lend themselves to several future applications.
The truncated functional formalism that we developed
is generally applicable to Fourier transforms of spherically symmetric functions $f(\abs{\vec{r}})$
(also known as Hankel transforms) in any number of dimensions,
with minimal modifications from the two-dimensional case that we presented.
Our results are useful in scenarios
where the test function is only known on a radial interval $0 < \abs{\vec{r}\,} < r_\mathrm{cut}$,
and provide a systematic expansion of the Fourier transform at large $\vec{k}$
in inverse powers of $r_\mathrm{cut} \abs{\vec{k}} \gg 1$.
In practical applications, e.g.\ in the fit we proposed in \sec{data},
one could consider averaging over several values of the cut parameter ($\btcut$ in our case)
to smoothen the residual oscillations,
which is equivalent to using a smoother function to switch off the integrand around the central $\btcut$.
The numerical stability of the functionals might also be further improved 
by moving part of the boundary term under the integral as a subtraction term.

For physics applications to TMDs, we note that a fit template linear in the nonperturbative parameter of interest can also be obtained in modifications of our setup.
For example, one could apply the exponential $b^*$ prescription introduced in \eq{bstar_exponential} to individual terms in the OPE,
and multiply the result by a similarly constructed ``model function'', e.g.\ $1 - \exp\bigl[ -(b_T/\bmax - a)^{-1} \bigr]$.
This yields
a $b_T$-space integrand that does not contain spurious power corrections,
but does fall off as $1/b_T$ at large $b_T$
and thus can be integrated over all $0 \leq b_T < \infty$.
This may be numerically advantageous due to the ability to use series acceleration
for the indefinite integral.

Finally, for the study of individual TMD PDFs,
an obvious next step is to study the gluon TMD PDF,
which we will address in a separate publication.
The gluon TMD PDF is an interesting test case
because it features polarization already at leading twist even within an unpolarized hadron~\cite{Mantry:2009qz}.

For polarized hadrons, it would be interesting to verify
in a model-independent way the observation of \refcite{Bacchetta:2013pqa}
that the cumulative helicity ($g_1$) and transversity ($h_1$) TMD PDFs are approximately
equal to their respective collinear counter parts for $\mu = \ktcut$ and large enough $\ktcut$.
This extension is straightforward in our framework
because the OPE for the helicity and transversity TMD PDFs
has the same general form as discussed in \secs{review_tmd_pdf_and_normalization}{single_tmdpdf_evolution_ope},
i.e., they both have a leading contribution from (polarized) leading-twist collinear PDFs,
so our setup carries over immediately.
For TMD PDFs whose OPE receives its first contribution at $\ord{\lqcd b_T}$ (at subleading twist),
our results imply that the cumulative distribution at large $\ktcut$ vanishes as $\lqcd/\ktcut$.
This has interesting consequences e.g.\ for the Burkardt sum rule for the Sivers function~\cite{Burkardt:2004ur},
whose possible violation at the level of renormalized quantities can be studied in our numerical setup as a function of $\ktcut$
including in particular the effect of (rapidity) evolution. Work on this is also in progress.

\paragraph{Note added:}
While finalizing this manuscript, we became aware that related techniques
for computing asymptotic expansions of integral transforms
were developed in applied mathematics several decades ago, see e.g.\ \refscite{SoniKusum1980EETi, WongR1980EBfA, Soni:1982xxx}.
The most closely related results were obtained in \refcite{Soni:1982xxx},
where Bessel integrals over an interval $0 \leq t \leq a$
were computed in terms of the test function and its derivatives at $t = a$,
and the result was then used to express full Bessel integrals over $0 \leq t < \infty$
in terms of the behavior of the function near the origin $t \to 0$.
Our approach in \sec{truncated_functionals} differs by instead using (numerical) test function data on a compact interval $0 \leq b_T \leq \btcut$
to approximate the full integral, which is better suited for our physics applications. This is the case because 
the analytic form of derivatives of our test functions as $b_T \to 0$
is not readily accessible for TMD PDFs or cross sections
once radiative and evolution effects are included.

\acknowledgments

We thank Markus Diehl, Andreas Metz, Jianwei Qiu, and Alexey Vladimirov for helpful comments and discussions.
We also thank Frank Tackmann for his leading role and many contributions
in developing the \texttt{SCETlib} numerical library used here.
This work was supported in part by the Office of Nuclear Physics of the U.S.\
Department of Energy under Contract No.\ DE-SC0011090,
and within the framework of the TMD Topical Collaboration.
I.S. was also supported in part by the Simons Foundation through the Investigator grant 327942.
Z.S. was also supported by a fellowship from the MIT Department of Physics.

\appendix
\section{Asymptotic expansion of \texorpdfstring{${}_1 F_{\,2}$}{1F2} hypergeometric functions}
\label{app:1F2_asymptotics}

\subsection{Bessel \texorpdfstring{$J_0$}{J0} integral (transverse momentum spectra)}
\label{app:1F2_asymptotics_spectrum}

In \eq{gbeta_S}, we defined the integral
\begin{align} \label{eq:gbeta_S_rep}
 g(\beta, x)
 \equiv \int_x^\infty \! dy \, J_0(y) \, y^{1+\beta}
 = \frac{2^{1+\beta} \, \Gamma \bigl( 1+ \frac{\beta}{2} \bigr) }
        { \Gamma\bigl( -\frac{\beta}{2} \bigr) }
   -  \frac{x^{2+\beta}}{2+\beta} \, \phantom{}_1F_{\,2}\Bigl( 1+ \frac{\beta}{2}; 1, 2+ \frac{\beta}{2}; -\frac{x^2}{4} \Bigr)
\,,\end{align}
which converges for $x > 0$ and $\beta < -1/2$.
Using the asymptotic behavior of the hypergeometric function
$\phantom{}_1 F_{\, 2} \bigl( a_1;  b_1, b_2; -x^2/4\bigr)$
as $x \to \infty$~\cite[\S 16.11]{NIST:DLMF},
we have
\begin{align}
 g(\beta, x)
 = \sqrt{\frac{2 x}{\pi}} x^\beta \sum_{k=0}^{n-1} \frac{c_k(\beta)}{x^k}
   \cos\Bigl(x + \frac{\pi}{4} - \frac{k\pi}{2} \Bigr) + \Ord{x^{\beta - n + \frac{1}{2}}}
\,.\end{align}
The coefficients $c_k(\beta)$ can be calculated recursively from the relation%
\footnote{Note that taking the limit $b_1 \to 1$ in the recursion relation requires care.}
\begin{align}
 c_0(\beta) = 1
\,,\qquad
 c_k(\beta) = - \frac{1}{4 k} \sum_{m=0}^{k-1} c_m(\beta) \, e_{k,m}(\beta)
\,,\end{align}
where the coefficients $e_{k,m}$ are given by
\begin{align}
 e_{k,m}(\beta) &
 = \frac{2 \beta}{\beta+2} \frac{\Gamma(k+\tfrac{5}{2})}{\Gamma (m+\tfrac{1}{2})}
   \bigl[ \psi(k+\tfrac{5}{2})- \psi(m+\tfrac{1}{2}) \bigr]
 \nn\\&\quad
   + \frac{4}{(\beta+2)^2} \biggl[ \frac{\Gamma (k+\tfrac{5}{2})}{\Gamma (m+\tfrac{1}{2})}
      -  \frac{\Gamma (k-\beta +\tfrac{1}{2})}{\Gamma (m-\beta -\tfrac{3}{2})} \biggr]
\,,\end{align}
and $\psi(z) = \Gamma'(z)/\Gamma(z)$ is the digamma function.
The first few coefficients are given by
\begin{align}
c_{0}(\beta) &= 1
\,,\nn\\
c_{1}(\beta) &= -\beta -\frac{3}{8}
\,,\nn\\
c_{2}(\beta) &= \beta^2-\frac{\beta}{8}-\frac{15}{128}
\,,\nn\\
c_{3}(\beta) &= -\beta^3+\frac{13 \beta^2}{8}-\frac{9 \beta}{128}-\frac{105}{1024}
\,,\nn\\
c_{4}(\beta) &= \beta^4-\frac{33 \beta^3}{8}+\frac{529 \beta^2}{128}-\frac{75 \beta}{1024}-\frac{4725}{32768}
\,,\nn\\
c_{5}(\beta) &= -\beta^5+\frac{61 \beta^4}{8}-\frac{2377 \beta^3}{128}+\frac{14887 \beta^2}{1024}-\frac{3675 \beta}{32768}-\frac{72765}{262144}
\,,\nn\\
c_{6}(\beta) &= \beta^6-\frac{97 \beta^5}{8}+\frac{6769 \beta^4}{128}-\frac{100459 \beta^3}{1024}+\frac{2147403 \beta^2}{32768}-\frac{59535 \beta}{262144}-\frac{2837835}{4194304}
\,,\nn\\
c_{7}(\beta) &= -\beta^7+\frac{141 \beta^6}{8}-\frac{15305 \beta^5}{128}+\frac{398295 \beta^4}{1024}-\frac{19828187 \beta^3}{32768}+\frac{94545267 \beta^2}{262144}
         \nn\\&\quad
         -\frac{2401245 \beta}{4194304}-\frac{66891825}{33554432}
\,,\nn\\
c_{8}(\beta) &= \beta^8-\frac{193 \beta^7}{8}+\frac{29969 \beta^6}{128}-\frac{1194155 \beta^5}{1024}+\frac{102673547 \beta^4}{32768}-\frac{1125610991 \beta^3}{262144}
         \nn\\&\quad
         +\frac{9835109013 \beta^2}{4194304}-\frac{57972915 \beta}{33554432}-\frac{14783093325}{2147483648}
\,,\nn\\
c_{9}(\beta) &= -\beta^9+\frac{253 \beta^8}{8}-\frac{53129 \beta^7}{128}+\frac{2992295 \beta^6}{1024}-\frac{389270747 \beta^5}{32768}+\frac{7286023811 \beta^4}{262144}
         \nn\\&\quad
         -\frac{144908427933 \beta^3}{4194304}+\frac{590164513695 \beta^2}{33554432}-\frac{13043905875 \beta}{2147483648}-\frac{468131288625}{17179869184}
\,.\end{align}

\subsection{Bessel \texorpdfstring{$J_1$}{J1} integral (cumulative distribution)}
\label{app:1F2_asymptotics_cumulant}

We define the integral
\begin{align} \label{eq:def_g_beta_x}
\tilde g(\beta,x) \equiv \int_0^x \! \df y \, J_1(y) \, y^\beta
= 2^\beta\frac{\Gamma(1 + \frac{\beta}{2})}{\Gamma(1 - \frac{\beta}{2})}
   - \frac{x^{2+\beta}}{2(2+\beta)}\, \phantom{}_1F_{\,2}\Bigl(1 + \frac{\beta}{2}; 2, 2 + \frac{\beta}{2}; -\frac{x^2}{4}\Bigr)
\,,\end{align}
which exists for any $\Re(\beta) < 1/2$.
Using the asymptotic behavior of the hypergeometric function
$\phantom{}_1 F_{\, 2} \bigl( a_1;  b_1, b_2; -x^2/4\bigr)$
as $x \to \infty$~\cite[\S 16.11]{NIST:DLMF},
we have
\begin{align}
 \tilde g(\beta, x)
 = \sqrt{\frac{2}{\pi x}} x^\beta \sum_{k=0}^{n-1} \frac{\tilde c_k}{x^k} \cos\left(x - \frac{\pi}{4} - \frac{k \pi}{2}  \right)
   +  \Ord{x^{\beta - n + \frac{1}{2}}}
\,.\end{align}
The coefficients $\tilde c_k(\beta)$ can be calculated recursively from the relation
\begin{align}
 \tilde c_0(\beta) = 1
\,,\qquad
 \tilde c_k(\beta) = - \frac{1}{4 k} \sum_{m=0}^{k-1} \tilde c_m(\beta) \, \tilde e_{k,m}(\beta)
\,,\end{align}
where the coefficients $\tilde e_{k,m}$ are given by
\begin{align}
 \tilde e_{k,m}(\beta) &
 = \frac{2-\beta}{\beta} \frac{\Gamma(k + \tfrac{3}{2})}{\Gamma(m-\tfrac{1}{2})}
 + \frac{\beta}{2+\beta} \frac{\Gamma(k + \tfrac{7}{2})}{\Gamma(m+\tfrac{3}{2})}
 - \frac{4}{\beta(2+\beta)} \frac{\Gamma(k - \beta + \tfrac{3}{2})}{\Gamma(m-\beta-\tfrac{1}{2})}
\,.\end{align}
The first few coefficients are given by
\begin{align}
 \tilde c_{0}(\beta) &= 1
\,, \\
 \tilde c_{1}(\beta) &= -\beta + \frac{1}{8}
\,,\nn\\
 \tilde c_{2}(\beta) &= \beta^2-\frac{13 \beta}{8}+\frac{9}{128}
\,,\nn\\
 \tilde c_{3}(\beta) &= -\beta^3+\frac{33 \beta^2}{8}-\frac{529 \beta}{128}+\frac{75}{1024}
\,,\nn\\
 \tilde c_{4}(\beta) &= \beta^4-\frac{61 \beta^3}{8}+\frac{2377 \beta^2}{128}-\frac{14887 \beta}{1024}+\frac{3675}{32768}
\,,\nn\\
 \tilde c_{5}(\beta) &= -\beta^5+\frac{97 \beta^4}{8}-\frac{6769 \beta^3}{128}+\frac{100459 \beta^2}{1024}-\frac{2147403 \beta}{32768}+\frac{59535}{262144}
\,,\nn\\
 \tilde c_{6}(\beta) &= \beta^6-\frac{141 \beta^5}{8}+\frac{15305 \beta^4}{128}-\frac{398295 \beta^3}{1024}+\frac{19828187 \beta^2}{32768}-\frac{94545267 \beta}{262144}+\frac{2401245}{4194304}
\,,\nn\\
 \tilde c_{7}(\beta) &= -\beta^7+\frac{193 \beta^6}{8}-\frac{29969 \beta^5}{128}+\frac{1194155 \beta^4}{1024}-\frac{102673547 \beta^3}{32768}+\frac{1125610991 \beta^2}{262144}
 \nn\\&\quad
 -\frac{9835109013 \beta}{4194304}+\frac{57972915}{33554432}
\,,\nn\\
 \tilde c_{8}(\beta) &= \beta^8-\frac{253 \beta^7}{8}+\frac{53129 \beta^6}{128}-\frac{2992295 \beta^5}{1024}+\frac{389270747 \beta^4}{32768}-\frac{7286023811 \beta^3}{262144}
 \nn\\&\quad
 +\frac{144908427933 \beta^2}{4194304}-\frac{590164513695 \beta}{33554432}+\frac{13043905875}{2147483648}
\,,\nn\\
 \tilde c_{9}(\beta) &= -\beta^9+\frac{321 \beta^8}{8}-\frac{87537 \beta^7}{128}+\frac{6605067 \beta^6}{1024}-\frac{1203174987 \beta^5}{32768}+\frac{33756434607 \beta^4}{262144}
 \nn\\&\quad
 -\frac{1135807666229 \beta^3}{4194304}+\frac{10443937613139 \beta^2}{33554432}-\frac{321062539355955 \beta}{2147483648}+\frac{418854310875}{17179869184}
\nn
\,.\end{align}
%

\section{Results for other light quark TMD PDFs}
\label{app:more_results_tmdpdf}

In this appendix, we show the normalized deviation of the transverse-momentum integral of the TMD PDFs from the collinear PDFs for other light quark flavors
$u, \bar{d}, \bar{u},$ and $s$.
The figures correspond to \fig{tmd_np_uncertainty_ktcut_x}, \fig{tmd_resum_uncertainty_ktcut_x}, \fig{tmd_mu_zeta_heatmap}, and \fig{zeta_evolution_scale_np}
in the main text, which are for the $d$ quark. 
We find that the transverse-momentum integral of the TMD PDF remains a percent-level deviation from the collinear PDF
for all the light quark flavors and agrees exactly for $x = 0.01$,
with a slightly larger deviation ($\sim 0.5 \%$ for the central value at $\ktcut = 10 \GeV$ and $x = 0.01$) for the strange quark.
The discussions about perturbative and nonperturbative uncertainties presented in \sec{single_tmdpdf}
remain unchanged.

\newpage
\begin{figure}
 \includegraphics[width=\WidthTwoSubfigs]{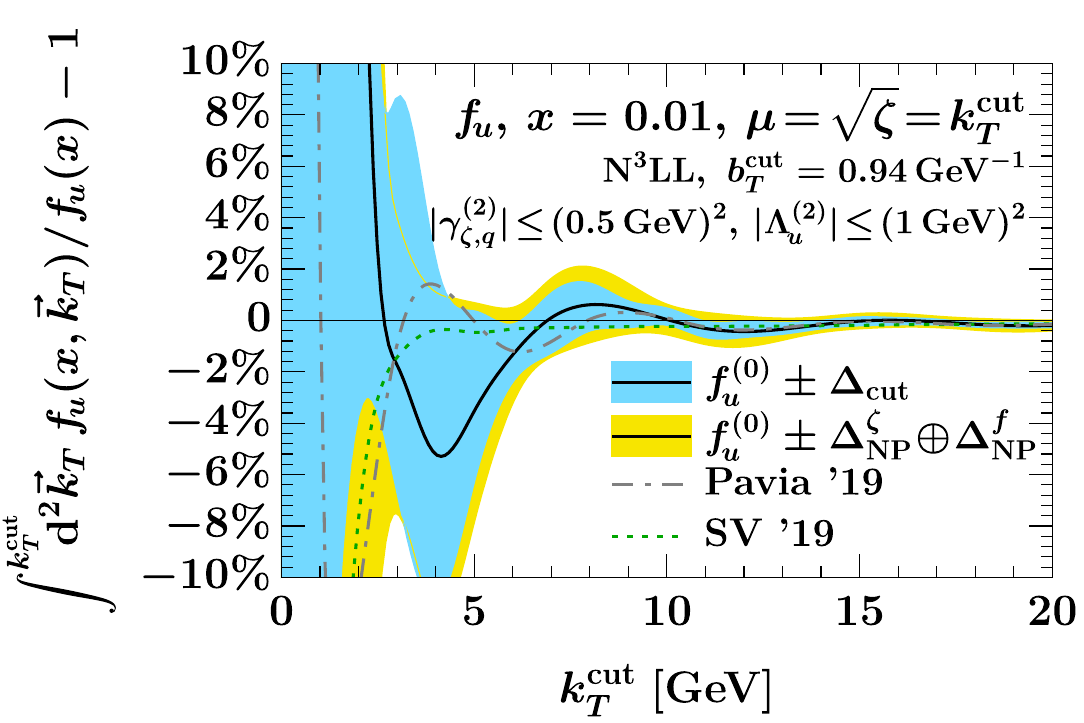}
 \hfill
 \includegraphics[width=\WidthTwoSubfigs]{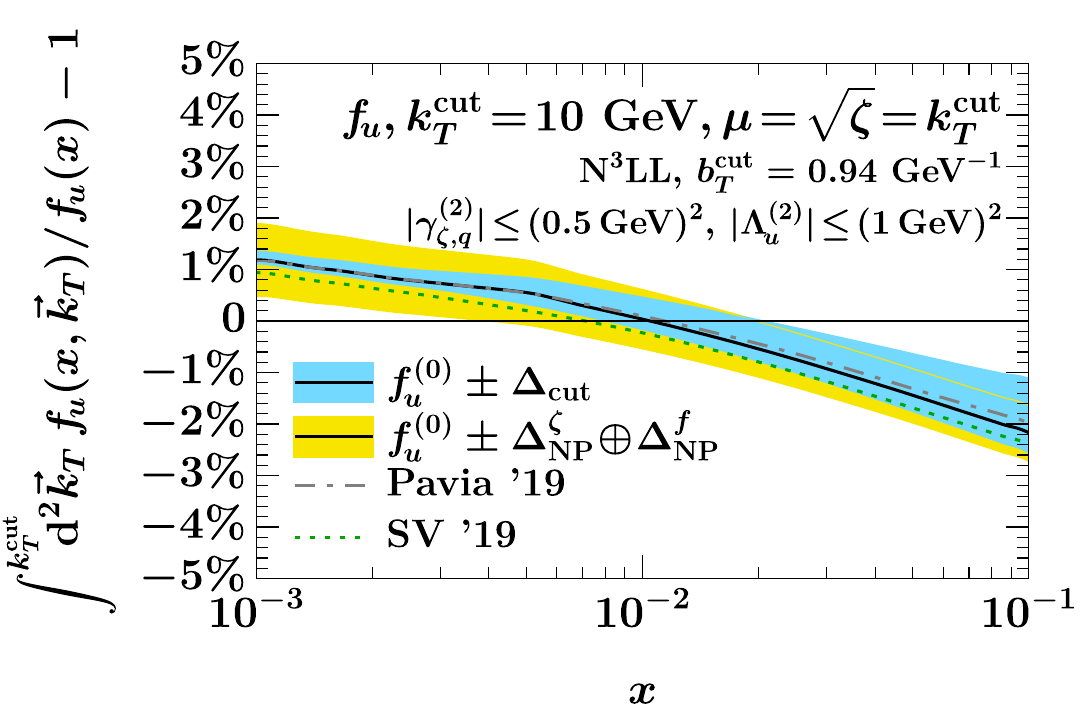}
 \\
 \includegraphics[width=\WidthTwoSubfigs]{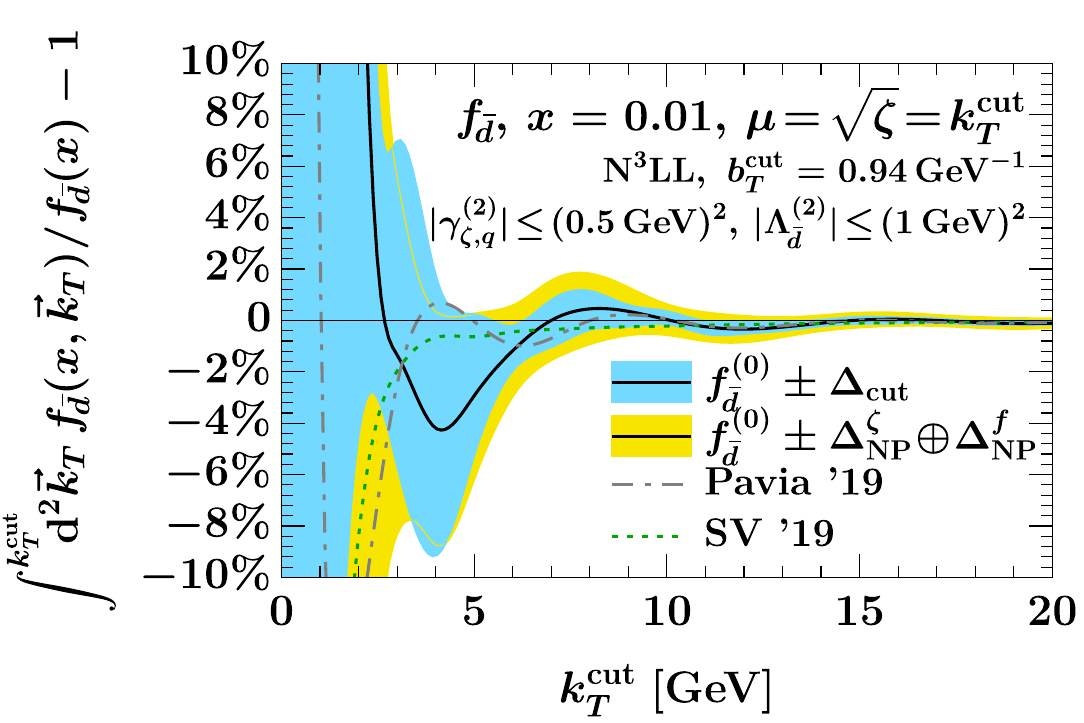}
 \hfill
 \includegraphics[width=\WidthTwoSubfigs]{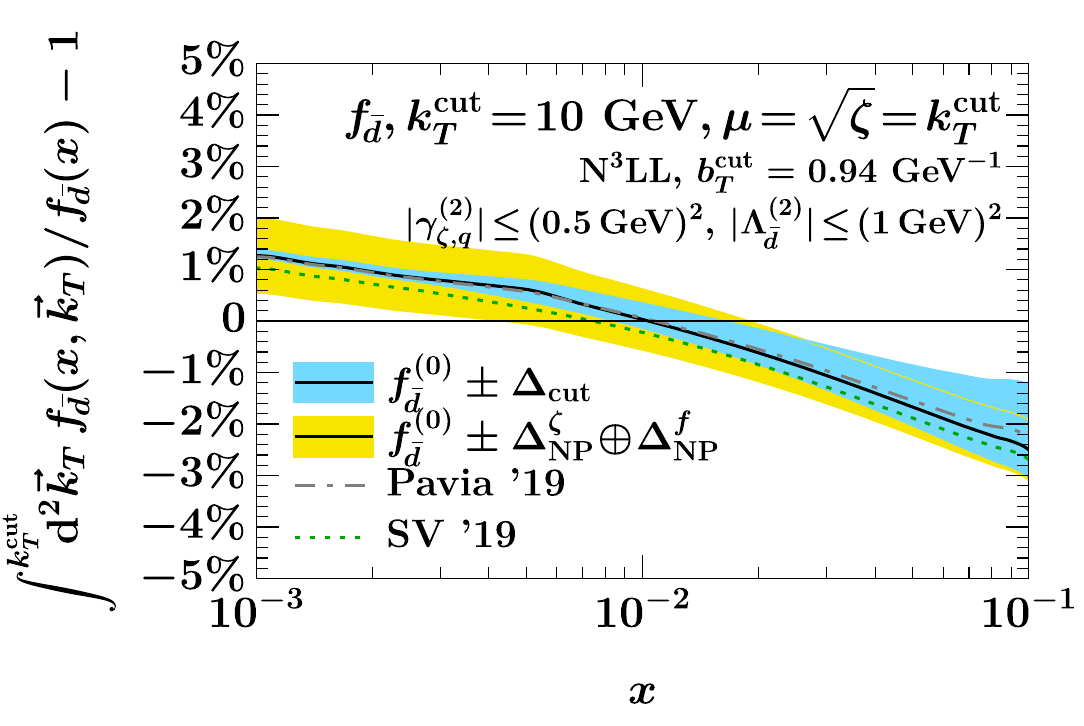}
 \\
 \includegraphics[width=\WidthTwoSubfigs]{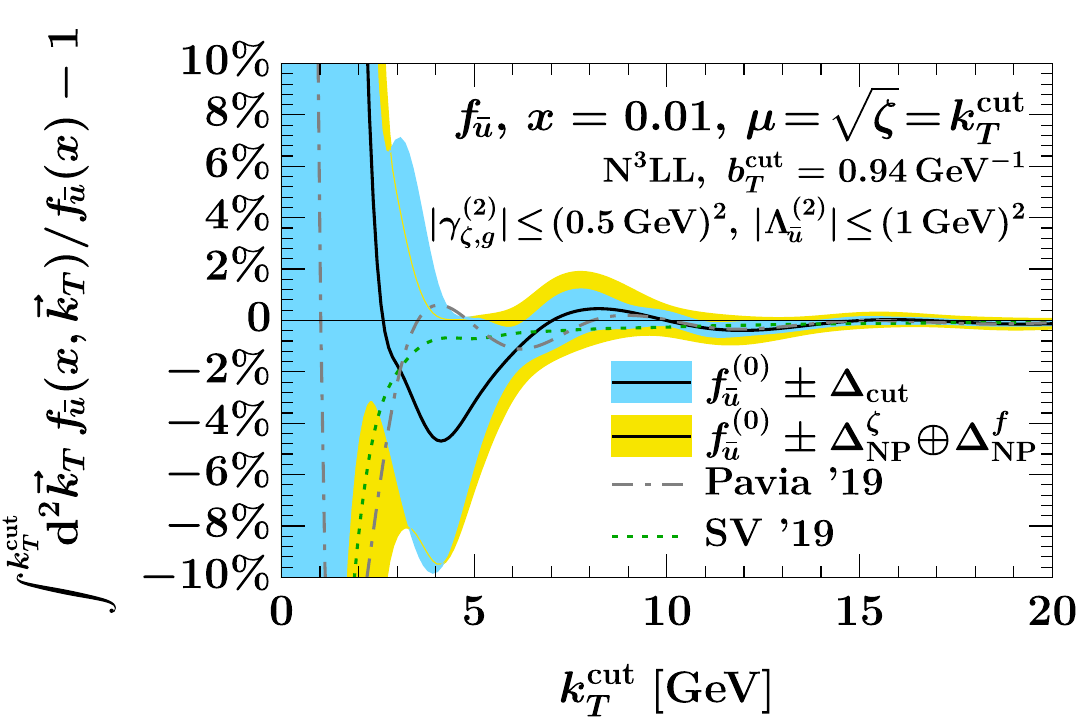}
 \hfill
 \includegraphics[width=\WidthTwoSubfigs]{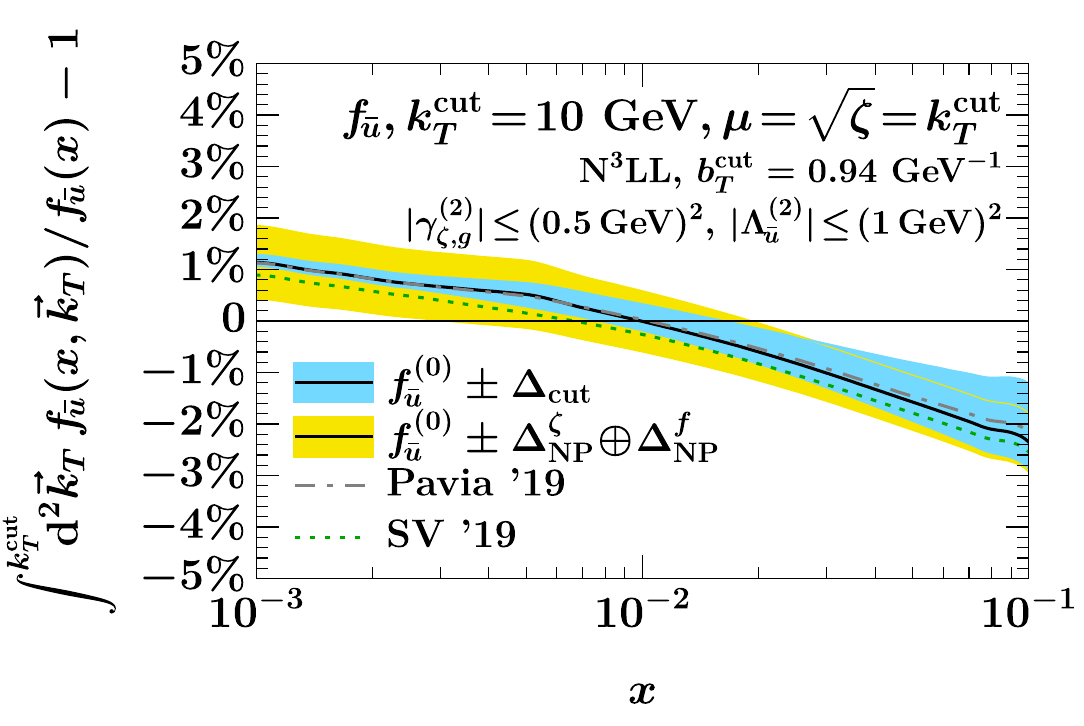}
 \\
 \includegraphics[width=\WidthTwoSubfigs]{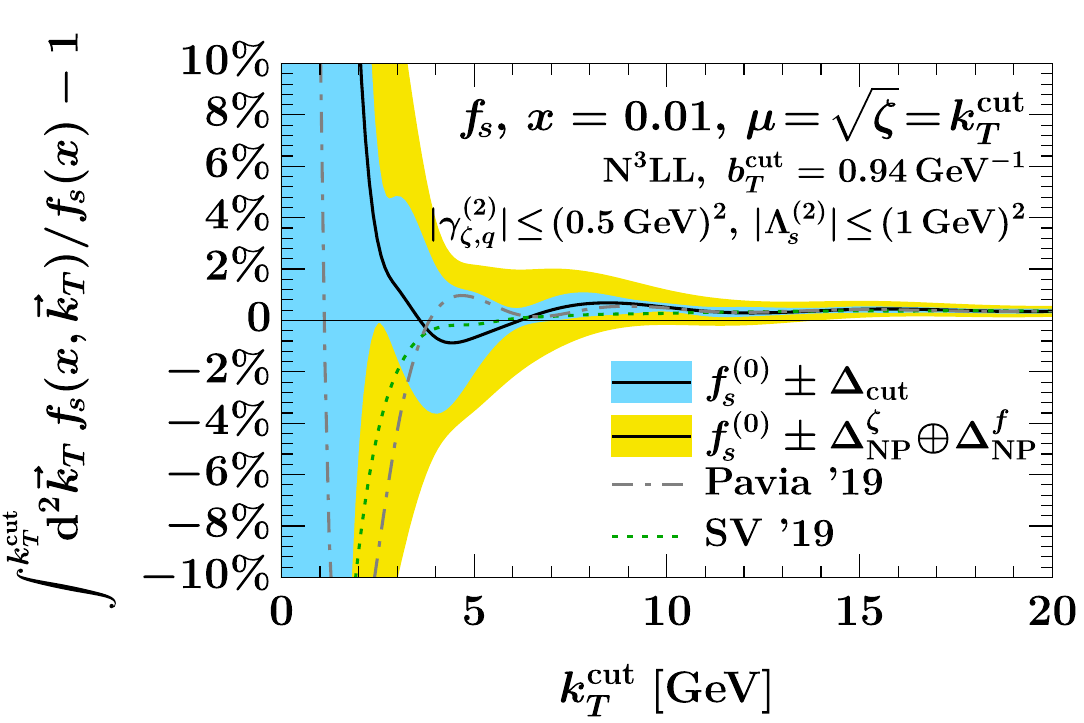}
 \hfill
 \includegraphics[width=\WidthTwoSubfigs]{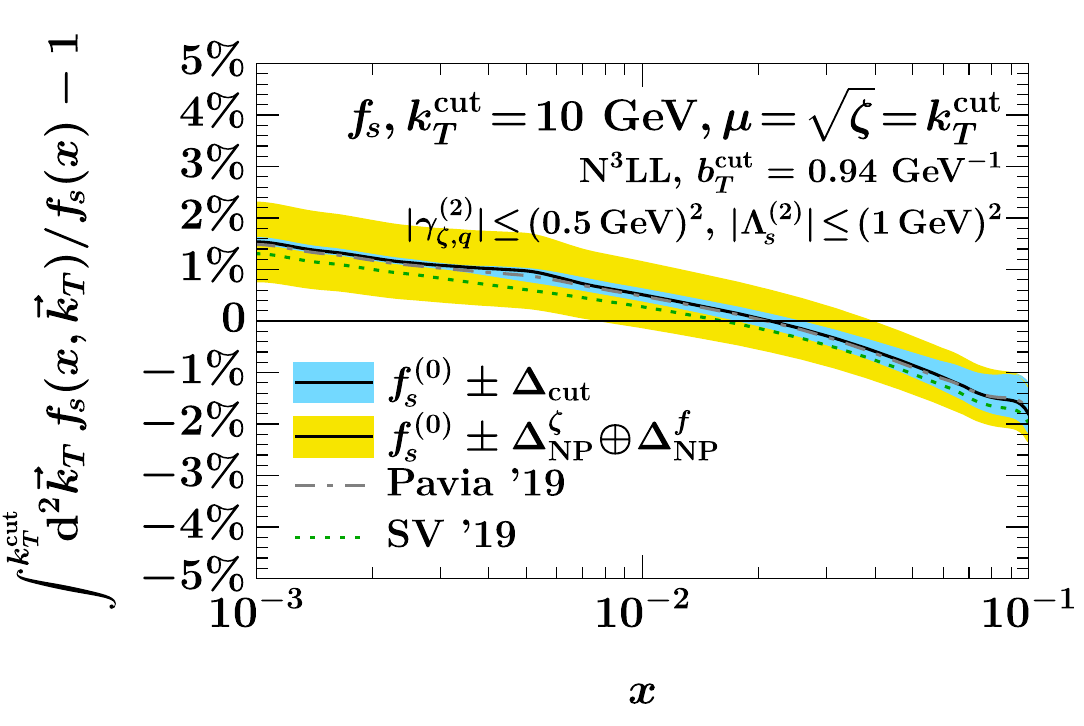}
 \caption{%
 Same as \fig{tmd_np_uncertainty_ktcut_x}, for $i = u, \bar{d}, \bar{u}, s$.
 }
\label{fig:more_np_uncertainty}
\end{figure}

\newpage
\begin{figure}
 \includegraphics[width=\WidthTwoSubfigs]{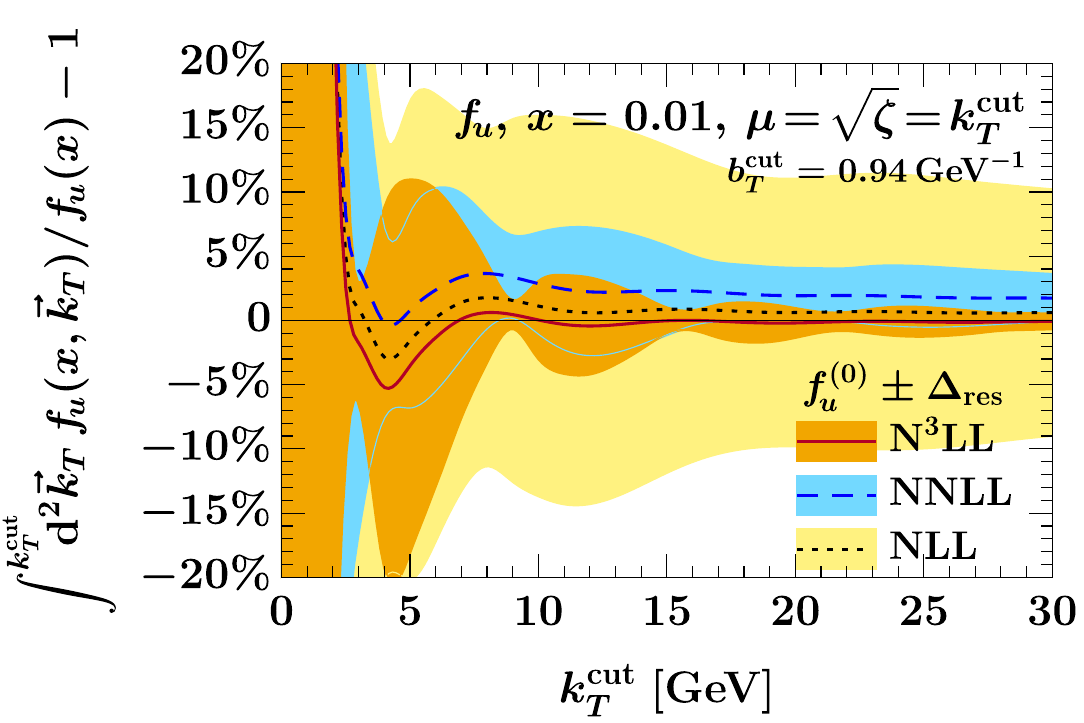}
 \hfill
 \includegraphics[width=\WidthTwoSubfigs]{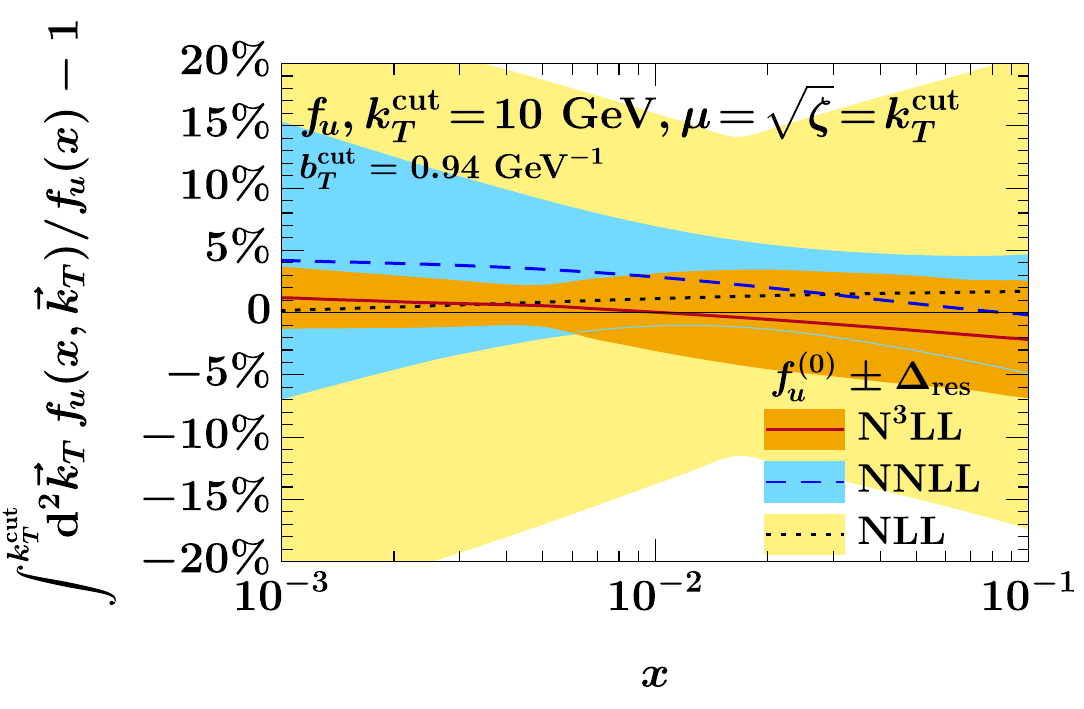}
 \\
  \includegraphics[width=\WidthTwoSubfigs]{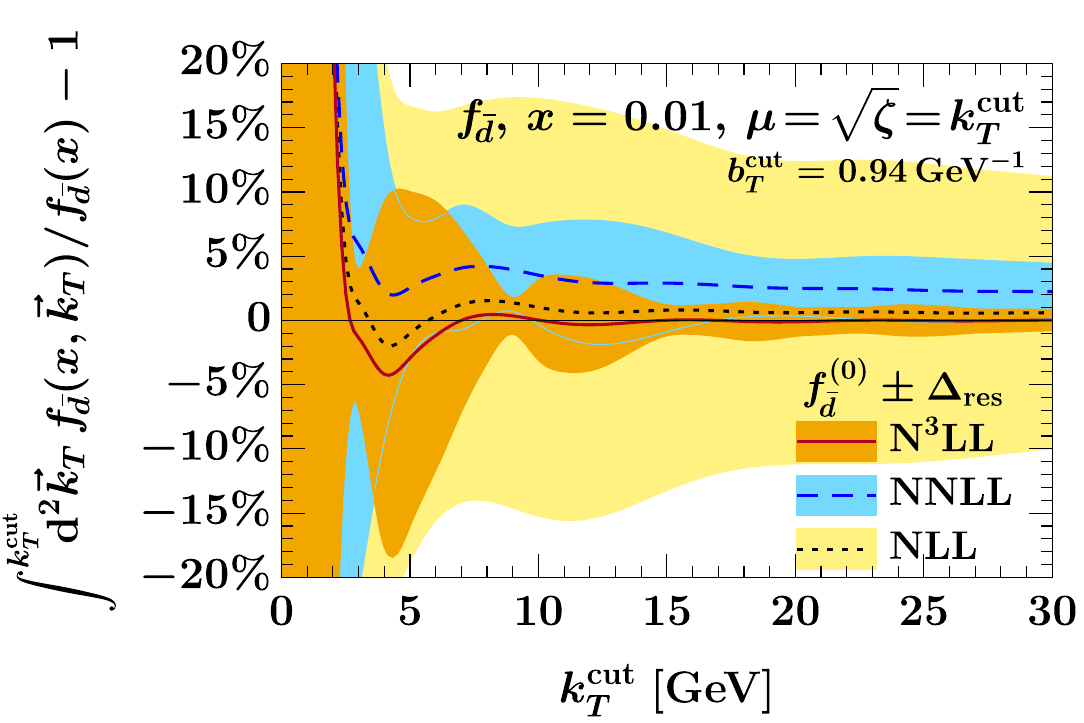}
 \hfill
 \includegraphics[width=\WidthTwoSubfigs]{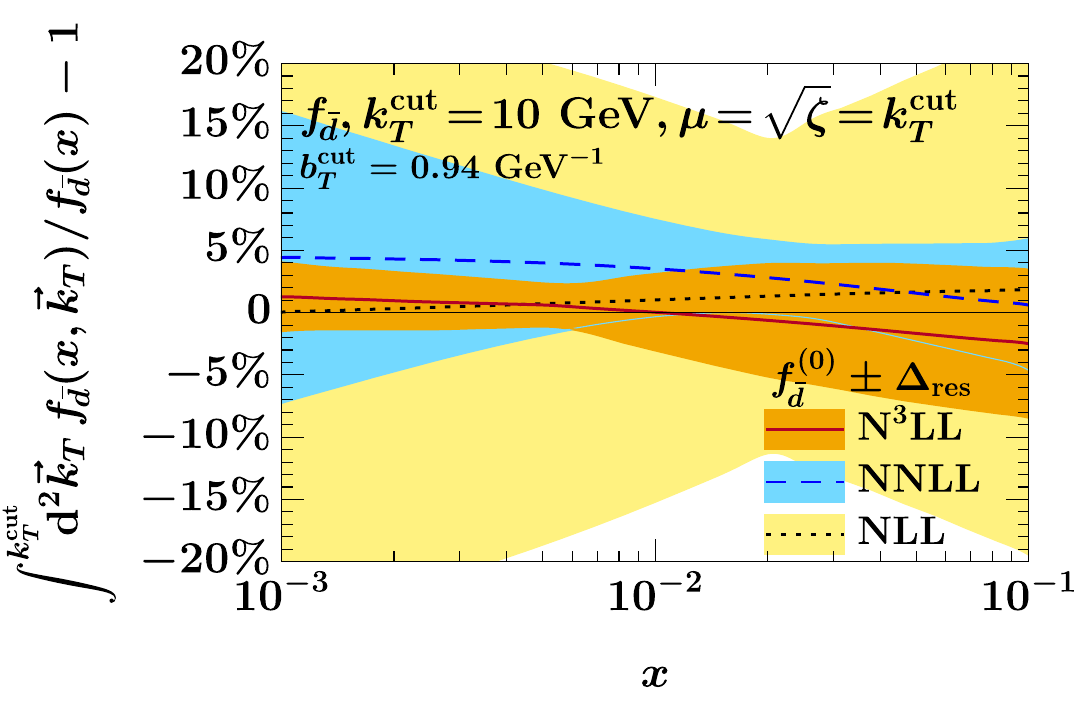}
 \\
  \includegraphics[width=\WidthTwoSubfigs]{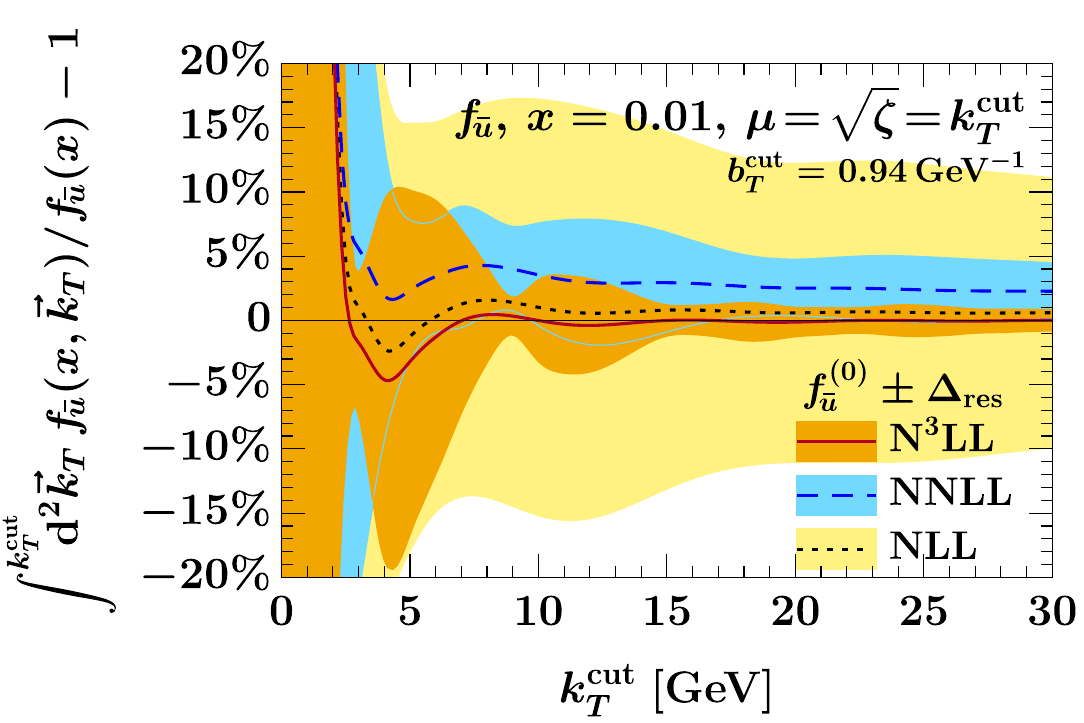}
 \hfill
 \includegraphics[width=\WidthTwoSubfigs]{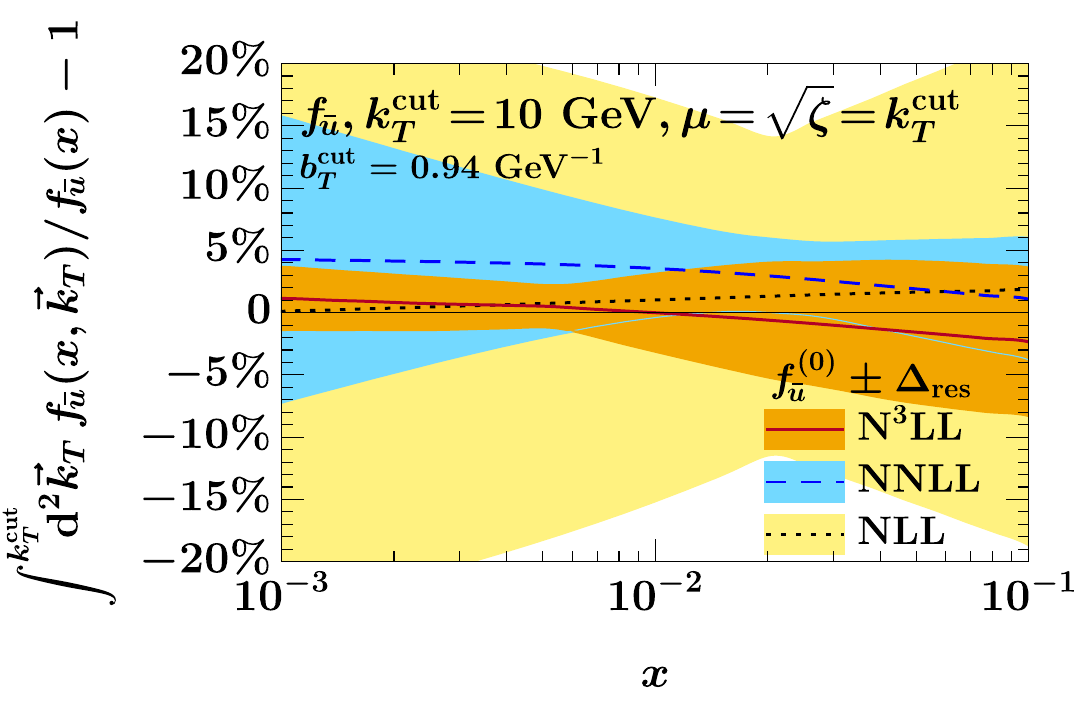}
 \\
  \includegraphics[width=\WidthTwoSubfigs]{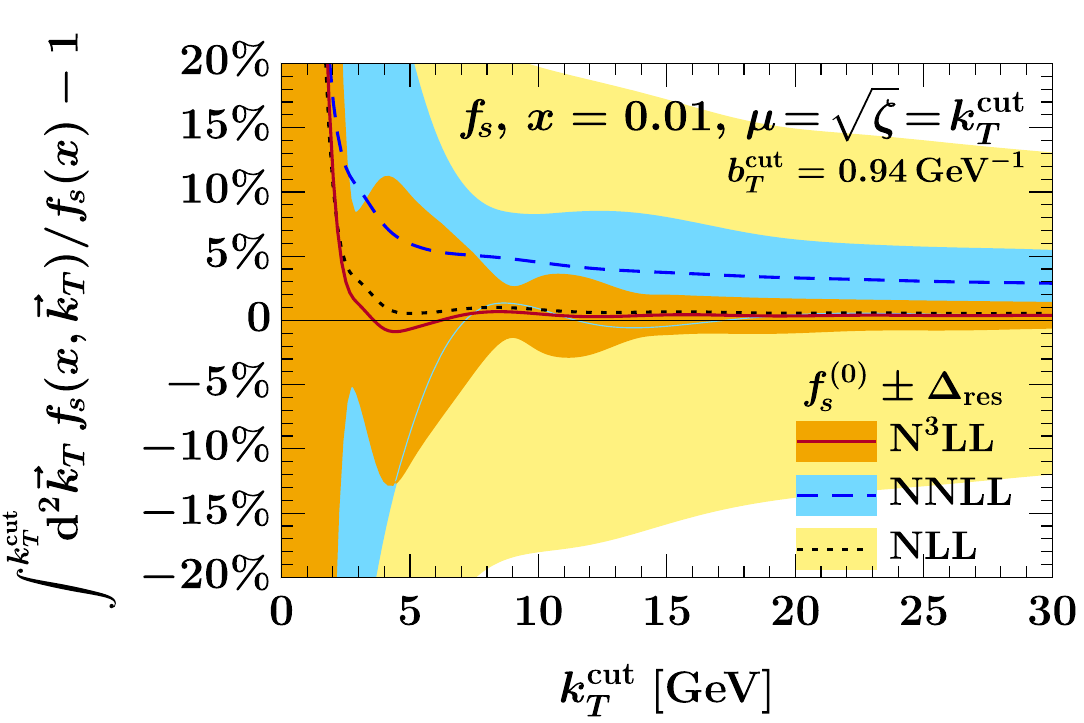}
 \hfill
 \includegraphics[width=\WidthTwoSubfigs]{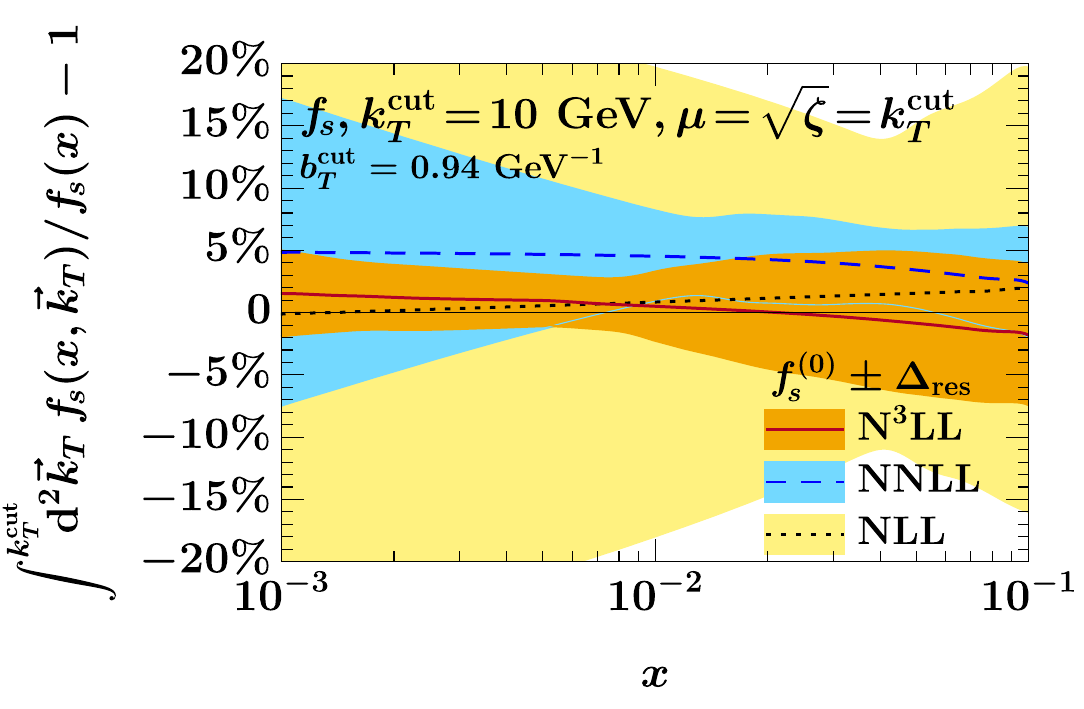}
\caption{%
 Same as \fig{tmd_resum_uncertainty_ktcut_x}, for $i = u, \bar{d}, \bar{u}, s$.
 }
\label{fig:more_scale_variation}
\end{figure}

\newpage
\begin{figure}
 \includegraphics[width=\WidthTwoSubfigs]{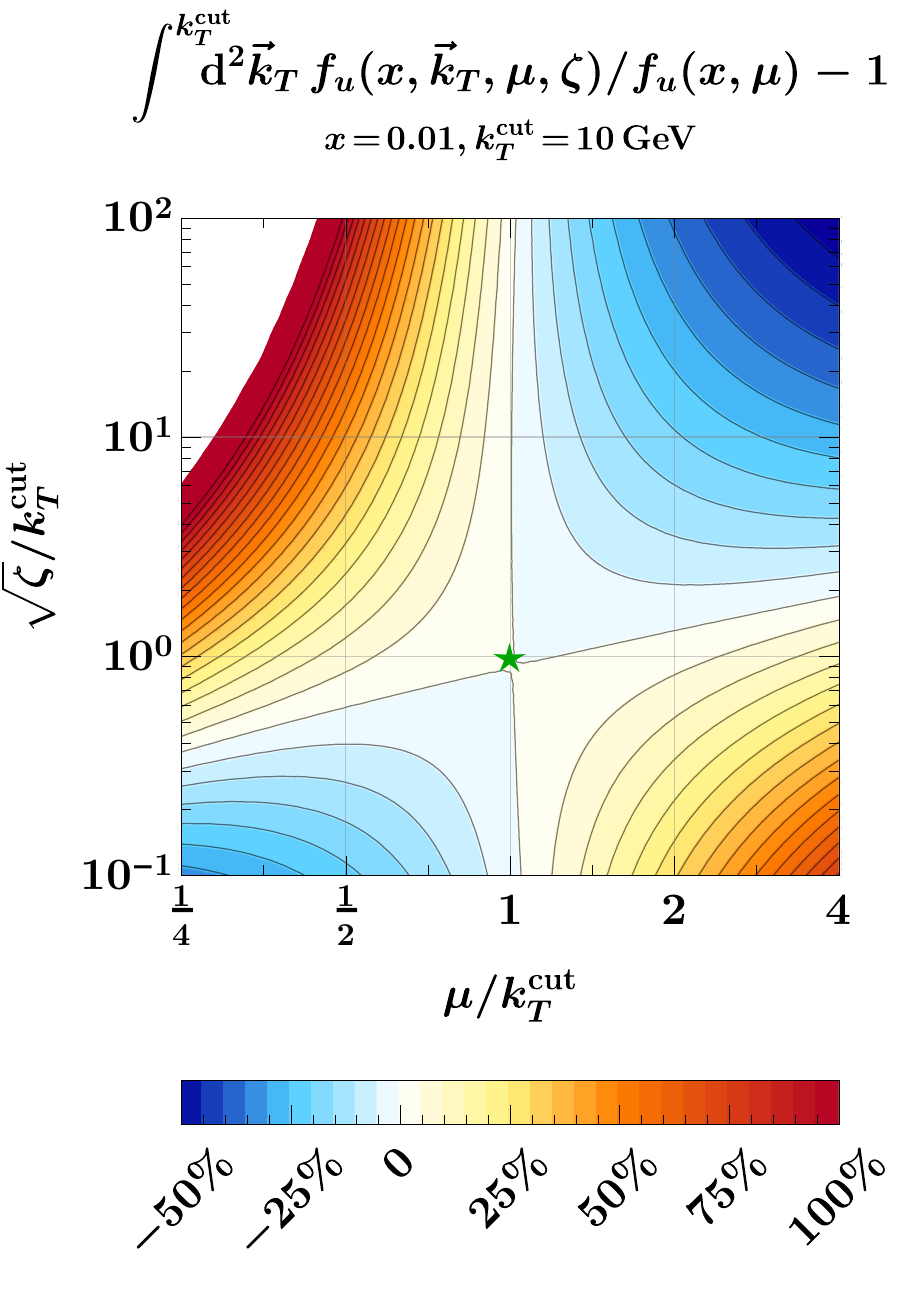}
 \hfill
 \includegraphics[width=\WidthTwoSubfigs]{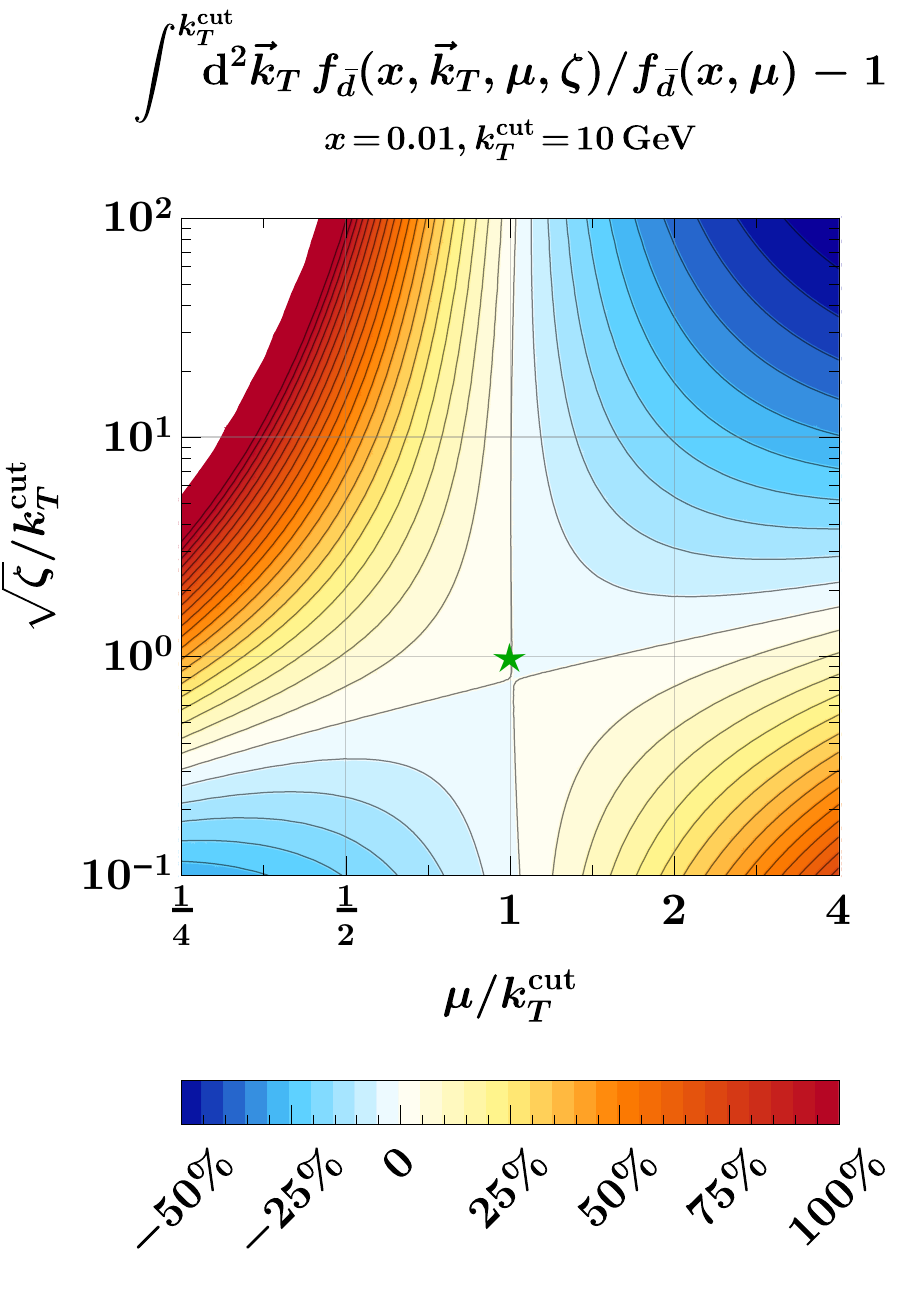}
 \\ 
 \includegraphics[width=\WidthTwoSubfigs]{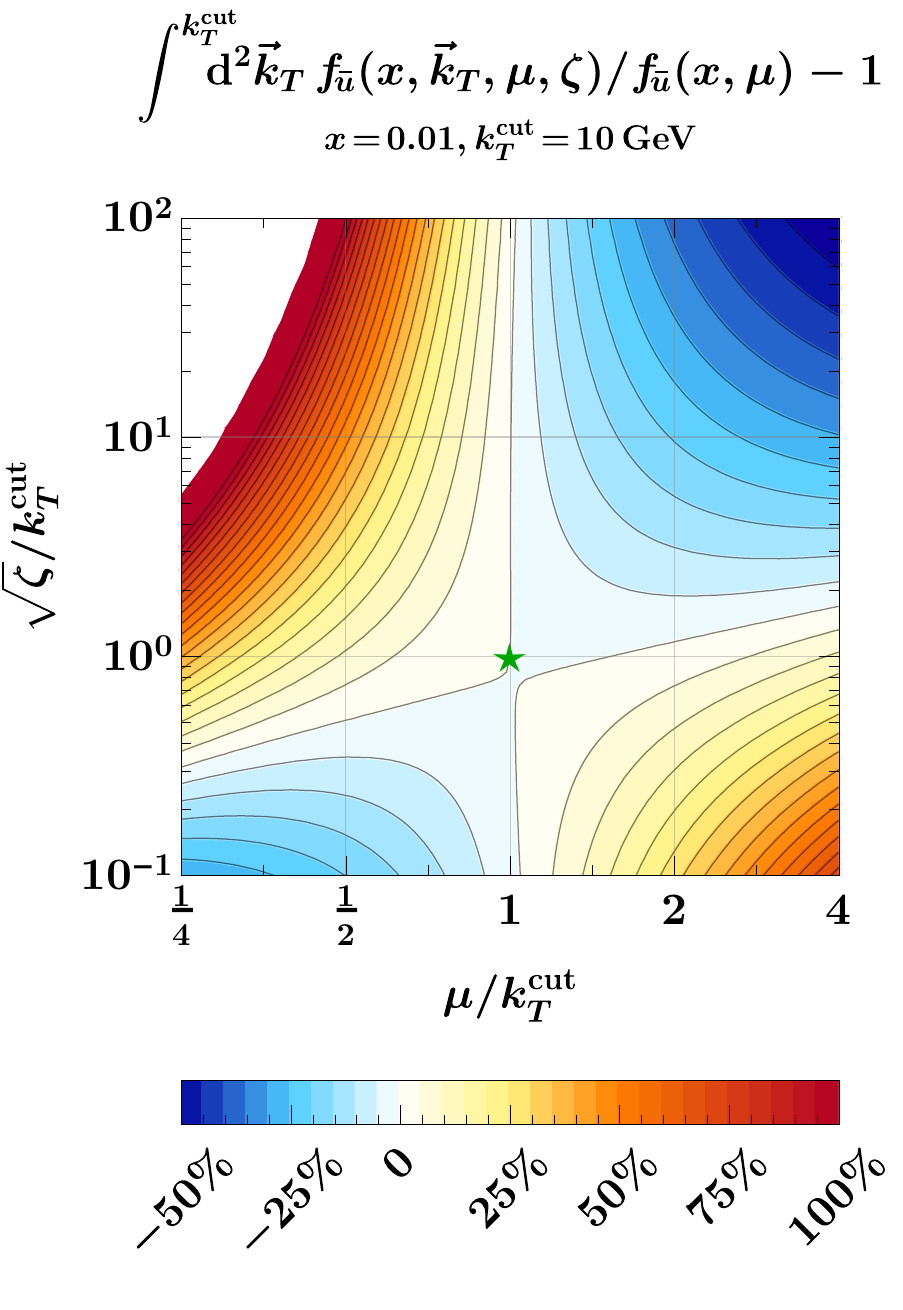}
 \hfill
 \includegraphics[width=\WidthTwoSubfigs]{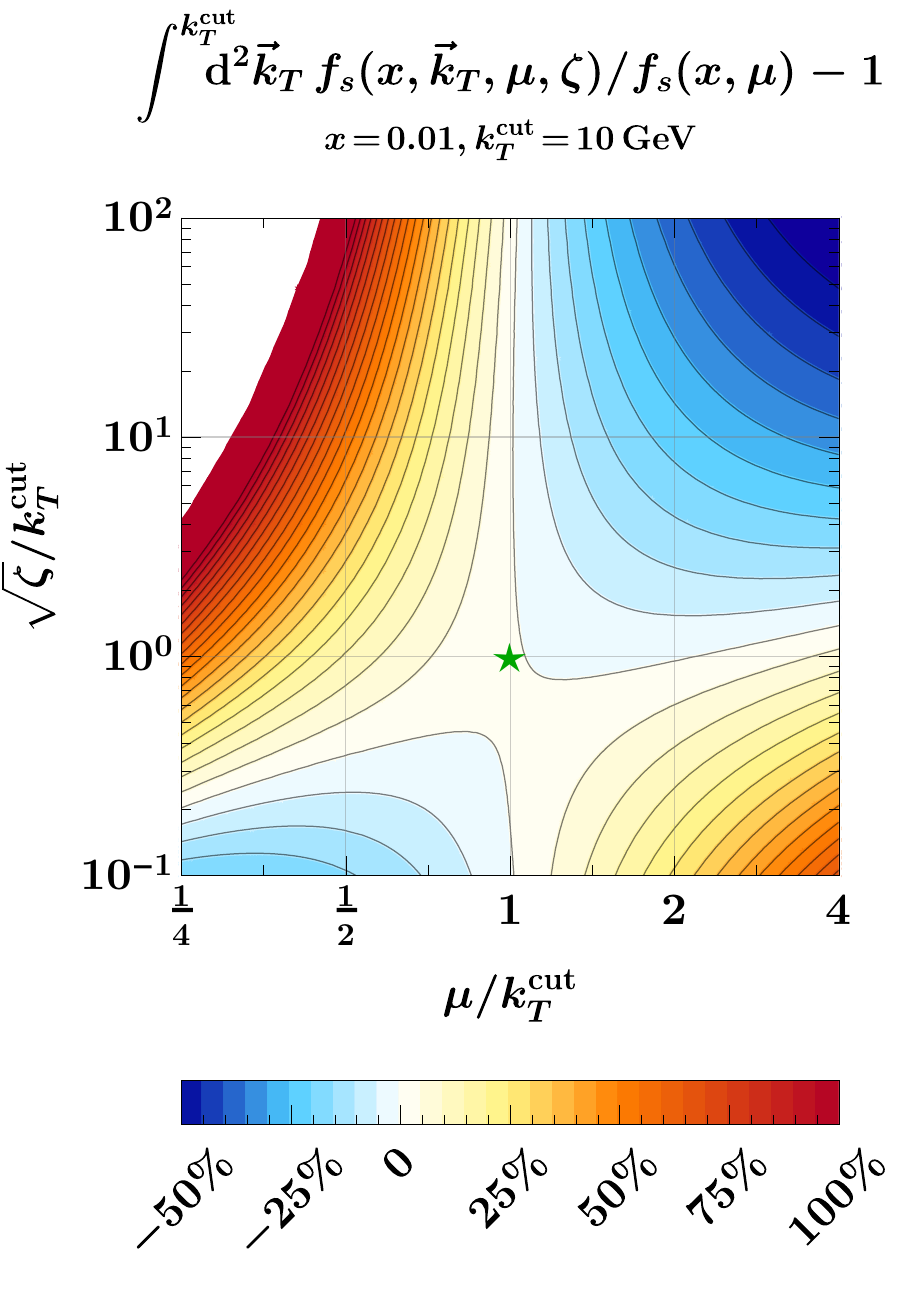}
 \caption{%
 Same as \fig{tmd_mu_zeta_heatmap}, for $i =u, \bar{d}, \bar{u}, s$.
 }
\label{fig:more_heatmap}
\end{figure}

\newpage
\begin{figure}
 \includegraphics[width=\WidthTwoSubfigs]{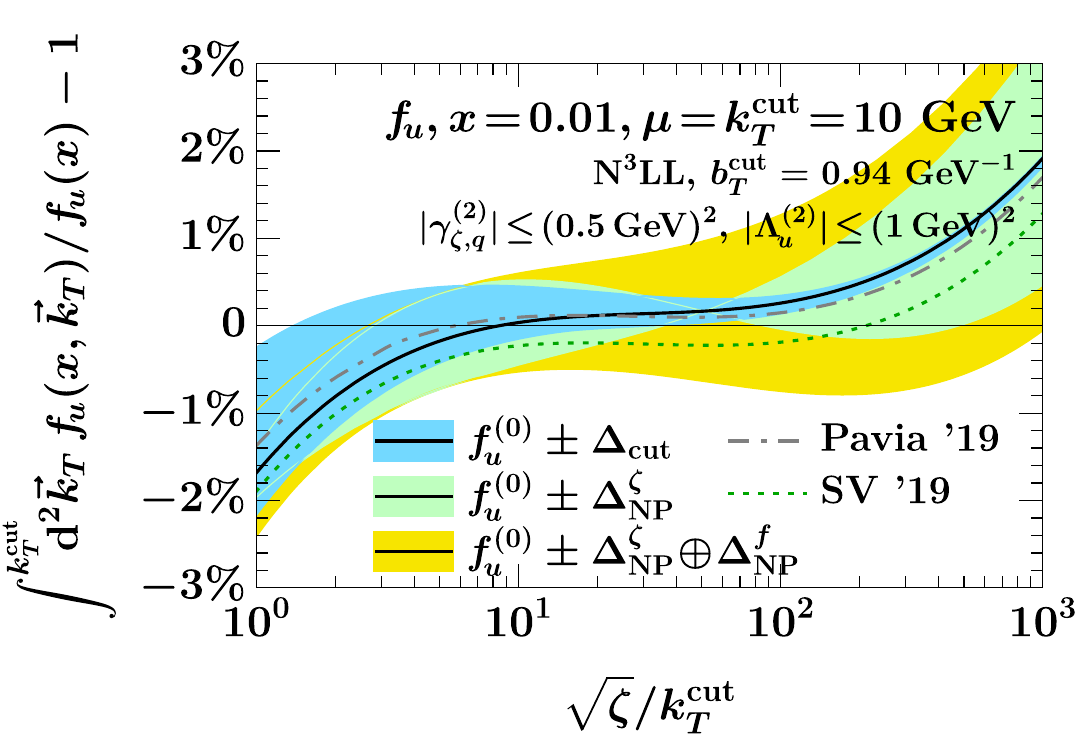}
  \hfill
 \includegraphics[width=\WidthTwoSubfigs]{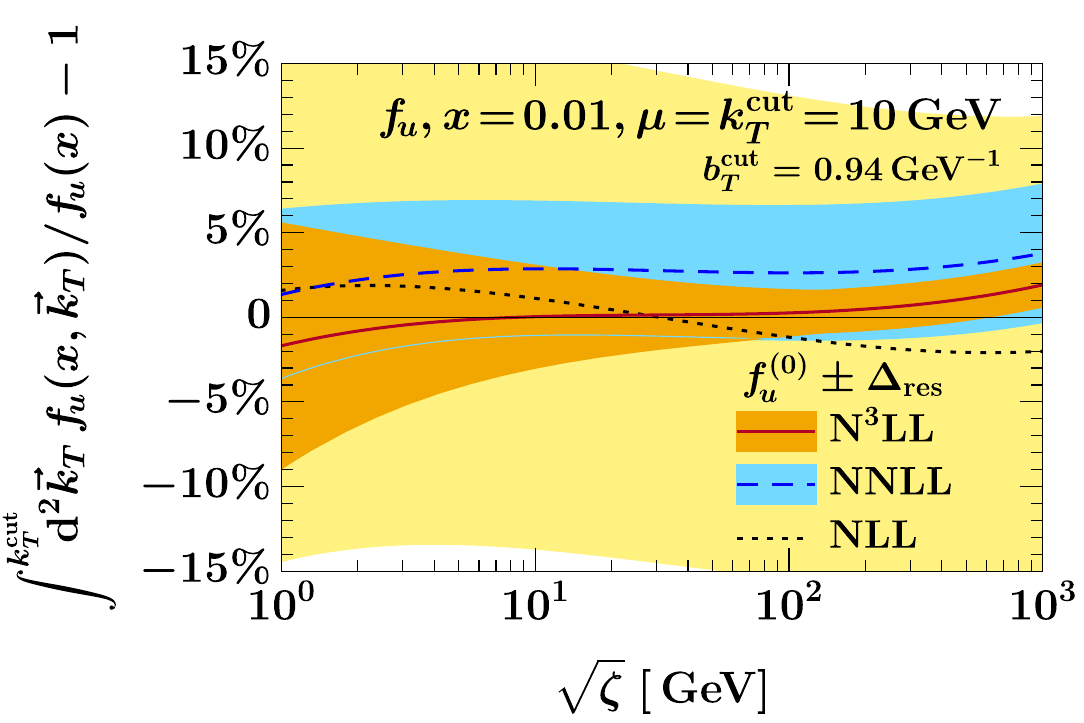}
 \\ 
 \includegraphics[width=\WidthTwoSubfigs]{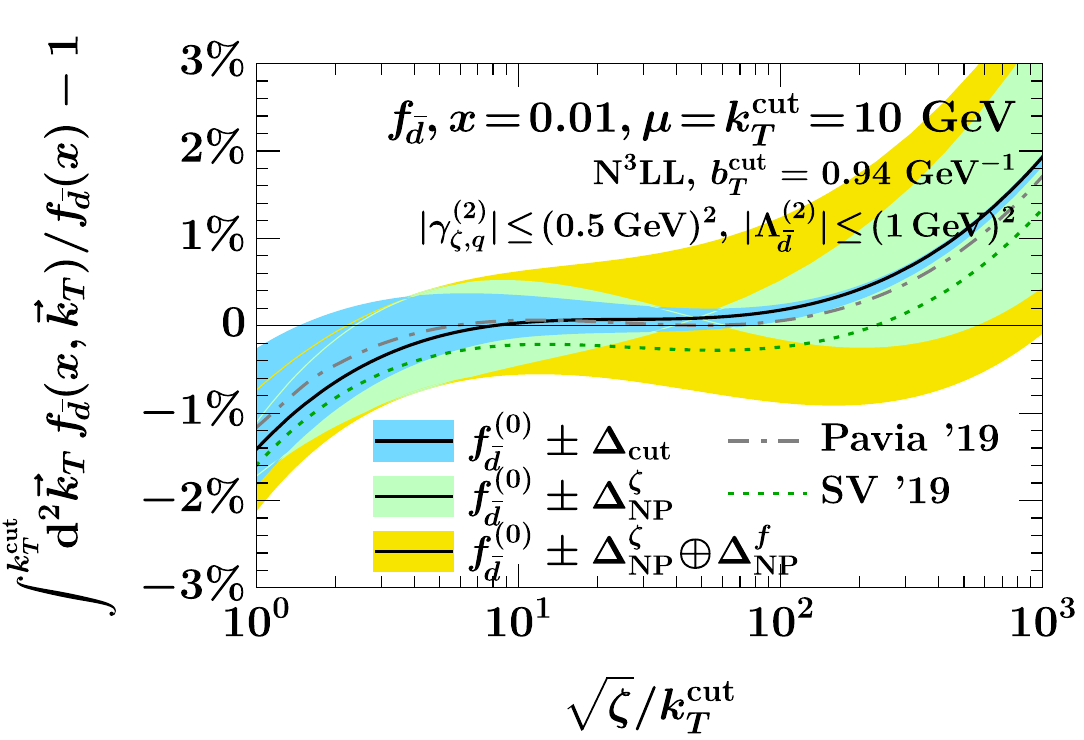}
  \hfill
 \includegraphics[width=\WidthTwoSubfigs]{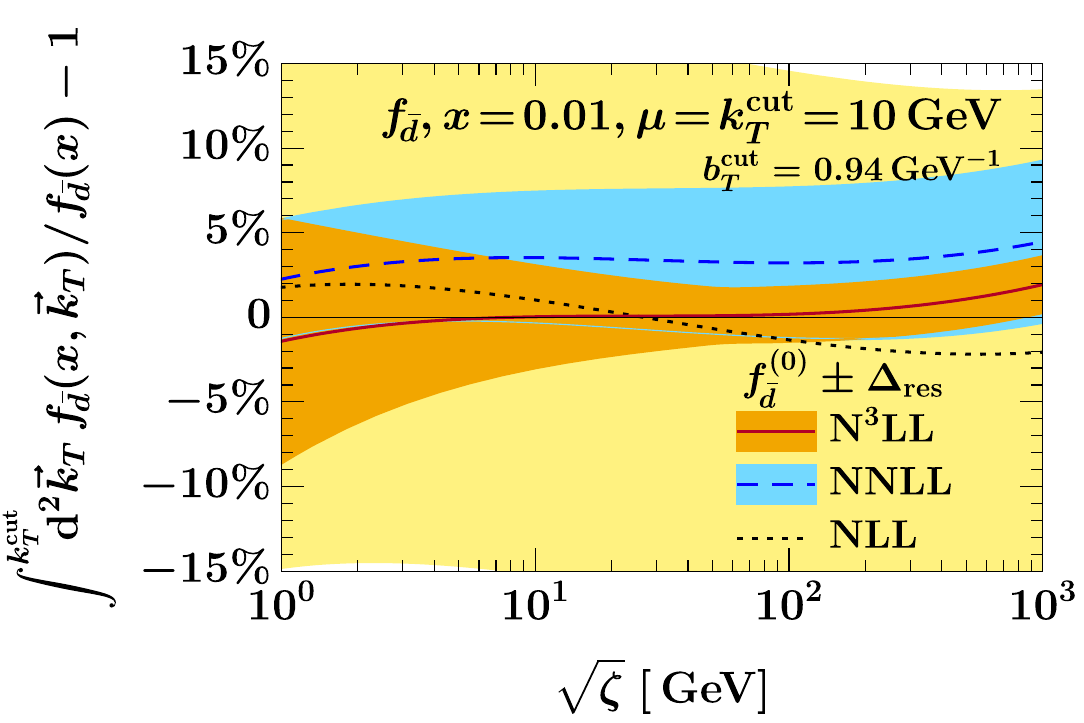} 
 \\
 \includegraphics[width=\WidthTwoSubfigs]{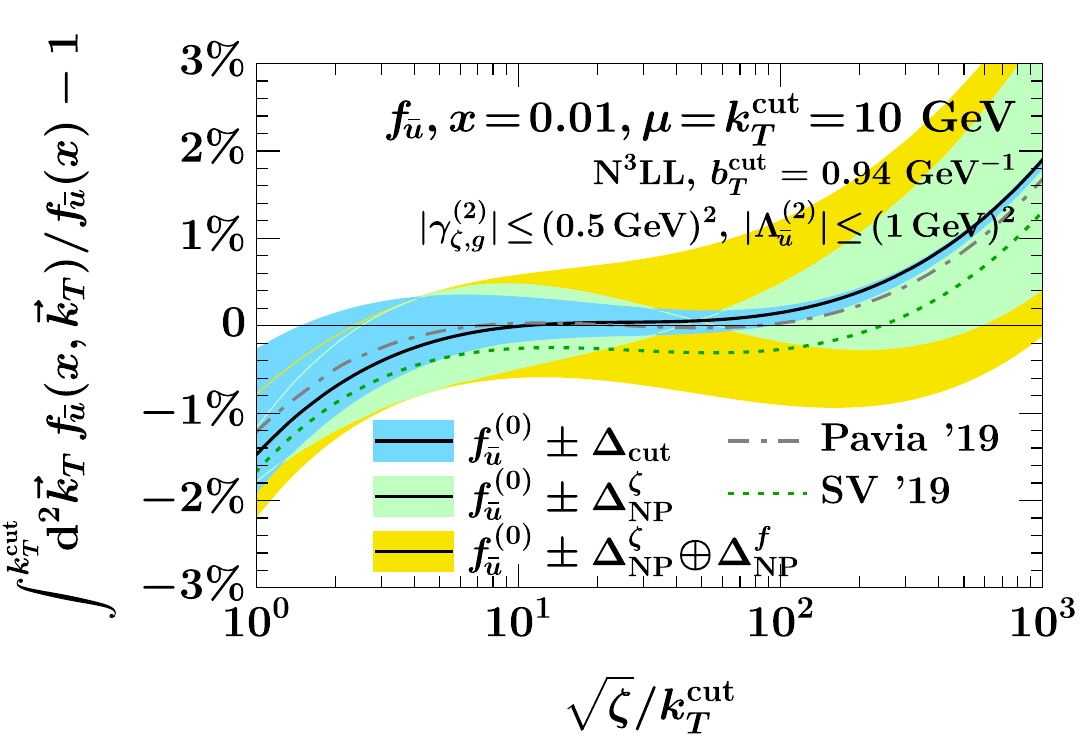}
  \hfill
 \includegraphics[width=\WidthTwoSubfigs]{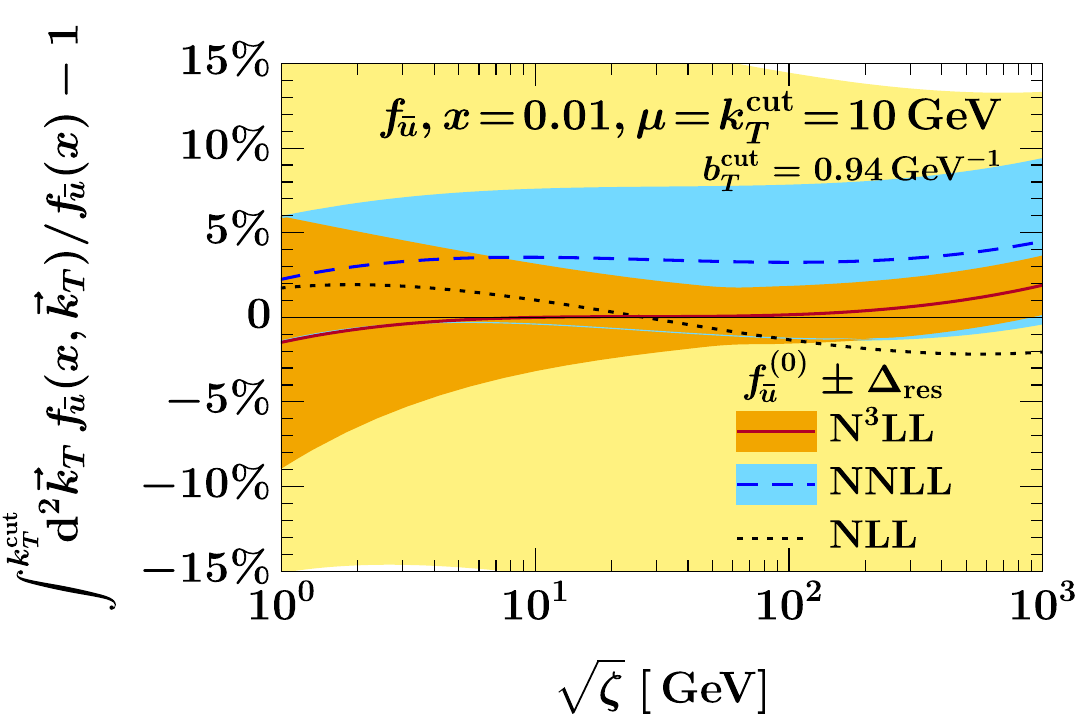}
 \\
 \includegraphics[width=\WidthTwoSubfigs]{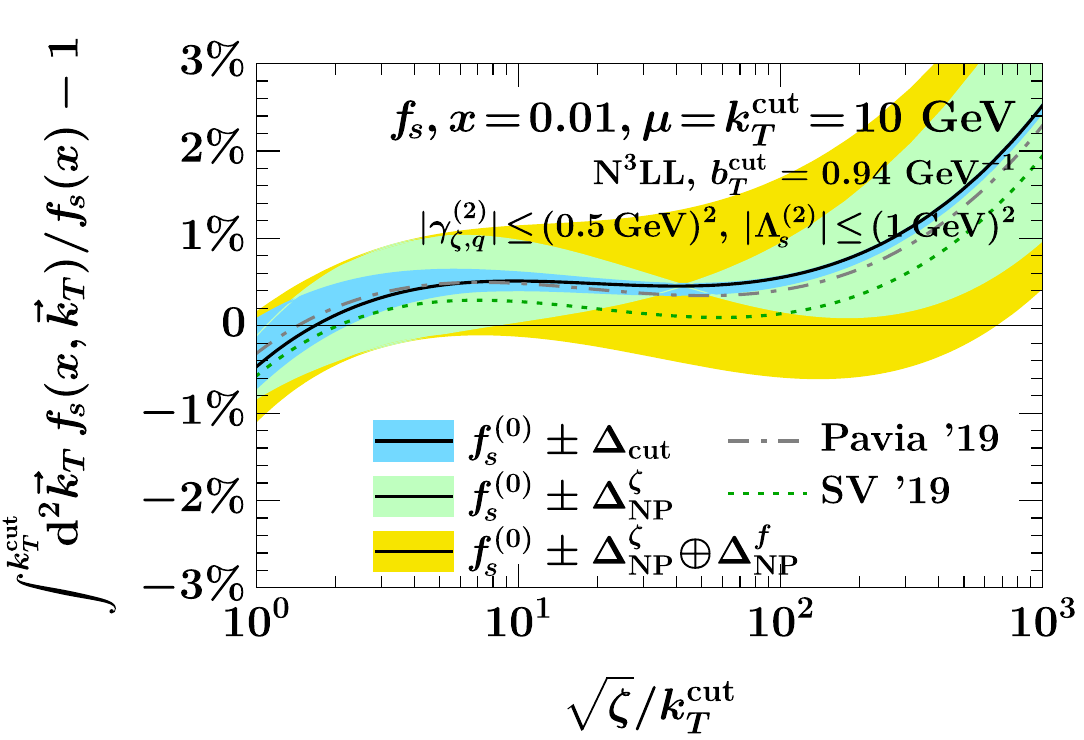}
  \hfill
  \includegraphics[width=\WidthTwoSubfigs]{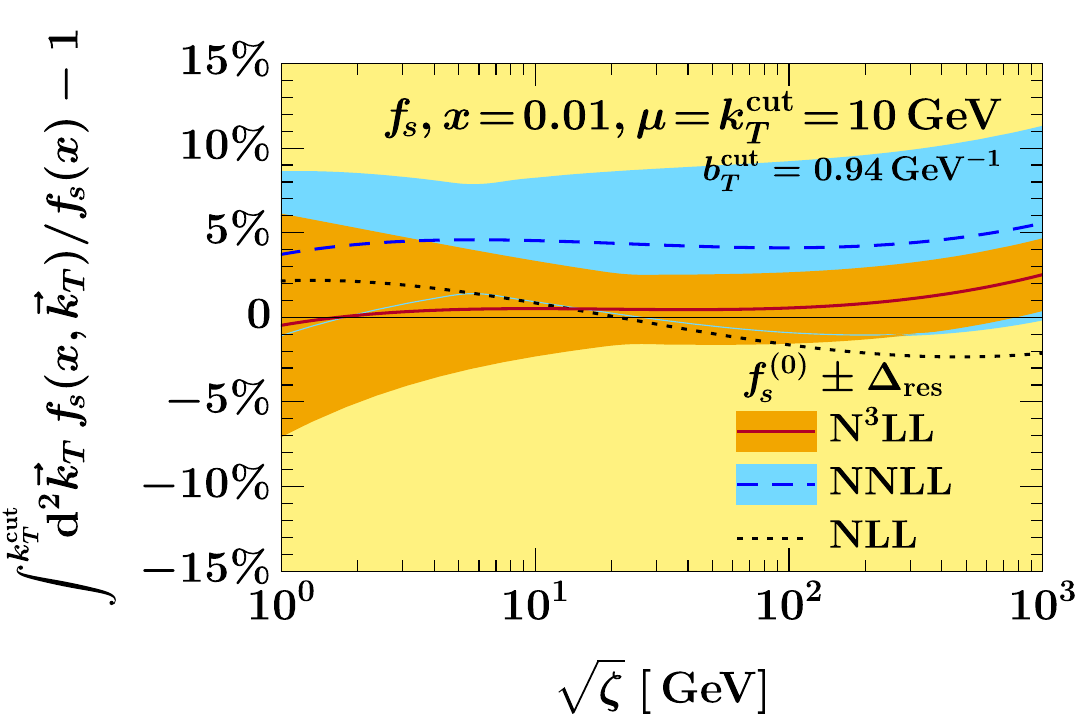}
 \\
 \caption{%
 Same as \fig{zeta_evolution_scale_np}, for $i = u, \bar{d}, \bar{u}, s$.
 }
\label{fig:more_zeta_evolution}
\end{figure}

\FloatBarrier
\addcontentsline{toc}{section}{References}
\bibliographystyle{jhep}
\bibliography{refs}

\end{document}